\documentclass[11pt,a4]{elsarticle}

\usepackage{setspace}
\usepackage{geometry}
\usepackage{url}
\usepackage{graphicx,graphics}
\usepackage{subfigure}
\usepackage{epsfig}
\usepackage{psfrag}
\usepackage{hyperref}
\usepackage{amsmath,amssymb,amsfonts,amsthm}
\usepackage{mathrsfs}
\usepackage{color}
\usepackage{float}
\usepackage{natbib}
\usepackage{tabularx}
\usepackage{multirow}

\usepackage{pgfplots}
\usetikzlibrary{plotmarks}

\theoremstyle{remark}

\newcommand{\Eref}[1]{Equation (\ref{#1})}
\newcommand{\fref}[1]{Figure (\ref{#1})}

\newcommand{\frefs}[1]{Figures~\ref{#1}}

\newcommand{\ff}{\mathbf{f}}
\newcommand{\KK}{\mathbf{K}}

\newcommand{\bveps}{\boldsymbol{\varepsilon}}
\newcommand{\bvsig}{\boldsymbol{\sigma}}

\begin{document}

\begin{frontmatter}


\title{Assessment of certain higher-order structural models based on global approach for bending analysis of curvilinear composite laminates}

\author[india,indiab]{Anand Venkatachari\fnref{fn1}}
\author[unsw]{S Natarajan \corref{cor1}\fnref{fn2}}
\author[indiab]{K Ramajeyathilagam}
\author[india]{M Ganapathi}

\cortext[cor1]{Corresponding author}

\address[india]{Tech Manhindra Ltd., Electronic City, Bangalore - 560100, India.}
\address[unsw]{School of Civil \& Environmental Engineering, The University of New South Wales, Sydney, NSW 2052, Australia}
\address[indiab]{School of Aeronautical Sciences, Hindustan University, Keelambakkam, Chennai-603103, India.}

\fntext[fn1]{Research Scholar, School of Aeronautical Sciences, Hindustan University, Keelambakkam, Chennai-603103, India.}
\fntext[fn2]{School of Civil and Environmental Engineering, University of New South Wales, Sydney, NSW 2052, Australia. Tel: +61 2 93855030, Email: snatarajan@cardiffalumni.org.uk; sundararajan.natarajan@gmail.com}

\begin{abstract}
In this paper, the performance of different structural models based on global approach in evaluating the static response of curvilinear fibre composite laminates is analyzed. A $\mathcal{C}^o$ shear flexible Quad-8 element developed based on higher-order structural theory is employed for the present study. The structural  theory accounting for the realistic variation of displacements through the thickness and the possible discontinuity in the slope at the interface is considered. Four alternate discrete structural models, deduced from the generic structural model by retaining various terms in the displacement functions are examined for their applicability. The accuracy of the present formulation is demonstrated considering the problems for which analytical solutions are available. A systematic numerical study, assuming different ply-angle and lay-up, is conducted in bringing out the influence of various structural models on the static response of composite laminates with curvilinear fibres.
\end{abstract}


\end{frontmatter}


\section{Introduction} The need for high strength-to and high stiffness-to-weight ratio materials has led to the development of laminated composite materials. This class of material has seen increasing utilization as structural elements and as primary structures of large-scale aerospace structures~\cite{botelhosilva2006}. This is because of the possibility to tailor the properties to optimize the structural response. Conventionally, the fibre reinforced composites have straight and unidirectional fibres. In this type of construction, the stiffness of the laminate does not vary in the domain. However, recently, the composite materials with varying stiffness has received greater interest, as they may lead to better and efficient design~\cite{hyerlee1991,setoodehabdalla2008,lopesgurdal2010,khaniijsselmuiden2011}. The stiffness of the composite material can be varied by: (a) using curvilinear fibres~\cite{hyerlee1991}; (b) varying the volume fraction or varying the fibre spacing~\cite{muruganflores2012}; (c) dropping or adding plies to the laminate~\cite{mukherjeevarughese2001} and (d) attaching discrete stiffeners to the laminates~\cite{mejdiatalla2012}. Among these concepts related to variable stiffness composite laminates (VSCL), the approach of curvilinear fibre does not introduce major geometry variations, it however does impose the constraint on the curvature of the fibre. In the following, we restrict ourselves to work related to composite laminates with curvilinear fibres. Laminates with curvilinear fibres offers a wider degree of possibilities than variations of rectilinear fibre volume fraction and provides a solution to the problem of continuity when manufacturing a structure with different fibre angles in adjoining element. Furthermore, it exhibits a way to diminish the stress concentration in cutouts. Hyer and Lee~\cite{hyerlee1991} introduced the concept of variable stiffness panels to improve the structural response of panels with holes. G\"urdal and Olmedo~\cite{gurdalolmedo1993} studied the in-plane response of VSCL. Although the concept of tailored fibre orientation was developed in the nineties, the design based on VSCL has recently spurred interest among researchers~\cite{abdallagurdal2009,akhavanribeiro2011,hondanarita2011,houmat2013,grohweaver2014,rajuwu2012,akhavanribeiro2013} due to the improvement in the manufacturing capability~\cite{kimpotter2012,kimweaver2014}. Kim \textit{et al.,}~\cite{kimpotter2012,kimweaver2014} have demonstrated a process to manufacture variable angle composites. When composite laminates are modelled as plate structures, with VSCL, the plate stiffness coefficients vary with spatial coordinates. Such laminates not only have variable in-plane stiffness, in general may possess variable bending and coupling stiffness. 

Abdallla \textit{et al.,}~\cite{abdallagurdal2009} by employing the classical lamination theory and the generalised reciprocal approximation, maximized the fundamental frequency of composite laminates. Their study showed that by employing VSCL significant increase in the fundamental frequency is achieved when compared to constant stiffness panels. This was later extended to study nonlinear dynamic response of VSCL. By using the $p-$version finite element, Akhavan and Ribeiro~\cite{akhavanribeiro2011} studied the fundamental frequencies and mode shapes of laminated composites with curvilinear fibres. It was inferred that VSCL introduces greater degree of flexibility in adjusting the frequencies and mode shapes. Honda and Narita~\cite{hondanarita2011,hondanarita2012}, studied the fundamental frequency of VSCL by employing classical plate theory. It was shown that the mode shapes of VSCL are significantly different from the constant stiffness laminated composites. Houmat~\cite{houmat2013} investigated the nonlinear free vibration of laminated rectangular plates with curvilinear fibres. Groh and Weaver~\cite{grohweaver2014} and Raju \textit{et al.,}~\cite{rajuwu2012,rajuwu2013} studied buckling of variable thickness curvilinear fibre panel using differential quadrature method. It was inferred that for VSCL shell-like description more accurately characterizes the buckling phenomenon than a plate-like description. A 2D analytical model based on equivalent single-layer formulation introduced recently in~\cite{grohweaver2013} in predicting the static response of curvilinear fibre laminates and the governing equations were solved by employing the differential quadrature method. Akhavan \textit{et al.,}~\cite{akhavanribeiro2013,akhavanribeiro2013a} employed the $p-$version finite element and studied the response of laminated composites with curvilinear fibres under static and dynamic loads. The plate kinematics was represented by third-order shear deformation theory. 

It is evident from the literature that the dynamic response of VCSL has received greater attention, whilst, the studies on the static characteristics of VSCL structural elements are limited. The existing studies employed either classical lamination theory, first order or the third order shear deformation theory. However, these theories have limitation when employed to study the response of curvilinear fibre laminated composites. More accurate analytical/numerical models based on the three-dimensional models can be computationally involved and expensive. A layer wise theory is a possible candidature for this purpose, but it may be computationally expensive as the number of unknowns to be solved increases with increasing number of mathematical or physical layers. Hence, among the researchers, there is a growing appreciation of the  importance of applying two-dimensional theories with new kinematics for accurate analysis. It is observed from the literature~\cite{murukami1986,alibhaskar1999,makhechaganapathi2001} that, for realistic structural analysis of laminated composites, higher-order theory with the inclusion of zig-zag function is necessary. The zig-zag function incorporated in the in-plane kinematics has been employed in~\cite{murukami1986,alibhaskar1999,makhechaganapathi2001,carrera2004} to study laminated composites. Makhecha \textit{et al.,}~\cite{makhechaganapathi2001} used higher-order accurate theory based on global approach for multilayered laminated composites by incorporating the realistic through the thickness approximations of the in-plane and transverse displacements by adding a zig-zag function and higher-order terms, respectively. A 8-noded quadrilateral element with 13 degrees of freedom per node was employed for the study. Recently, some researchers have attempted to combine the single layer and discrete layer theories to overcome the limitations of each one. Carrera~\cite{carrera2003} derived a series of axiomatic approaches for the general description of two-dimensional formulations for multilayered plates and shells. 
The formulation is a valuable tool for gaining a deep insight into the complex mechanics of laminated structures. In the same spirit, the main objective of this paper is to study the static response of laminated composites with curvilinear fibres. From the higher-order accurate theory, four alternate structural models are deduced by deleting the appropriate degree of freedom. A $\mathcal{C}^o$ shear flexbile quadrilateral 8-noded element is used for the present study.

The paper commences with a discussion on the variable stiffness composite laminates. Section \ref{highertheory} presents an overview of the higher-order accurate theory to describe the plate kinematics. The various structural models deduced from the higher-order accurate theory and the element employed to discretize the plate is discussed in Section \ref{eledes}. The influence of various structural models on the deflection and through thickness stress distribution is numerically studied for three layered symmetric and four layer anti-symmetric composite laminate with curvilinear fibers in Section \ref{numexamples}, followed by concluding remarks in the last section.

\section{Curvilinear fibre composite laminate}
\label{cfiber}
\begin{figure}[htpb]
\centering
\includegraphics[scale=0.6]{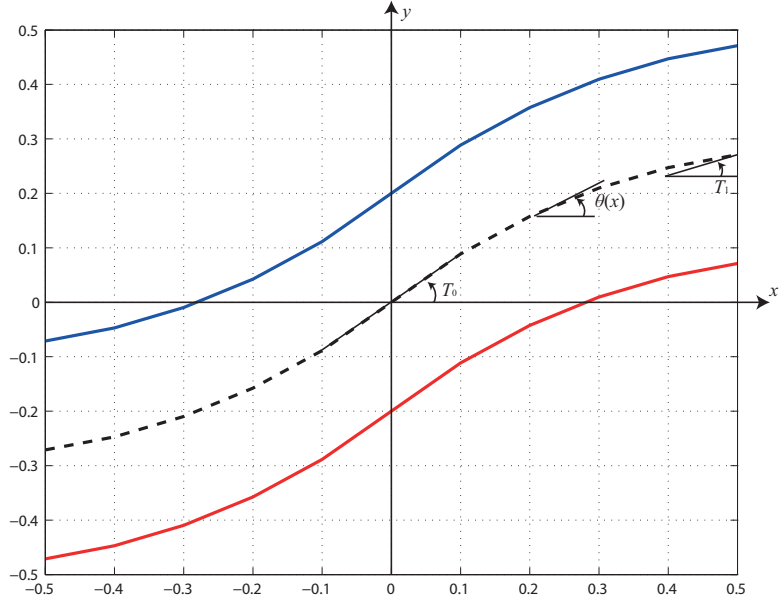}
\caption{Reference fibre path ('dashed line') and the shifted fibres ('solid line') for a laminae.}
\label{fig:reffibpath}
\end{figure}

In the conventional laminated composites, the fibres are straight and the stiffness of the laminate is constant in the in-plane direction. However in the current study we employ curvilinear fibres. In this type of construction, the fibre path is not straight, instead it is a function of the spatial coordinates, i.e., the fibre path cannot be described by a single orientation. The main advantage is that the resulting stiffness is a function of the spatial coordinate as well. It is assumed that the fibre path varies linearly with $x$ from a orientation $T_o$ at the center to $T_1$ at a distance $a/2$ from the center, where $a$ is the total length of the laminate. The orientation of a single fibre is denoted by $\phi \pm \langle T_o/T_1 \rangle$, where $\phi$ denotes the rotation of the fibre path with respect to the $x-$ axis. The reference fibre path is given by~\cite{houmat2013}:
\begin{equation}
\renewcommand{\arraystretch}{2}
y = \left\{ \begin{array}{ccc} \frac{a}{2(T_1-T_o)} \left\{ - {\rm ln} [\cos T_o] + {\rm ln} \left[ \cos \left(T_o - \frac{2(T_1-T_o)x}{a} \right) \right] \right\} & {\rm for} & -a/2 \le x \le 0 \\ \frac{a}{2(T_1-T_o)} \left\{ {\rm ln} [\cos T_o] - {\rm ln} \left[ \cos \left(T_o + \frac{2(T_1-T_o)x}{a} \right) \right] \right\} & {\rm for} & 0 \le x \le a/2 \end{array} \right.
\end{equation}
\fref{fig:reffibpath} shows a reference fibre path based on the above equation and the shifted fibres. In this study, the fibre orientation is assumed to be just a function of $x$ and is given by~\cite{houmat2013}:
\begin{equation}
\renewcommand{\arraystretch}{2}
\theta(x) = \left\{ \begin{array}{ccc} \frac{-2}{a}(T_1 - T_o)x + T_o & ~~\textup{for} & ~~ -a/2 \le x \le 0 \\
\frac{2}{a}(T_1 - T_o)x + T_o & ~~\textup{for} & ~~ 0 \le x \le a/2
\end{array} \right.
\label{eqn:anglevari}
\end{equation}
where $T_0$ and $T_1$ are the fibre angles at the ply centre, $x=0$ and at the ends $x=\pm a/2$, respectively and is denoted by $<T_0,T_1>$.

The curvature $\kappa$ is given by:
\begin{equation}
\kappa = \frac{ \frac{d^2y}{dx^2}}{ \left[ 1 + \left( \frac{dy}{dx} \right)^2 \right]^{3/2}}
\label{eqn:cur}
\end{equation}

\begin{figure}[htpb]
\centering
\includegraphics[scale=0.6]{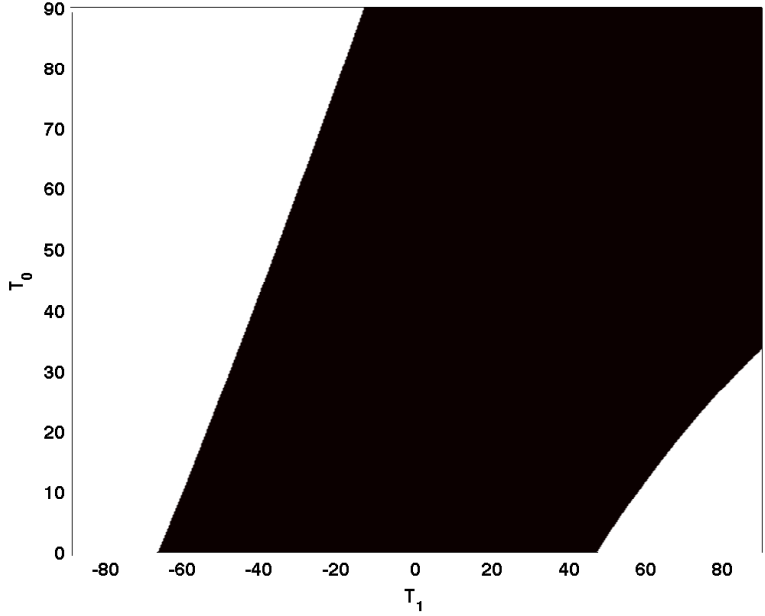}
\caption{Curvilinear fibre composite laminate: design space. The `dark' shaded region depicts the fibre path that satisfies the curvature constraint. The values of fibre orientation, $T_o$ and $T_1$ are in degrees.}
\label{fig:curvConst}
\end{figure}

The curvature of the fibre path in the positive $x$ axis is found by substituting the first and the second derivatives of the fibre path in~\Eref{eqn:cur}. At each point along the fibre path, the curvature must be less than the maximum prescribed value of 3.28 m$^{-1}$~\cite{waldhar1996,houmat2013}. \fref{fig:curvConst} shows the design space for curvilinear fibre composite laminates. If at any spatial coordinate, the fibre curvature constraint is not satisfied, the corresponding fiber orientation is discarded.

\section{Higher order accurate theory} \label{highertheory}
A curvilinear fibre composite plate is considered with the coordinates $x,y$ along the in-plane directions and $z$ along the thickness direction. The in-plane displacements $u^sk$ and $v^k$, and the transverse displacement $w^k$ for the $k^{th}$ layer, are assumed as~\cite{alibhaskar1999,makhechaganapathi2001}:
\begin{align}
u^k(x,y,z,t) &= u_o(x,y,t) + z \theta_x(x,y,t) + z^2 \beta_x(x,y,t) + z^3 \phi_x(x,y,t) + S^k \psi_x(x,y,t)  \nonumber \\
v^k(x,y,z,t) &= v_o(x,y,t) + z \theta_y(x,y,t) + z^2 \beta_y(x,y,t) + z^3 \phi_y(x,y,t) + S^k \psi_y(x,y,t)  \nonumber \\
w^k(x,y,z,t) &= w_o(x,y,t) + z w_1(x,y,t) + z^2 \Gamma(x,y,t)
\label{eqn:dispField}
\end{align}
The terms with even powers of $z$ in the in-plane displacements and odd powers of $z$ occurring in the expansion for $w^k$ correspond to the stretching problem. However, the terms with odd powers of $z$ in the in-plane displacements and the even ones in the expression for $w^k$ represent the flexure problem. $u_o, v_o$ and $w_o$ are the displacements of a generic point on the reference surface; $\theta_x$ and $\theta_y$ are the rotations of the normal to the reference surface about the $y$ and $x$ axes, respectively; $w_1,\beta_x,\beta_y,\Gamma,\phi_x$ and $\phi_y$ are the higher-order terms in the Taylor`s series expansions, defined at the reference surface. $\psi_x$ and $\psi_y$ are generalized variables associated with the zigzag function, $S^k$. The zigzag function, $S^k$, as given in~\cite{murukami1986,makhechaganapathi2001,roderiguesroque2011}, is defined by
\begin{equation}
S^k = 2(-1)^k \frac{z_k}{h_k}
\end{equation}
where $z_k$ is the local transverse coordinate with the origin at the center of the $k^{th}$ layer and $h_k$ is the corresponding layer thickness. Thus, the zigzag function is piecewise linear with values of -1 and 1 alternatively at different interfaces. The `zig-zag' function, as defined above, takes care of the inclusion of the slope discontinuities of $u$ and $v$ at the interfaces of the laminated plate as observed in the exact three-dimensional elasticity solutions of thick laminated composite plates. The main advantage of using such formulation is that such a function is more economical than a discrete layer approach~\cite{nosierkapania1993,ferreira2005}. The strains in terms of mid-plane deformation, rotations of normal and higher-order terms associated with displacements are:
\begin{equation}
\bveps = \left \{ \begin{array}{c} \bveps_{\rm bm} \\ \bveps_s \end{array} \right\} - \left \{ \begin{array}{c} \overline{\bveps}_t \\ \mathbf{0} \end{array} \right\}
\label{eqn:strain}
\end{equation}
The vector  $\bveps_{\rm bm}$ includes the bending and membrane terms of the strain components, vector $\bveps_s$ contains the transverse shear strain terms and $\overline{\bveps}_t$ contains the vector of thermal strains. These strain vectors are defined as:
\begin{eqnarray}
\bveps_{\rm bm} &=& \left\{ \begin{array}{c} \varepsilon_{xx} \\ \varepsilon_{yy} \\ \varepsilon_{zz} \\ \gamma_{xy} \end{array} \right\} + \left\{ \begin{array}{c} u_{,x} \\ v_{,y} \\ w_{,z} \\ u_{,y} + v_{,x} \end{array} \right\} \nonumber \\
&=& \bveps_0 + z \bveps_1 + z^2 \bveps_2 + z^3 \bveps_3 + S^k \bveps_4 
\end{eqnarray}
\begin{eqnarray}
\bveps_s &=& \left\{ \begin{array}{c} \gamma_{xz} \\ \gamma_{yz} \end{array} \right\} = \left\{ \begin{array}{c} u_{,z} + w_{,x} \\ v_{,z} + w_{,y} \end{array} \right\} \nonumber \\
&=& \gamma_o + z \gamma_1 + z^2 \gamma_2 + S^k_{,z} \gamma_3
\end{eqnarray}
where
\begin{align}
\bveps_o &= \left\{ \begin{array}{c} u_{o,x} \\ v_{o,y} \\ w_1 \\ u_{o,y} + v_{o,x} \end{array} \right\},\hspace{1cm} \bveps_1 = \left\{ \begin{array}{c} \theta_{x,x} \\ \theta_{y,y} \\ 2\Gamma \\ \theta_{x,y} + \theta_{y,x} \end{array} \right\}, \nonumber \\
\bveps_2 &= \left\{ \begin{array}{c} \beta_{x,x} \\ \beta_{y,y} \\ 0 \\ \beta_{x,y} + \beta_{y,x} \end{array} \right\}, \hspace{1cm} \bveps_3 = \left\{ \begin{array}{c} \phi_{x,x} \\ \phi_{y,y} \\ 0 \\ \phi_{x,y} + \phi_{y,x} \end{array} \right\}, \nonumber \\
\bveps_4 &= \left\{ \begin{array}{c} \psi_{x,x} \\ \psi_{y,y} \\ 0 \\ \psi_{x,y} + \psi_{y,x} \end{array} \right\}.
\end{align}
and,
\begin{align}
\gamma_o &= \left\{ \begin{array}{c} \theta_x + w_{o,x} \\ \theta_y + w_{o,y} \end{array} \right\}, \hspace{1cm} \gamma_1 = \left\{ \begin{array}{c} 2\beta_x + w_{1,x} \\ 2\beta_y + w_{1,y} \end{array} \right\}, \nonumber \\
\gamma_2 &= \left\{ \begin{array}{c} 3\phi_x + \Gamma_{,x} \\ 3\phi_y + \Gamma_{,y} \end{array} \right\}, \hspace{1cm} \gamma_3 = \left\{ \begin{array}{c} \psi_x S_{,z}^k \\ \psi_y S_{,z}^k \end{array} \right\} .
\end{align}
The subscript comma denotes partial derivatives with respect to the spatial coordinate succeeding it. The thermal strain vector, $\overline{\bveps}_t$ is given by:
\begin{align}
\overline{\bveps}_t &= \left\{ \begin{array}{cccccc} \varepsilon_{xx} & \varepsilon_{yy} & \varepsilon_{zz} & \varepsilon_{xy} & \varepsilon_{yz} & \varepsilon_{xz} \end{array} \right\}^{\rm T} \nonumber \\
&= \Delta T \left\{ \begin{array}{cccccc} \alpha_{xx} & \alpha_{yy} & \alpha_{zz} & \alpha_{xy} & 0 & 0 \end{array} \right\}^{\rm T}
\end{align}
where $\Delta T$ is the rise in the temperature and is generally represented as a function of the spatial coordinates, $\alpha_{xx}, \alpha_{yy}, \alpha_{zz}$ and $\alpha_{xy}$ are thermal expansion coefficients in the plate coordinates and can be related to the thermal expansion coefficients in the material principal directions. The constitutive relations for an arbitrary layer $k$ can be expressed as:
\begin{eqnarray}
\bvsig &=& \left\{ \begin{array}{cccccc} \sigma_{xx} & \sigma_{yy} & \sigma_{zz} & \tau_{xz} & \tau_{yz} & \tau_{xy} \end{array} \right\}^{\rm T} \nonumber \\
&=& \bf{\overline{Q}}^k \left\{ \begin{array}{cc} \bveps_{\rm bm} & \bveps_s \end{array} \right\}^{\rm T} - \bf{\overline{Q}}^k \overline{\bveps}_t
\end{eqnarray}
where the terms of $\bf{\overline{Q}}^k $ of $k^{th}$ ply are referred to the laminate axes and can be obtained from the $\bf{Q}^k$ matrix corresponding to the fibre directions with the appropriate transformations. The transformed stiffnesses are constant in a straight fibre laminate whilst for a curvilinear fibre laminate, it is a function of spatial coordinates $(x,y)$. The strain energy function $U$ is given by
\begin{equation}
U( \boldsymbol{\delta}) = \frac{1}{2} \iint \left[ \sum_{k=1}^n \int\limits_{h_k}^{h_{k+1}} \bvsig^{\rm T} \bveps ~dz \right] dx dy - \iint {\bf q} w~dxdy
\label{eqn:potential}
\end{equation}
where $\boldsymbol{\delta}$ is the vector of degrees of freedom and ${\bf q}$ is the distributed force acting on the top surface of the plate. By following the Galerkin procedure outlined in~\cite{hughes2000}, one obtains the following governing equations for static deflection 

\begin{equation}
\KK \boldsymbol{\delta} = \ff_m + \ff_t
\label{eqn:staticdefl}
\end{equation}
where $\KK$ is the global stiffness matrix and $\ff_m$ and $\ff_t$ are the global mechanical and thermal load vectors, respectively.

\section{Element description}
\label{eledes}
In this paper, $\mathcal{C}^o$ continuous, eight-noded serendipity quadrilateral shear flexible plate element is used. The finite element represented as per the kinematics based on \Eref{eqn:dispField} is referred to as HSDT13 with cubic variation. The 13 dofs are: $(u_o,v_o,w_o,\theta_x,\theta_y,w_1,\beta_x,\beta_y,\Gamma,\phi_x,\phi_y,\psi_x,\psi_y)$. Four more alternate discrete models are proposed to study the influence of higher-order terms in the displacement functions, whose displacement fields are deduced from the original element by deleting the appropriate degrees of freedom. These structural models, and the corresponding degrees of freedom are listed in Table \ref{table:alternatemodels}

\begin{table} [htpb]
\renewcommand\arraystretch{1.5}
\caption{Structural models employed in this study for describing the plate kinematics.}
\centering
\begin{tabular}{ll}
\hline
Finite element model & Degrees of freedom per node  \\
\hline
HSDT13 & $u_o,v_o,w_o,\theta_x,\theta_y,w_1,\beta_x,\beta_y,\Gamma,\phi_x,\phi_y,\psi_x,\psi_y$ \\
HSDT11A & $u_o,v_o,w_o,\theta_x,\theta_y,\beta_x,\beta_y,\phi_x,\phi_y,\psi_x,\psi_y$ \\
HSDT11B & $u_o,v_o,w_o,\theta_x,\theta_y,w_1,\beta_x,\beta_y,\Gamma,\phi_x,\phi_y$ \\
TSDT7 & $u_o,v_o,w_o,\theta_x,\theta_y,\beta_x,\beta_y$ \\
FSDT5 & $u_o,v_o,w_o,\theta_x,\theta_y$ \\
\hline
\end{tabular}
\label{table:alternatemodels}
\end{table}


\section{Numerical results and discussion}
\label{numexamples}
In this section, the static bending response of curvilinear fibre composite laminated plates using the eight-noded shear flexible quadrilateral element is presented. The effect of various parameters, such as the plate thickness, the ply-angle and the lay-up sequence on the global response is numerically studied considering mechanical and thermal loads. Here, the laminate is assumed to be simply supported with movable boundary conditions. For the higher-order structural theory (HSDT13), the boundary conditions are given by:

\begin{eqnarray}
u_o = w_o = \theta_x = w_1 = \Gamma = \beta_x = \phi_x = \psi_x = 0, \hspace{0.2cm} ~\textup{on} ~ y = 0,b \nonumber \\
v_o = w_o = \theta_y = w_1 = \Gamma = \beta_y = \phi_y = \psi_y = 0, \hspace{0.2cm} ~\textup{on} ~ x= 0,a
\end{eqnarray}
where $a$ and $b$ refer to the length and width of the plate, respectively. The transverse shear stresses are evaluated by integrating the three-dimensional equilibrium equations for all types of elements. The material properties, unless otherwise specified, used in the present analysis are:

\begin{eqnarray}
E_L/E_T = 25, \hspace{10pt} G_{LT}/E_T = 0.5, \hspace{10pt} G_{TT}/E_T = 0.2 \nonumber \\
\nu_{LT}=0.25, \hspace{10pt} \alpha_T/\alpha_L = 1125, \hspace{10pt} \alpha_L = 10^{-6} \nonumber \\
E_T = 10^9 {\rm GPa}
\end{eqnarray}
where $E,G$ and $\nu$ are Young's modulus, shear modulus and Poisson's ratio, respectively. $L$ and $T$ are the longitudinal and transverse directions, respectively, with respect to fibres. All the layers are of equal thickness and the ply angle is measures with respect to the $x$-axis. For the present study, the following two types of loadings are considered:
\begin{itemize}
\item Mechanical loading: $q(x,y) = q_o \sin \frac{\pi x}{a} \sin \frac{\pi y}{b}$,
\item Thermal loading: $\Delta T(x,y) = \overline{T}_o \left( \frac{2z}{h} \right) \sin \frac{\pi x}{a} \sin \frac{\pi y}{b}$
\end{itemize}
where $q_o$ and $\overline{T}_o$ are the amplitude of the mechanical load and the thermal load, respectively. The physical quantities are non-dimensionalized by relations, unless stated otherwise:
\begin{align}
\renewcommand{\arraystretch}{2}
(\overline{u},\overline{v}) &= \frac{100E_{_T}}{q_o h S^3}(u,v) \nonumber \\
\overline{w} &= \frac{100E_T}{q_o h S^4}w \nonumber \\
\overline{\sigma}_{xx} &= \frac{\sigma_{xx}}{q_oS^2} \nonumber \\
\overline{\tau}_{xz} &= \frac{\tau_{xz}}{q_oS}
\end{align}
for the applied mechanical load and by
\begin{align}
\renewcommand{\arraystretch}{2}
(\overline{u},\overline{v}) &= \frac{E_T}{ h \alpha_L \overline{T}_o S}(u,v) \nonumber \\
\overline{w} &= \frac{E_T}{h \alpha_L \overline{T}_o h S^2}w\nonumber \\
\overline{\sigma}_{xx} &= \frac{\sigma_{xx}}{ E_T \alpha_L \overline{T}_o} \nonumber \\
\overline{\tau}_{xz} &= \frac{\tau_{xz}}{ E_T \alpha_L \overline{T}_o}
\end{align}
for the applied thermal load, where the $E_{T}, \alpha_{L}$ are the Young's modulus and the co-efficient of thermal expansion of the laminated composite in the transverse and longitudinal directions, respectively, and $S=a/h$. Considering different types of four-layered curvilinear fibre laminates, the convergence study is carried out by decreasing the mesh size and the results are presented in Table \ref{table:curvfibMeshconv} for both thick and thin plates ($a/h=$ 5 and $a/h=$ 50) by employing both the first- and higher-order (FSDT5, HSDT13) structural models. A very good convergence of the results is observed with decreasing element size. For the problem considered here, a structured quadrilateral mesh with 6$\times$6 mesh is found to be adequate to model the full plate.

\begin{table} [htpb]
\renewcommand\arraystretch{1.2}
\caption{Convergence study with mesh size for a four-layered curvilinear fibre composite plate under sinusoidally distributed pressure load.}
\centering
\begin{tabular}{ccrrrrrrrr}
\hline
$a/h$ & $T_o/T_1$ & Mesh size & \multicolumn{6}{c}{Structural model}\\
\cline{4-10}
 &  $(^\circ)$&  & \multicolumn{3}{c}{HSDT13} && \multicolumn{3}{c}{FSDT5} \\
\cline{4-6}\cline{8-10}
 &  &  & $\overline{w}$ & $\overline{\tau}_{xz}$ & $\overline{\sigma}_{xx}$ && $\overline{w}$ & $\overline{\tau}_{xz}$ & $\overline{\sigma}_{xx}$ \\
\hline
\multirow{15}{*}{5} & \multirow{5}{*}{$[0 \pm \langle 45/0 \rangle]_s$} & 2$\times$2 & 1.039 & 0.524 & 0.155 && 0.960 & 0.493 & 0.176 \\
&& 4$\times$4 & 1.277 & 0.296 & 0.360 && 1.189 & 0.333 & 0.252 \\
&& 6$\times$6 & 1.300 & 0.234 & 0.372 && 1.202 & 0.259 & 0.263 \\
&& 8$\times$8 & 1.305 & 0.198 & 0.378 && 1.204 & 0.210 & 0.268 \\
&& 16 $\times$16 & 1.305 & 0.196 & 0.379 && 1.204 & 0.210 & 0.270 \\
\cline{3-10}
& \multirow{5}{*}{$[0 \pm \langle 0/45 \rangle]_s$} & 2$\times$2 & 0.851 & -0.161 & 0.539 && 0.850 & -0.134 & 0.393 \\
&& 4$\times$4 & 1.179 & 0.153 & 0.372 && 1.125 & 0.254 & 0.321 \\
&& 6$\times$6 & 1.204 & 0.182 & 0.408 && 1.138 & 0.248 & 0.314 \\
&& 8$\times$8 & 1.210 & 0.228 & 0.414 && 1.141 & 0.298 & 0.313 \\
&& 16 $\times$16 & 1.211 & 0.232 & 0.416 && 1.142 & 0.302 & 0.311\\
\cline{3-10}
& \multirow{5}{*}{$[0 \pm \langle 90/0 \rangle]_s$} & 2$\times$2 &0.918 & 0.750 & 0.065 && 0.861 & 0.692 & 0.053 \\
&& 4$\times$4 & 1.267 & -0.037 & 0.096 && 1.204 & 0.022 & 0.093 \\
&& 6$\times$6 & 1.307 & -0.070 & 0.102 && 1.239 & 0.033 & 0.097 \\
&& 8$\times$8 & 1.318 & -0.080 & 0.103 && 1.248 & 0.032 & 0.095 \\
&& 16 $\times$16 & 1.319 & -0.085 & 0.104 && 1.248 & 0.031 & 0.092 \\
\cline{2-10} 
\multirow{15}{*}{50} & \multirow{5}{*}{$[0 \pm \langle 45/0 \rangle]_s$} & 2$\times$2 & 0.466 & 0.417 & 0.149 && 0.434 & 0.374 & 0.244 \\
&& 4$\times$4 & 0.444 & 0.289 & 0.301 && 0.443 & 0.299 & 0.275 \\
&& 6$\times$6 & 0.445 & 0.285 & 0.277 && 0.443 & 0.283 & 0.277 \\
&& 8$\times$8 & 0.445 & 0.273 & 0.287 && 0.444 & 0.275 & 0.281 \\
&& 16 $\times$16 & 0.445 & 0.270 & 0.291 && 0.444 & 0.272 & 0.280 \\
\cline{3-10}
& \multirow{5}{*}{$[0 \pm \langle 0/45\rangle]_s$} & 2$\times$2 & 0.296 & 0.150 & 0.426 && 0.287 & 0.154 & 0.407 \\
&& 4$\times$4 &  0.322 & 0.207 & 0.384 && 0.319 & 0.260 & 0.390 \\
&& 6$\times$6 & 0.324 & 0.235 & 0.391 && 0.321 & 0.241 & 0.381 \\
&& 8$\times$8 & 0.325 & 0.249 & 0.376 && 0.322 & 0.299 & 0.378 \\
&& 16 $\times$16 & 0.325 & 0.252 & 0.372 && 0.323 & 0.303 & 0.375 \\
\cline{3-10}
& \multirow{5}{*}{$[0 \pm \langle 90/0 \rangle]_s$} & 2$\times$2 & 0.413 & 0.475 & 0.036 && 0.388 & 0.428 & 0.033 \\
&& 4$\times$4 &  0.445 & 0.068 & 0.047 && 0.439 & 0.073 & 0.049 \\
&& 6$\times$6 & 0.454 & 0.097 & 0.062 && 0.451 & 0.087 & 0.061 \\
&& 8$\times$8 & 0.454 & 0.057 & 0.064 && 0.452 & 0.064 & 0.063 \\
&& 16 $\times$16 & 0.455 & 0.055 & 0.065 && 0.452 & 0.060 & 0.064 \\
\hline
\end{tabular}
\label{table:curvfibMeshconv}
\end{table}

\begin{table} [htpb]
\renewcommand\arraystretch{1.2}
\caption{Comparison of deflection and stresses for a three-layered cross ply simply supported square plate subjected to sinusoidally distributed mechanical load against elasticity solution~\cite{bhaskarvaradan1996}.}
\centering
\begin{tabular}{crrrrrr}
\hline
$a/h$ &  & $\overline{u}(-a/2,0,h/2)$ & $\overline{v}(0,-a/2,h/2)$ & $\overline{w}(0,0,h/2)$ & $\overline{\sigma}_{xx}(0,0,h/2)$ & $\overline{\tau}_{xz}(-a/2,0,0)$ \\
\hline
\multirow{4}{*}{4} & HSDT13 & -0.977 & -2.275 & 2.094 & 0.825 & 0.245 \\
& HSDT11A & -0.977 & -2.332 & 2.027 & 0.805 & 0.247 \\
& HSDT11B & -0.967 & -2.178 & 1.993 & 0.816 & 0.270 \\
& Elasticity~\cite{bhaskarvaradan1996}& -0.969 & -2.281 & 2.006 & 0.755 & 0.256 \\
\cline{2-7}
\multirow{4}{*}{10} & HSDT13 & -0.745 & -1.103 & 0.752 & 0.611 & 0.344 \\
& HSDT11A & -0.750 & -1.112 & 0.755 & 0.613 & 0.345 \\
& HSDT11B & -0.736 & -1.049 & 0.715 & 0.604 & 0.353 \\
& Elasticity~\cite{bhaskarvaradan1996}& -0.735 & -1.099 & 0.753 & 0.590 & 0.357 \\
\cline{2-7}
\multirow{4}{*}{20} & HSDT13 & -0.699 & -0.796 & 0.516 & 0.569 & 0.373 \\
& HSDT11A & -0.700 & -0.798 & 0.517 & 0.570 & 0.373 \\
& HSDT11B & -0.696 & -0.779 & 0.505 & 0.567 & 0.376 \\
& Elasticity~\cite{bhaskarvaradan1996}& -0.693 & -0.794 & 0.516 & 0.553 & 0.383 \\
\cline{2-7}
\multirow{4}{*}{50} & HSDT13 & -0.682 & -0.697 & 0.445 & 0.555 & 0.384 \\
& HSDT11A & -0.682 & -0.697 & 0.445 & 0.555 & 0.384 \\
& HSDT11B & -0.681 & -0.694 & 0.443 & 0.554 & 0.385 \\
& Elasticity~\cite{bhaskarvaradan1996}& -0.680 & -0.697 & 0.445 & 0.541 & 0.393 \\
\hline
\end{tabular}
\label{table:3L_mech_validation}
\end{table}

\begin{table} [htpb]
\renewcommand\arraystretch{1.2}
\caption{Comparison of deflection and stresses for a three-layered cross ply simply supported square plate subjected to sinusoidally distributed thermal load against elasticity solution~\cite{bhaskarvaradan1996}.}
\centering
\begin{tabular}{crrrrrr}
\hline
$a/h$ &  & $\overline{u}(-a/2,0,h/2)$ & $\overline{v}(0,-a/2,h/2)$ & $\overline{w}(0,0,h/2)$ & $\overline{\sigma}_{xx}(0,0,h/2)$ & $\overline{\tau}_{xz}(-a/2,0,-h/6)$ \\
\hline
\multirow{4}{*}{4} & HSDT13 & -18.069 & -81.010 & 42.331 & 1216.860 & 82.926 \\
& HSDT11A & -14.457 & -77.187 & 26.222 & 919.206 & 85.649 \\
& HSDT11B &  -18.396 & -75.714 & 42.040 & 1239.820 & 87.841 \\
& Elasticity~\cite{bhaskarvaradan1996}& -18.110 & -81.830 & 42.330 & 1183.000 & 84.810 \\
\cline{2-7}
\multirow{4}{*}{10} & HSDT13 & -16.738 & -31.687 & 17.374 & 1067.370 & 58.920 \\
& HSDT11A & -16.046 & -30.991 & 14.656 & 1009.980 & 58.621 \\
& HSDT11B & -16.689 & -29.978 & 16.901 & 1062.050 & 60.113 \\
& Elasticity~\cite{bhaskarvaradan1996}& -16.610 & -31.950 & 17.390 & 1026.000 & 60.540 \\
\cline{2-7}
\multirow{4}{*}{20} & HSDT13 & -16.255 & -20.183 & 12.116 & 1018.480 & 33.164 \\
& HSDT11A & -16.076 & -20.003 & 11.428 & 1003.610 & 32.855 \\
& HSDT11B & -16.229 & -19.694 & 11.961 & 1016.050 & 33.365 \\
& Elasticity~\cite{bhaskarvaradan1996}& -16.170 & -20.340 & 12.120 & 982.00 & 33.980 \\
\cline{2-7}
\multirow{4}{*}{50} & HSDT13 & -16.039 & -16.651 & 10.493 & 998.252 & 13.807 \\
& HSDT11A & -16.010 & -16.622 & 10.383 & 995.838 & 13.663 \\
& HSDT11B & -16.032 & -16.569 & 10.466 & 997.641 & 13.823 \\
& Elasticity~\cite{bhaskarvaradan1996}& -16.020 & -16.710 & 10.500 & 967.500 & 14.070 \\
\hline
\end{tabular}
\label{table:3L_therm_validation}
\end{table}

\begin{table}[htpb]
\centering
\caption{Comparison of different structural models in predicting the deflection of a four-layered curvilinear fibres composite laminate $([0 \pm \langle 45/0 \rangle]_s)$ under uniform pressure load against available solutions~\cite{grohweaver2013}.}
\begin{tabular}{rrrrr}
\hline
 & \multicolumn{4}{c}{$a/h$} \\
 \cline{2-5}
 & 5 & 10 & 20 & 50 \\
\hline
HSDT13 & -3.823 & -2.393 & -1.995 & -1.882 \\
HSDT11A & -3.730 & -2.317 & -1.930 & -1.814 \\
HSDT11B & -3.822 & -2.393 & -1.995 & -1.882 \\
TSDT7 & -3.368 & -2.214 & -1.899 & -1.807 \\
FSDT5 &  -3.671 & -2.298 & -1.922 & -1.811 \\
2D equivalent~\cite{grohweaver2013} & -4.002 & -2.396 & -1.943 & -1.805 \\
FEM S8R~\cite{grohweaver2013} & -3.761 & -2.311 & -1.913 & -1.796 \\
FEM C3D20R~\cite{grohweaver2013} & -4.630 & -2.520 & -2.000 & -1.820 \\
\hline
\end{tabular}
\label{table:curvFibLamValida}
\end{table}

Before proceeding with the detailed investigation, the efficacy of the present formulation is validated considering the static response of cross-ply straight fibre laminates subjected to mechanical and thermal loads. The results obtained with the present formulation are shown in Tables \ref{table:3L_mech_validation} - \ref{table:3L_therm_validation} for different plate side to thickness ratios. It is observed that the results from the present formulation agree very well with the results available in the literature~\cite{bhaskarvaradan1996}. Next, the various structural models proposed here are tested considering curvilinear fibre composite laminates subjected to mechanical load for which results are available in the literature~\cite{grohweaver2013}. The maximum transverse displacements calculated are tabulated in Table \ref{table:curvFibLamValida} for various plate side to thickness ratios. It can be observed from this Table that the present results are found to be in excellent agreement with those reported in the literature~\cite{grohweaver2013} and it may be further inferred that the performance of the present HSDT13 structural model is better than the analytical model and 2D element employed in~\cite{grohweaver2013}.

A systematic parametric study is further conducted to examine the suitability of an appropriate structural theory using different structural models deduced from the present formulation as given in Table \ref{table:alternatemodels}. The computed results pertaining to maximum displacements and stresses for different plate side to thickness ratios and curvilinear fibre angles $( \langle T_o/T_1 \rangle )$ are highlighted in Tables \ref{table:Table6symmMech} - \ref{table:Table7symmTherm}. For the present study, a three-layered symmetric laminate configuration is considered. It is observed from the Tables \ref{table:Table6symmMech} - \ref{table:Table7symmTherm} that, for the mechanical load case, the higher-order model, HSDT11A, is in close agreement with HDST13 in predicting the displacements, whereas, for the thermal loading case, the higher-order model HSDT11B is in close agreement with the HSDT13. However, depending on the curvilinear ply-angle and the plate aspect ratio, the stress values predicted either by HSDT11A or HSDT11B are close to the values of full model, HSDT13. Also, it may be opined that the influence of lower-order theories, HSDT7 and FSDT5 underestimates the deflection and are somewhat different from those of HSDT13 and HSDT11 (HSDT11A and HSDT11B). The displacements and the stress distribution through the thickness are shown in \frefs{fig:3LayerMechLoad} - \ref{fig:3LayerThermLoad} for thick laminate with curvilinear ply-angle $( \langle$ 45$^\circ$/-45$^\circ \rangle)$. The in-plane displacement is continuous through the thickness whereas the transverse displacement varies quadratically through the thickness. Also, it can be seen that the slope of the in-plane displacements shows discontinuity at the interface as expected. From Tables \ref{table:3L_mech_validation}-\ref{table:3L_therm_validation} and Tables \ref{table:Table6symmMech} - \ref{table:Table7symmTherm}, one can opine that, for the straight fibre laminates, either HSDT11A or HSDT11B may in general be employed depending on the loading types (mechanical or thermal), however, for curvilinear case, HSDT13 may have to be used for accurate results. Furthermore, it is observed that the differences in the performance among various theories are more for curvilinear case compared to the straight fibre case. It can be further noted that the lower order models (HSDT7 and FSDT5) cannot predict the through thickness variation of displacements.

\begin{table}[htpb]
\centering
\caption{Maximum non-dimensionalized deflections and stresses for a three-layered symmetric curvilinear fibre composite panles under sinusoidally distributed mechanical load. The influence of plate thickness ratio, the fibre orientation and structural model is illustrated.} 
\begin{tabular}{crrrrrr}
\hline
 $a/h$ & Fibre angle $(^\circ)$ & Structural & $\overline{u}(-a/2,0,h/2)$ & $\overline{w}(0,0,h/2)$ & $\overline{\sigma}_{xx}(0,0,h/2)$ & $\overline{\tau}_{xz}(-a/2,0,0)$ \\
\hline
\multirow{15}{*}{5} & \multirow{5}{*}{$[0 \pm \langle -45/45 \rangle]_s$} &	HSDT13	&	-1.236	&	1.458	&	0.352	&	-0.086	\\
&&	HSDT11A	&	-1.235	&	1.448	&	0.345	&	-0.091	\\
&&	HSDT11B	&	-1.170	&	1.424	&	0.350	&	-0.068	\\
&&	TSDT7	&	-0.961	&	1.260	&	0.247	&	-0.054	\\
&&	FSDT5	&	-0.993	&	1.399	&	0.243	&	-0.074	\\
\cline{3-7}
&\multirow{5}{*}{$[0 \pm \langle -45/30 \rangle]_s$} &	HSDT13	&	-1.265	&	1.467	&	0.394	&	-0.107	\\
&&	HSDT11A	&	-1.264	&	1.458	&	0.387	&	-0.110	\\
&&	HSDT11B	&	-1.214	&	1.441	&	0.394	&	-0.148	\\
&&	TSDT7	&	-0.995	&	1.269	&	0.278	&	-0.139	\\
&&	FSDT5	&	-1.024	&	1.403	&	0.274	&	-0.167	\\
\cline{3-7}
&\multirow{5}{*}{$[0 \pm \langle -45/15 \rangle]_s$} &	HSDT13	&	-1.229	&	1.451	&	0.416	&	-0.047	\\
&&	HSDT11A	&	-1.227	&	1.442	&	0.410	&	-0.051	\\
&&	HSDT11B	&	-1.198	&	1.432	&	0.416	&	-0.126	\\
&&	TSDT7	&	-0.976	&	1.252	&	0.298	&	-0.112	\\
&&	FSDT5	&	-1.003	&	1.384	&	0.294	&	-0.135	\\
\cline{2-7}
\multirow{15}{*}{10} & \multirow{5}{*}{$[0 \pm \langle -45/45 \rangle]_s$} &	HSDT13	&	-0.829	&	0.757	&	0.328	&	0.028	\\
&&	HSDT11A	&	-0.831	&	0.762	&	0.329	&	0.023	\\
&&	HSDT11B	&	-0.803	&	0.740	&	0.327	&	0.064	\\
&&	TSDT7	&	-0.747	&	0.692	&	0.280	&	0.081	\\
&&	FSDT5	&	-0.770	&	0.738	&	0.275	&	0.067	\\
\cline{3-7}
&\multirow{5}{*}{$[0 \pm \langle -45/30 \rangle]_s$} &	HSDT13	&	-0.894	&	0.788	&	0.362	&	0.023	\\
&&	HSDT11A	&	-0.897	&	0.794	&	0.362	&	0.021	\\
&&	HSDT11B	&	-0.872	&	0.773	&	0.358	&	0.027	\\
&&	TSDT7	&	-0.811	&	0.722	&	0.310	&	0.060	\\
&&	FSDT5	&	-0.830	&	0.765	&	0.306	&	0.037	\\
\cline{3-7}
&\multirow{5}{*}{$[0 \pm \langle -45/15 \rangle]_s$} &	HSDT13	&	-0.899	&	0.791	&	0.371	&	0.055	\\
&&	HSDT11A	&	-0.901	&	0.797	&	0.372	&	0.054	\\
&&	HSDT11B	&	-0.886	&	0.779	&	0.367	&	0.023	\\
&&	TSDT7	&	-0.822	&	0.726	&	0.323	&	0.051	\\
&&	FSDT5	&	-0.838	&	0.767	&	0.320	&	0.032	\\
\cline{2-7}
\multirow{15}{*}{50} & \multirow{5}{*}{$[0 \pm \langle -45/45 \rangle]_s$} &	HSDT13	&	-0.608	&	0.447	&	0.333	&	0.179	\\
&&	HSDT11A	&	-0.608	&	0.448	&	0.333	&	0.177	\\
&&	HSDT11B	&	-0.604	&	0.445	&	0.331	&	0.198	\\
&&	TSDT7	&	-0.603	&	0.442	&	0.329	&	0.204	\\
&&	FSDT5	&	-0.605	&	0.444	&	0.328	&	0.197	\\
\cline{3-7}
&\multirow{5}{*}{$[0 \pm \langle -45/30 \rangle]_s$} &	HSDT13	&	-0.695	&	0.497	&	0.354	&	0.227	\\
&&	HSDT11A	&	-0.695	&	0.497	&	0.354	&	0.226	\\
&&	HSDT11B	&	-0.691	&	0.494	&	0.350	&	0.244	\\
&&	TSDT7	&	-0.688	&	0.491	&	0.350	&	0.254	\\
&&	FSDT5	&	-0.689	&	0.493	&	0.349	&	0.247	\\
\cline{3-7}
&\multirow{5}{*}{$[0 \pm \langle -45/15 \rangle]_s$} &	HSDT13	&	-0.732	&	0.524	&	0.352	&	0.190	\\
&&	HSDT11A	&	-0.732	&	0.525	&	0.352	&	0.190	\\
&&	HSDT11B	&	-0.730	&	0.523	&	0.349	&	0.189	\\
&&	TSDT7	&	-0.726	&	0.520	&	0.348	&	0.190	\\
&&	FSDT5	&	-0.728	&	0.522	&	0.348	&	0.188	\\
\hline
\end{tabular}
\label{table:Table6symmMech}
\end{table}

\begin{table}[htpb]
\centering
\caption{Maximum non-dimensionalized deflections and stresses for a three-layered symmetric curvilinear fibre composite panels under sinusoidally distributed thermal load. The influence of plate thickness ratio, the fibre orientation and structural model is illustrated.}
\begin{tabular}{crrrrrr}
\hline
 $a/h$ & Fibre angle $(^\circ)$ & Structural & $\overline{u}(-a/2,0,h/2)$ & $\overline{w}(0,0,h/2)$ & $\overline{\sigma}_{xx}(0,0,h/2)$ & $\overline{\tau}_{xz}(-a/2,0,-h/6)$ \\
\hline
\multirow{15}{*}{5} & \multirow{5}{*}{$[0 \pm \langle -45/45 \rangle]_s$} &	HSDT13	&	-27.950	&	34.367	&	6.221	&	-244.644	\\
&&	HSDT11A	&	-25.290	&	23.968	&	-90.977	&	-244.895	\\
&&	HSDT11B	&	-26.522	&	33.741	&	-14.033	&	-231.722	\\
&&	TSDT7	&	-21.954	&	21.679	&	-203.089	&	-227.654	\\
&&	FSDT5	&	-22.120	&	22.513	&	-223.965	&	-234.282	\\
\cline{3-7}
&\multirow{5}{*}{$[0 \pm \langle -45/30 \rangle]_s$} &	HSDT13	&	-29.665	&	35.595	&	160.573	&	-315.784	\\
&&	HSDT11A	&	-27.032	&	25.227	&	60.893	&	-314.015	\\
&&	HSDT11B	&	-28.365	&	35.105	&	149.233	&	-321.485	\\
&&	TSDT7	&	-23.465	&	22.699	&	-84.625	&	-303.010	\\
&&	FSDT5	&	-23.639	&	23.600	&	-107.980	&	-315.750	\\
\cline{3-7}
&\multirow{5}{*}{$[0 \pm \langle -45/15 \rangle]_s$} &	HSDT13	&	-28.904	&	35.332	&	252.264	&	-273.893	\\
&&	HSDT11A	&	-26.331	&	24.978	&	151.906	&	-269.167	\\
&&	HSDT11B	&	-27.907	&	35.083	&	244.267	&	-274.903	\\
&&	TSDT7	&	-23.315	&	22.501	&	-5.351	&	-250.441	\\
&&	FSDT5	&	-23.528	&	23.427	&	-28.973	&	-262.651	\\
\cline{2-7}
\multirow{15}{*}{10} & \multirow{5}{*}{$[0 \pm \langle -45/45 \rangle]_s$} &	HSDT13	&	-20.642	&	19.893	&	87.577	&	-88.406	\\
&&	HSDT11A	&	-19.915	&	17.226	&	61.182	&	-90.070	\\
&&	HSDT11B	&	-19.997	&	19.521	&	94.325	&	-81.138	\\
&&	TSDT7	&	-18.448	&	15.908	&	-40.625	&	-80.442	\\
&&	FSDT5	&	-18.949	&	16.578	&	-62.509	&	-84.812	\\
\cline{3-7}
&\multirow{5}{*}{$[0 \pm \langle -45/30 \rangle]_s$} &	HSDT13	&	-22.462	&	20.916	&	210.949	&	-100.217	\\
&&	HSDT11A	&	-21.751	&	18.257	&	184.324	&	-101.110	\\
&&	HSDT11B	&	-21.866	&	20.560	&	209.106	&	-99.162	\\
&&	TSDT7	&	-20.205	&	16.856	&	73.668	&	-90.195	\\
&&	FSDT5	&	-20.667	&	17.499	&	54.317	&	-98.292	\\
\cline{3-7}
&\multirow{5}{*}{$[0 \pm \langle -45/15 \rangle]_s$} &	HSDT13	&	-22.372	&	20.941	&	260.079	&	-79.427	\\
&&	HSDT11A	&	-21.675	&	18.285	&	233.383	&	-79.320	\\
&&	HSDT11B	&	-21.998	&	20.676	&	256.181	&	-80.453	\\
&&	TSDT7	&	-20.482	&	16.978	&	133.370	&	-70.373	\\
&&	FSDT5	&	-20.856	&	17.546	&	118.235	&	-77.477	\\
\cline{2-7}
\multirow{15}{*}{50} & \multirow{5}{*}{$[0 \pm \langle -45/45 \rangle]_s$} &	HSDT13	&	-15.276	&	11.527	&	196.482	&	-7.135	\\
&&	HSDT11A	&	-15.252	&	11.423	&	195.945	&	-7.318	\\
&&	HSDT11B	&	-15.179	&	11.472	&	191.699	&	-6.768	\\
&&	TSDT7	&	-15.122	&	11.269	&	183.538	&	-6.786	\\
&&	FSDT5	&	-15.165	&	11.330	&	178.293	&	-7.200	\\
\cline{3-7}
&\multirow{5}{*}{$[0 \pm \langle -45/30 \rangle]_s$} &	HSDT13	&	-17.259	&	12.915	&	263.560	&	-3.884	\\
&&	HSDT11A	&	-17.234	&	12.811	&	262.923	&	-4.063	\\
&&	HSDT11B	&	-17.158	&	12.851	&	255.328	&	-3.656	\\
&&	TSDT7	&	-17.060	&	12.652	&	251.140	&	-3.400	\\
&&	FSDT5	&	-17.104	&	12.704	&	247.428	&	-3.846	\\
\cline{3-7}
&\multirow{5}{*}{$[0 \pm \langle -45/15 \rangle]_s$} &	HSDT13	&	-18.077	&	13.730	&	266.365	&	-3.164	\\
&&	HSDT11A	&	-18.051	&	13.625	&	265.507	&	-3.317	\\
&&	HSDT11B	&	-18.034	&	13.687	&	258.147	&	-3.145	\\
&&	TSDT7	&	-17.934	&	13.505	&	254.992	&	-3.126	\\
&&	FSDT5	&	-17.971	&	13.545	&	252.778	&	-3.303	\\
\hline
\end{tabular}
\label{table:Table7symmTherm}
\end{table}

\begin{figure}
\centering
\newlength\figureheight 
\newlength\figurewidth 
\setlength\figureheight{10cm} 
\setlength\figurewidth{12cm}
\subfigure[$\overline{u} = \overline{u}(-a/2,0,z)$]{\scalebox{0.5}{
%
%
%
%
\begin{tikzpicture}

\begin{axis}[%
width=\figurewidth,
height=\figureheight,
scale only axis,
xmin=-1.5,
xmax=1.5,
xlabel={$\bar{u}$},
ymin=-0.5,
ymax=0.5,
ylabel={Normalized thickness, z/h},
legend style={draw=black,fill=white,legend cell align=left}
]
\addplot [
color=blue,
mark size=2.5pt,
only marks,
mark=o,
mark options={solid,draw=black}
]
table[row sep=crcr]{
1.12064 -0.483336\\
1.05004 -0.466669\\
0.982879 -0.450002\\
0.919037 -0.433336\\
0.858386 -0.416669\\
0.800802 -0.400002\\
0.74616 -0.383335\\
0.694335 -0.366669\\
0.645203 -0.350002\\
0.598639 -0.333335\\
0.554518 -0.316668\\
0.512715 -0.300002\\
0.473107 -0.283335\\
0.435567 -0.266668\\
0.399972 -0.250001\\
0.366197 -0.233335\\
0.334118 -0.216668\\
0.303608 -0.200001\\
0.274544 -0.183334\\
0.246802 -0.166668\\
0.246802 -0.166668\\
0.218992 -0.150001\\
0.192253 -0.133334\\
0.166462 -0.116667\\
0.141493 -0.100001\\
0.117221 -0.0833338\\
0.093523 -0.066667\\
0.0702728 -0.0500003\\
0.0473462 -0.0333335\\
0.0246186 -0.0166668\\
0.00196509 -3.03577e-17\\
-0.0207388 0.0166667\\
-0.0436179 0.0333335\\
-0.0667968 0.0500002\\
-0.0904003 0.066667\\
-0.114553 0.0833337\\
-0.13938 0.1\\
-0.165005 0.116667\\
-0.191553 0.133334\\
-0.219149 0.150001\\
-0.247918 0.166667\\
-0.247918 0.166668\\
-0.276721 0.183334\\
-0.306945 0.200001\\
-0.338716 0.216668\\
-0.372159 0.233334\\
-0.407398 0.250001\\
-0.444557 0.266668\\
-0.483762 0.283335\\
-0.525137 0.300001\\
-0.568806 0.316668\\
-0.614896 0.333335\\
-0.663529 0.350002\\
-0.714831 0.366668\\
-0.768927 0.383335\\
-0.825941 0.400002\\
-0.885998 0.416669\\
-0.949223 0.433335\\
-1.01574 0.450002\\
-1.08567 0.466669\\
-1.15915 0.483336\\
};
\addlegendentry{HSDT13};

\addplot [
color=black,
dashed,
line width=1.2pt
]
table[row sep=crcr]{
1.23466 -0.500003\\
1.15844 -0.483336\\
1.08583 -0.466669\\
1.01669 -0.450002\\
0.950902 -0.433336\\
0.888345 -0.416669\\
0.828891 -0.400002\\
0.772418 -0.383335\\
0.718802 -0.366669\\
0.667917 -0.350002\\
0.61964 -0.333335\\
0.573846 -0.316668\\
0.530412 -0.300002\\
0.489213 -0.283335\\
0.450125 -0.266668\\
0.413024 -0.250001\\
0.377785 -0.233335\\
0.344285 -0.216668\\
0.3124 -0.200001\\
0.282004 -0.183334\\
0.252975 -0.166668\\
0.252975 -0.166668\\
0.224138 -0.150001\\
0.196419 -0.133334\\
0.169694 -0.116667\\
0.143838 -0.100001\\
0.118726 -0.0833338\\
0.094236 -0.066667\\
0.0702423 -0.0500003\\
0.0466213 -0.0333335\\
0.0232485 -0.0166668\\
-8.00439e-18 -3.03577e-17\\
-0.0232485 0.0166667\\
-0.0466213 0.0333335\\
-0.0702423 0.0500002\\
-0.094236 0.066667\\
-0.118726 0.0833337\\
-0.143838 0.1\\
-0.169694 0.116667\\
-0.196419 0.133334\\
-0.224138 0.150001\\
-0.252975 0.166667\\
-0.252975 0.166668\\
-0.282004 0.183334\\
-0.3124 0.200001\\
-0.344285 0.216668\\
-0.377785 0.233334\\
-0.413024 0.250001\\
-0.450125 0.266668\\
-0.489213 0.283335\\
-0.530412 0.300001\\
-0.573846 0.316668\\
-0.61964 0.333335\\
-0.667917 0.350002\\
-0.718802 0.366668\\
-0.772418 0.383335\\
-0.828891 0.400002\\
-0.888345 0.416669\\
-0.950902 0.433335\\
-1.01669 0.450002\\
-1.08583 0.466669\\
-1.15844 0.483336\\
-1.23466 0.500002\\
};
\addlegendentry{HSDT11A};

\addplot [
color=blue,
mark size=4.3pt,
only marks,
mark=diamond,
mark options={solid,draw=black}
]
table[row sep=crcr]{
1.05985 -0.483336\\
0.994141 -0.466669\\
0.931534 -0.450002\\
0.871915 -0.433336\\
0.815171 -0.416669\\
0.761191 -0.400002\\
0.709861 -0.383335\\
0.661069 -0.366669\\
0.614703 -0.350002\\
0.570649 -0.333335\\
0.528796 -0.316668\\
0.48903 -0.300002\\
0.45124 -0.283335\\
0.415312 -0.266668\\
0.381135 -0.250001\\
0.348595 -0.233335\\
0.31758 -0.216668\\
0.287978 -0.200001\\
0.259675 -0.183334\\
0.232561 -0.166668\\
0.232561 -0.166668\\
0.206521 -0.150001\\
0.181443 -0.133334\\
0.157216 -0.116667\\
0.133726 -0.100001\\
0.11086 -0.0833338\\
0.0885067 -0.066667\\
0.0665531 -0.0500003\\
0.0448867 -0.0333335\\
0.0233949 -0.0166668\\
0.00196509 -3.03577e-17\\
-0.0195151 0.0166667\\
-0.0411584 0.0333335\\
-0.0630772 0.0500002\\
-0.085384 0.066667\\
-0.108192 0.0833337\\
-0.131612 0.1\\
-0.155759 0.116667\\
-0.180743 0.133334\\
-0.206679 0.150001\\
-0.233677 0.166667\\
-0.233677 0.166668\\
-0.261852 0.183334\\
-0.291315 0.200001\\
-0.322179 0.216668\\
-0.354556 0.233334\\
-0.38856 0.250001\\
-0.424302 0.266668\\
-0.461895 0.283335\\
-0.501451 0.300001\\
-0.543084 0.316668\\
-0.586906 0.333335\\
-0.633029 0.350002\\
-0.681566 0.366668\\
-0.732629 0.383335\\
-0.78633 0.400002\\
-0.842784 0.416669\\
-0.902101 0.433335\\
-0.964395 0.450002\\
-1.02978 0.466669\\
-1.09836 0.483336\\
};
\addlegendentry{HSDT11B};

\addplot [
color=black,
solid,
line width=1.2pt
]
table[row sep=crcr]{
0.961301 -0.500003\\
0.929257 -0.483336\\
0.897214 -0.466669\\
0.865171 -0.450002\\
0.833127 -0.433336\\
0.801084 -0.416669\\
0.769041 -0.400002\\
0.736997 -0.383335\\
0.704954 -0.366669\\
0.672911 -0.350002\\
0.640867 -0.333335\\
0.608824 -0.316668\\
0.57678 -0.300002\\
0.544737 -0.283335\\
0.512694 -0.266668\\
0.48065 -0.250001\\
0.448607 -0.233335\\
0.416564 -0.216668\\
0.38452 -0.200001\\
0.352477 -0.183334\\
0.320434 -0.166668\\
0.320434 -0.166668\\
0.28839 -0.150001\\
0.256347 -0.133334\\
0.224304 -0.116667\\
0.19226 -0.100001\\
0.160217 -0.0833338\\
0.128173 -0.066667\\
0.0961301 -0.0500003\\
0.0640867 -0.0333335\\
0.0320434 -0.0166668\\
2.63378e-17 -3.03577e-17\\
-0.0320434 0.0166667\\
-0.0640867 0.0333335\\
-0.0961301 0.0500002\\
-0.128173 0.066667\\
-0.160217 0.0833337\\
-0.19226 0.1\\
-0.224304 0.116667\\
-0.256347 0.133334\\
-0.28839 0.150001\\
-0.320434 0.166667\\
-0.320434 0.166668\\
-0.352477 0.183334\\
-0.38452 0.200001\\
-0.416564 0.216668\\
-0.448607 0.233334\\
-0.48065 0.250001\\
-0.512694 0.266668\\
-0.544737 0.283335\\
-0.57678 0.300001\\
-0.608824 0.316668\\
-0.640867 0.333335\\
-0.672911 0.350002\\
-0.704954 0.366668\\
-0.736997 0.383335\\
-0.769041 0.400002\\
-0.801084 0.416669\\
-0.833127 0.433335\\
-0.865171 0.450002\\
-0.897214 0.466669\\
-0.929257 0.483336\\
-0.961301 0.500002\\
};
\addlegendentry{TSDT7};

\addplot [
color=blue,
mark size=2.5pt,
only marks,
mark=x,
mark options={solid,draw=black}
]
table[row sep=crcr]{
0.960169 -0.483336\\
0.92706 -0.466669\\
0.893951 -0.450002\\
0.860842 -0.433336\\
0.827732 -0.416669\\
0.794623 -0.400002\\
0.761514 -0.383335\\
0.728404 -0.366669\\
0.695295 -0.350002\\
0.662186 -0.333335\\
0.629077 -0.316668\\
0.595967 -0.300002\\
0.562858 -0.283335\\
0.529749 -0.266668\\
0.496639 -0.250001\\
0.46353 -0.233335\\
0.430421 -0.216668\\
0.397311 -0.200001\\
0.364202 -0.183334\\
0.331093 -0.166668\\
0.331093 -0.166668\\
0.297984 -0.150001\\
0.264874 -0.133334\\
0.231765 -0.116667\\
0.198656 -0.100001\\
0.165546 -0.0833338\\
0.132437 -0.066667\\
0.0993279 -0.0500003\\
0.0662186 -0.0333335\\
0.0331093 -0.0166668\\
1.24688e-17 -3.03577e-17\\
-0.0331093 0.0166667\\
-0.0662186 0.0333335\\
-0.0993279 0.0500002\\
-0.132437 0.066667\\
-0.165546 0.0833337\\
-0.198656 0.1\\
-0.231765 0.116667\\
-0.264874 0.133334\\
-0.297984 0.150001\\
-0.331093 0.166667\\
-0.331093 0.166668\\
-0.364202 0.183334\\
-0.397311 0.200001\\
-0.430421 0.216668\\
-0.46353 0.233334\\
-0.496639 0.250001\\
-0.529749 0.266668\\
-0.562858 0.283335\\
-0.595967 0.300001\\
-0.629077 0.316668\\
-0.662186 0.333335\\
-0.695295 0.350002\\
-0.728404 0.366668\\
-0.761514 0.383335\\
-0.794623 0.400002\\
-0.827732 0.416669\\
-0.860842 0.433335\\
-0.893951 0.450002\\
-0.92706 0.466669\\
-0.960169 0.483336\\
};
\addlegendentry{FSDT5};

\end{axis}
\end{tikzpicture}%
}}\hspace{60pt}
\subfigure[$\overline{w} = \overline{w}(0,0,z)$]{\scalebox{0.5}{
%
%
%
%
\begin{tikzpicture}

\begin{axis}[%
width=\figurewidth,
height=\figureheight,
scale only axis,
xmin=1.34,
xmax=1.46,
xlabel={$\bar{w}$},
ymin=-0.5,
ymax=0.5,
ylabel={Normalized thickness, z/h},
legend style={at={(0.157232142857143,0.693928571428572)},anchor=south west,draw=black,fill=white,legend cell align=left}
]
\addplot [
color=blue,
mark size=2.5pt,
only marks,
mark=o,
mark options={solid,draw=black}
]
table[row sep=crcr]{
1.39081 -0.483336\\
1.39261 -0.466669\\
1.39438 -0.450002\\
1.39613 -0.433336\\
1.39786 -0.416669\\
1.39957 -0.400002\\
1.40126 -0.383335\\
1.40292 -0.366669\\
1.40456 -0.350002\\
1.40618 -0.333335\\
1.40777 -0.316668\\
1.40934 -0.300002\\
1.41089 -0.283335\\
1.41242 -0.266668\\
1.41392 -0.250001\\
1.4154 -0.233335\\
1.41686 -0.216668\\
1.41829 -0.200001\\
1.41971 -0.183334\\
1.4211 -0.166668\\
1.4211 -0.166668\\
1.42246 -0.150001\\
1.42381 -0.133334\\
1.42513 -0.116667\\
1.42643 -0.100001\\
1.4277 -0.0833338\\
1.42896 -0.066667\\
1.43019 -0.0500003\\
1.4314 -0.0333335\\
1.43258 -0.0166668\\
1.43375 -3.03577e-17\\
1.43489 0.0166667\\
1.436 0.0333335\\
1.4371 0.0500002\\
1.43817 0.066667\\
1.43922 0.0833337\\
1.44025 0.1\\
1.44125 0.116667\\
1.44223 0.133334\\
1.44319 0.150001\\
1.44413 0.166667\\
1.44413 0.166668\\
1.44504 0.183334\\
1.44593 0.200001\\
1.4468 0.216668\\
1.44764 0.233334\\
1.44846 0.250001\\
1.44926 0.266668\\
1.45004 0.283335\\
1.45079 0.300001\\
1.45153 0.316668\\
1.45224 0.333335\\
1.45292 0.350002\\
1.45358 0.366668\\
1.45423 0.383335\\
1.45484 0.400002\\
1.45544 0.416669\\
1.45601 0.433335\\
1.45656 0.450002\\
1.45709 0.466669\\
1.45759 0.483336\\
};
\addlegendentry{HSDT13};

\addplot [
color=black,
dashed,
line width=1.2pt
]
table[row sep=crcr]{
1.44818 -0.500003\\
1.44818 -0.483336\\
1.44818 -0.466669\\
1.44818 -0.450002\\
1.44818 -0.433336\\
1.44818 -0.416669\\
1.44818 -0.400002\\
1.44818 -0.383335\\
1.44818 -0.366669\\
1.44818 -0.350002\\
1.44818 -0.333335\\
1.44818 -0.316668\\
1.44818 -0.300002\\
1.44818 -0.283335\\
1.44818 -0.266668\\
1.44818 -0.250001\\
1.44818 -0.233335\\
1.44818 -0.216668\\
1.44818 -0.200001\\
1.44818 -0.183334\\
1.44818 -0.166668\\
1.44818 -0.166668\\
1.44818 -0.150001\\
1.44818 -0.133334\\
1.44818 -0.116667\\
1.44818 -0.100001\\
1.44818 -0.0833338\\
1.44818 -0.066667\\
1.44818 -0.0500003\\
1.44818 -0.0333335\\
1.44818 -0.0166668\\
1.44818 -3.03577e-17\\
1.44818 0.0166667\\
1.44818 0.0333335\\
1.44818 0.0500002\\
1.44818 0.066667\\
1.44818 0.0833337\\
1.44818 0.1\\
1.44818 0.116667\\
1.44818 0.133334\\
1.44818 0.150001\\
1.44818 0.166667\\
1.44818 0.166668\\
1.44818 0.183334\\
1.44818 0.200001\\
1.44818 0.216668\\
1.44818 0.233334\\
1.44818 0.250001\\
1.44818 0.266668\\
1.44818 0.283335\\
1.44818 0.300001\\
1.44818 0.316668\\
1.44818 0.333335\\
1.44818 0.350002\\
1.44818 0.366668\\
1.44818 0.383335\\
1.44818 0.400002\\
1.44818 0.416669\\
1.44818 0.433335\\
1.44818 0.450002\\
1.44818 0.466669\\
1.44818 0.483336\\
1.44818 0.500002\\
};
\addlegendentry{HSDT11A};

\addplot [
color=blue,
mark size=4.3pt,
only marks,
mark=diamond,
mark options={solid,draw=black}
]
table[row sep=crcr]{
1.35648 -0.483336\\
1.35816 -0.466669\\
1.35982 -0.450002\\
1.36146 -0.433336\\
1.36308 -0.416669\\
1.36469 -0.400002\\
1.36628 -0.383335\\
1.36784 -0.366669\\
1.36939 -0.350002\\
1.37092 -0.333335\\
1.37244 -0.316668\\
1.37393 -0.300002\\
1.3754 -0.283335\\
1.37686 -0.266668\\
1.3783 -0.250001\\
1.37972 -0.233335\\
1.38112 -0.216668\\
1.3825 -0.200001\\
1.38387 -0.183334\\
1.38521 -0.166668\\
1.38521 -0.166668\\
1.38654 -0.150001\\
1.38785 -0.133334\\
1.38914 -0.116667\\
1.39041 -0.100001\\
1.39166 -0.0833338\\
1.3929 -0.066667\\
1.39412 -0.0500003\\
1.39531 -0.0333335\\
1.39649 -0.0166668\\
1.39765 -3.03577e-17\\
1.3988 0.0166667\\
1.39992 0.0333335\\
1.40102 0.0500002\\
1.40211 0.066667\\
1.40318 0.0833337\\
1.40423 0.1\\
1.40526 0.116667\\
1.40627 0.133334\\
1.40727 0.150001\\
1.40824 0.166667\\
1.40824 0.166668\\
1.4092 0.183334\\
1.41014 0.200001\\
1.41106 0.216668\\
1.41196 0.233334\\
1.41284 0.250001\\
1.41371 0.266668\\
1.41455 0.283335\\
1.41538 0.300001\\
1.41619 0.316668\\
1.41698 0.333335\\
1.41775 0.350002\\
1.41851 0.366668\\
1.41924 0.383335\\
1.41996 0.400002\\
1.42066 0.416669\\
1.42134 0.433335\\
1.422 0.450002\\
1.42264 0.466669\\
1.42327 0.483336\\
};
\addlegendentry{HSDT11B};

\addplot [
color=blue,
mark size=2.5pt,
only marks,
mark=x,
mark options={solid,draw=black}
]
table[row sep=crcr]{
1.39867 -0.483336\\
1.39867 -0.466669\\
1.39867 -0.450002\\
1.39867 -0.433336\\
1.39867 -0.416669\\
1.39867 -0.400002\\
1.39867 -0.383335\\
1.39867 -0.366669\\
1.39867 -0.350002\\
1.39867 -0.333335\\
1.39867 -0.316668\\
1.39867 -0.300002\\
1.39867 -0.283335\\
1.39867 -0.266668\\
1.39867 -0.250001\\
1.39867 -0.233335\\
1.39867 -0.216668\\
1.39867 -0.200001\\
1.39867 -0.183334\\
1.39867 -0.166668\\
1.39867 -0.166668\\
1.39867 -0.150001\\
1.39867 -0.133334\\
1.39867 -0.116667\\
1.39867 -0.100001\\
1.39867 -0.0833338\\
1.39867 -0.066667\\
1.39867 -0.0500003\\
1.39867 -0.0333335\\
1.39867 -0.0166668\\
1.39867 -3.03577e-17\\
1.39867 0.0166667\\
1.39867 0.0333335\\
1.39867 0.0500002\\
1.39867 0.066667\\
1.39867 0.0833337\\
1.39867 0.1\\
1.39867 0.116667\\
1.39867 0.133334\\
1.39867 0.150001\\
1.39867 0.166667\\
1.39867 0.166668\\
1.39867 0.183334\\
1.39867 0.200001\\
1.39867 0.216668\\
1.39867 0.233334\\
1.39867 0.250001\\
1.39867 0.266668\\
1.39867 0.283335\\
1.39867 0.300001\\
1.39867 0.316668\\
1.39867 0.333335\\
1.39867 0.350002\\
1.39867 0.366668\\
1.39867 0.383335\\
1.39867 0.400002\\
1.39867 0.416669\\
1.39867 0.433335\\
1.39867 0.450002\\
1.39867 0.466669\\
1.39867 0.483336\\
};
\addlegendentry{FSDT5};

\end{axis}
\end{tikzpicture}%
}}
\subfigure[$\overline{\sigma}_{xx} = \overline{\sigma}_{xx}(0,0,z)$]{\scalebox{0.5}{
%
%
%
%
\begin{tikzpicture}

\begin{axis}[%
width=\figurewidth,
height=\figureheight,
scale only axis,
xmin=-0.4,
xmax=0.4,
xlabel={$\bar{\sigma}_{xx}$},
ymin=-0.5,
ymax=0.5,
ylabel={Normalized thickness, z/h},
legend style={draw=black,fill=white,legend cell align=left},
legend pos = south east
]
\addplot [
color=blue,
mark size=2.5pt,
only marks,
mark=o,
mark options={solid,draw=black}
]
table[row sep=crcr]{
-0.30608 -0.483336\\
-0.280696 -0.466669\\
-0.256934 -0.450002\\
-0.234734 -0.433336\\
-0.214039 -0.416669\\
-0.194788 -0.400002\\
-0.176923 -0.383335\\
-0.160385 -0.366669\\
-0.145114 -0.350002\\
-0.131051 -0.333335\\
-0.118139 -0.316668\\
-0.106316 -0.300002\\
-0.0955245 -0.283335\\
-0.0857053 -0.266668\\
-0.0767992 -0.250001\\
-0.0687471 -0.233335\\
-0.06149 -0.216668\\
-0.0549688 -0.200001\\
-0.0491244 -0.183334\\
-0.0438977 -0.166668\\
-0.275027 -0.166668\\
-0.248322 -0.150001\\
-0.221424 -0.133334\\
-0.194354 -0.116667\\
-0.167133 -0.100001\\
-0.139779 -0.0833338\\
-0.112314 -0.066667\\
-0.0847573 -0.0500003\\
-0.0571295 -0.0333335\\
-0.0294506 -0.0166668\\
-0.00174071 -3.03577e-17\\
0.0259799 0.0166667\\
0.0536911 0.0333335\\
0.0813727 0.0500002\\
0.109005 0.066667\\
0.136567 0.0833337\\
0.164038 0.1\\
0.1914 0.116667\\
0.218631 0.133334\\
0.245712 0.150001\\
0.272622 0.166667\\
0.0370575 0.166668\\
0.0429578 0.183334\\
0.04954 0.200001\\
0.0568632 0.216668\\
0.0649864 0.233334\\
0.0739687 0.250001\\
0.0838692 0.266668\\
0.0947469 0.283335\\
0.106661 0.300001\\
0.119671 0.316668\\
0.133835 0.333335\\
0.149212 0.350002\\
0.165862 0.366668\\
0.183844 0.383335\\
0.203217 0.400002\\
0.224039 0.416669\\
0.246371 0.433335\\
0.27027 0.450002\\
0.295797 0.466669\\
0.323009 0.483336\\
};
\addlegendentry{HSDT13};

\addplot [
color=black,
dashed,
line width=1.2pt
]
table[row sep=crcr]{
-0.344519 -0.500003\\
-0.31579 -0.483336\\
-0.288866 -0.466669\\
-0.263686 -0.450002\\
-0.240186 -0.433336\\
-0.218306 -0.416669\\
-0.197982 -0.400002\\
-0.179153 -0.383335\\
-0.161755 -0.366669\\
-0.145727 -0.350002\\
-0.131007 -0.333335\\
-0.117532 -0.316668\\
-0.10524 -0.300002\\
-0.0940691 -0.283335\\
-0.0839563 -0.266668\\
-0.0748398 -0.250001\\
-0.0666571 -0.233335\\
-0.0593462 -0.216668\\
-0.0528446 -0.200001\\
-0.0470903 -0.183334\\
-0.0420208 -0.166668\\
-0.277293 -0.166668\\
-0.250012 -0.150001\\
-0.22259 -0.133334\\
-0.195042 -0.116667\\
-0.167384 -0.100001\\
-0.139631 -0.0833338\\
-0.111799 -0.066667\\
-0.0839044 -0.0500003\\
-0.0559625 -0.0333335\\
-0.0279891 -0.0166668\\
4.69282e-18 -3.03577e-17\\
0.0279891 0.0166667\\
0.0559625 0.0333335\\
0.0839044 0.0500002\\
0.111799 0.066667\\
0.139631 0.0833337\\
0.167384 0.1\\
0.195042 0.116667\\
0.22259 0.133334\\
0.250012 0.150001\\
0.277293 0.166667\\
0.0420208 0.166668\\
0.0470903 0.183334\\
0.0528446 0.200001\\
0.0593462 0.216668\\
0.0666571 0.233334\\
0.0748398 0.250001\\
0.0839563 0.266668\\
0.0940691 0.283335\\
0.10524 0.300001\\
0.117532 0.316668\\
0.131007 0.333335\\
0.145727 0.350002\\
0.161755 0.366668\\
0.179153 0.383335\\
0.197982 0.400002\\
0.218306 0.416669\\
0.240186 0.433335\\
0.263686 0.450002\\
0.288866 0.466669\\
0.31579 0.483336\\
0.344519 0.500002\\
};
\addlegendentry{HSDT11A};

\addplot [
color=blue,
mark size=4.3pt,
only marks,
mark=diamond,
mark options={solid,draw=black}
]
table[row sep=crcr]{
-0.30525 -0.483336\\
-0.280759 -0.466669\\
-0.25777 -0.450002\\
-0.236231 -0.433336\\
-0.216085 -0.416669\\
-0.197278 -0.400002\\
-0.179755 -0.383335\\
-0.163462 -0.366669\\
-0.148343 -0.350002\\
-0.134344 -0.333335\\
-0.12141 -0.316668\\
-0.109485 -0.300002\\
-0.0985159 -0.283335\\
-0.0884468 -0.266668\\
-0.0792231 -0.250001\\
-0.07079 -0.233335\\
-0.0630926 -0.216668\\
-0.0560761 -0.200001\\
-0.0496856 -0.183334\\
-0.0438663 -0.166668\\
-0.333653 -0.166668\\
-0.297233 -0.150001\\
-0.261837 -0.133334\\
-0.227351 -0.116667\\
-0.193658 -0.100001\\
-0.160645 -0.0833338\\
-0.128195 -0.066667\\
-0.0961954 -0.0500003\\
-0.0645296 -0.0333335\\
-0.0330831 -0.0166668\\
-0.00174071 -3.03577e-17\\
0.0296124 0.0166667\\
0.0610912 0.0333335\\
0.0928108 0.0500002\\
0.124886 0.066667\\
0.157432 0.0833337\\
0.190564 0.1\\
0.224396 0.116667\\
0.259044 0.133334\\
0.294623 0.150001\\
0.331248 0.166667\\
0.0370261 0.166668\\
0.043519 0.183334\\
0.0506473 0.200001\\
0.0584657 0.216668\\
0.0670292 0.233334\\
0.0763926 0.250001\\
0.0866107 0.266668\\
0.0977384 0.283335\\
0.10983 0.300001\\
0.122942 0.316668\\
0.137127 0.333335\\
0.152442 0.350002\\
0.16894 0.366668\\
0.186677 0.383335\\
0.205707 0.400002\\
0.226085 0.416669\\
0.247867 0.433335\\
0.271107 0.450002\\
0.295859 0.466669\\
0.32218 0.483336\\
};
\addlegendentry{HSDT11B};

\addplot [
color=black,
solid,
line width=1.2pt
]
table[row sep=crcr]{
-0.246611 -0.500003\\
-0.238391 -0.483336\\
-0.23017 -0.466669\\
-0.22195 -0.450002\\
-0.21373 -0.433336\\
-0.205509 -0.416669\\
-0.197289 -0.400002\\
-0.189069 -0.383335\\
-0.180848 -0.366669\\
-0.172628 -0.350002\\
-0.164407 -0.333335\\
-0.156187 -0.316668\\
-0.147967 -0.300002\\
-0.139746 -0.283335\\
-0.131526 -0.266668\\
-0.123306 -0.250001\\
-0.115085 -0.233335\\
-0.106865 -0.216668\\
-0.0986445 -0.200001\\
-0.0904241 -0.183334\\
-0.0822037 -0.166668\\
-0.385899 -0.166668\\
-0.347309 -0.150001\\
-0.308719 -0.133334\\
-0.270129 -0.116667\\
-0.231539 -0.100001\\
-0.192949 -0.0833338\\
-0.15436 -0.066667\\
-0.11577 -0.0500003\\
-0.0771798 -0.0333335\\
-0.0385899 -0.0166668\\
-2.39863e-17 -3.03577e-17\\
0.0385899 0.0166667\\
0.0771798 0.0333335\\
0.11577 0.0500002\\
0.15436 0.066667\\
0.192949 0.0833337\\
0.231539 0.1\\
0.270129 0.116667\\
0.308719 0.133334\\
0.347309 0.150001\\
0.385899 0.166667\\
0.0822037 0.166668\\
0.0904241 0.183334\\
0.0986445 0.200001\\
0.106865 0.216668\\
0.115085 0.233334\\
0.123306 0.250001\\
0.131526 0.266668\\
0.139746 0.283335\\
0.147967 0.300001\\
0.156187 0.316668\\
0.164407 0.333335\\
0.172628 0.350002\\
0.180848 0.366668\\
0.189069 0.383335\\
0.197289 0.400002\\
0.205509 0.416669\\
0.21373 0.433335\\
0.22195 0.450002\\
0.23017 0.466669\\
0.238391 0.483336\\
0.246611 0.500002\\
};
\addlegendentry{TSDT7};

\addplot [
color=blue,
mark size=2.5pt,
only marks,
mark=x,
mark options={solid,draw=black}
]
table[row sep=crcr]{
-0.234609 -0.483336\\
-0.226519 -0.466669\\
-0.218429 -0.450002\\
-0.210339 -0.433336\\
-0.202249 -0.416669\\
-0.194159 -0.400002\\
-0.186069 -0.383335\\
-0.177979 -0.366669\\
-0.169889 -0.350002\\
-0.161799 -0.333335\\
-0.153709 -0.316668\\
-0.145619 -0.300002\\
-0.137529 -0.283335\\
-0.12944 -0.266668\\
-0.12135 -0.250001\\
-0.11326 -0.233335\\
-0.10517 -0.216668\\
-0.0970796 -0.200001\\
-0.0889897 -0.183334\\
-0.0808997 -0.166668\\
-0.399057 -0.166668\\
-0.359152 -0.150001\\
-0.319246 -0.133334\\
-0.27934 -0.116667\\
-0.239434 -0.100001\\
-0.199529 -0.0833338\\
-0.159623 -0.066667\\
-0.119717 -0.0500003\\
-0.0798115 -0.0333335\\
-0.0399057 -0.0166668\\
-4.7064e-17 -3.03577e-17\\
0.0399057 0.0166667\\
0.0798115 0.0333335\\
0.119717 0.0500002\\
0.159623 0.066667\\
0.199529 0.0833337\\
0.239434 0.1\\
0.27934 0.116667\\
0.319246 0.133334\\
0.359152 0.150001\\
0.399057 0.166667\\
0.0808997 0.166668\\
0.0889897 0.183334\\
0.0970796 0.200001\\
0.10517 0.216668\\
0.11326 0.233334\\
0.12135 0.250001\\
0.12944 0.266668\\
0.137529 0.283335\\
0.145619 0.300001\\
0.153709 0.316668\\
0.161799 0.333335\\
0.169889 0.350002\\
0.177979 0.366668\\
0.186069 0.383335\\
0.194159 0.400002\\
0.202249 0.416669\\
0.210339 0.433335\\
0.218429 0.450002\\
0.226519 0.466669\\
0.234609 0.483336\\
};
\addlegendentry{FSDT5};

\end{axis}

\begin{axis}[%
width=\figurewidth,
height=\figureheight,
scale only axis,
xmin=0,
xmax=1,
ymin=0,
ymax=1,
hide axis,
axis x line*=bottom,
axis y line*=left
]
\addplot [
color=black,
solid,
line width=0.0pt,
mark=square*,
mark options={solid,fill=black,draw=white},
forget plot
]
table[row sep=crcr]{
0.5 0.988095238095238\\
};
\addplot [
color=black,
solid,
line width=0.0pt,
mark=square*,
mark options={solid,fill=black,draw=white},
forget plot
]
table[row sep=crcr]{
0.991071428571429 0.5\\
};
\addplot [
color=black,
solid,
line width=0.0pt,
mark=square*,
mark options={solid,fill=black,draw=white},
forget plot
]
table[row sep=crcr]{
0.5 0.0119047619047619\\
};
\addplot [
color=black,
solid,
line width=0.0pt,
mark=square*,
mark options={solid,fill=black,draw=white},
forget plot
]
table[row sep=crcr]{
0.00892857142857143 0.5\\
};
\addplot [
color=black,
solid,
line width=0.0pt,
mark=square*,
mark options={solid,fill=black,draw=white},
forget plot
]
table[row sep=crcr]{
0.00892857142857143 0.988095238095238\\
};
\addplot [
color=black,
solid,
line width=0.0pt,
mark=square*,
mark options={solid,fill=black,draw=white},
forget plot
]
table[row sep=crcr]{
0.991071428571429 0.0119047619047619\\
};
\addplot [
color=black,
solid,
line width=0.0pt,
mark=square*,
mark options={solid,fill=black,draw=white},
forget plot
]
table[row sep=crcr]{
0.991071428571429 0.988095238095238\\
};
\addplot [
color=black,
solid,
line width=0.0pt,
mark=square*,
mark options={solid,fill=black,draw=white},
forget plot
]
table[row sep=crcr]{
0.00892857142857143 0.0119047619047619\\
};
\end{axis}
\end{tikzpicture}%
}}\hspace{60pt}
\subfigure[$\overline{\tau}_{yz} = \overline{\tau}_{yz}(-a/2,0,z)$]{\scalebox{0.5}{
%
%
%
%
\begin{tikzpicture}

\begin{axis}[%
width=\figurewidth,
height=\figureheight,
scale only axis,
xmin=-0.35,
xmax=0,
xlabel={$\bar{\tau}_{yz}$},
ymin=-0.5,
ymax=0.5,
ylabel={Normalized thickness, z/h},
legend style={draw=black,fill=white,legend cell align=left},
legend pos = north west
]
\addplot [
color=blue,
mark size=2.5pt,
only marks,
mark=o,
mark options={solid,draw=black}
]
table[row sep=crcr]{
-0.0199093 -0.483336\\
-0.0388316 -0.466669\\
-0.0567927 -0.450002\\
-0.0738173 -0.433336\\
-0.0899295 -0.416669\\
-0.105153 -0.400002\\
-0.119509 -0.383335\\
-0.133021 -0.366669\\
-0.145708 -0.350002\\
-0.157591 -0.333335\\
-0.16869 -0.316668\\
-0.179022 -0.300002\\
-0.188606 -0.283335\\
-0.197458 -0.266668\\
-0.205595 -0.250001\\
-0.213031 -0.233335\\
-0.219782 -0.216668\\
-0.225861 -0.200001\\
-0.231281 -0.183334\\
-0.236053 -0.166668\\
-0.236053 -0.166668\\
-0.254129 -0.150001\\
-0.270136 -0.133334\\
-0.284134 -0.116667\\
-0.296175 -0.100001\\
-0.306305 -0.0833338\\
-0.314562 -0.066667\\
-0.320977 -0.0500003\\
-0.325578 -0.0333335\\
-0.328382 -0.0166668\\
-0.329402 -3.03577e-17\\
-0.328644 0.0166667\\
-0.326105 0.0333335\\
-0.321779 0.0500002\\
-0.315652 0.066667\\
-0.307701 0.0833337\\
-0.2979 0.1\\
-0.286214 0.116667\\
-0.272603 0.133334\\
-0.257018 0.150001\\
-0.239405 0.166667\\
-0.239405 0.166668\\
-0.233188 0.183334\\
-0.226391 0.200001\\
-0.219006 0.216668\\
-0.211028 0.233334\\
-0.202447 0.250001\\
-0.193257 0.266668\\
-0.183447 0.283335\\
-0.173007 0.300001\\
-0.161927 0.316668\\
-0.150194 0.333335\\
-0.137797 0.350002\\
-0.124722 0.366668\\
-0.110955 0.383335\\
-0.0964805 0.400002\\
-0.0812837 0.416669\\
-0.0653478 0.433335\\
-0.0486555 0.450002\\
-0.0311888 0.466669\\
-0.0129288 0.483336\\
};
\addlegendentry{HSDT13};

\addplot [
color=black,
dashed,
line width=1.2pt
]
table[row sep=crcr]{
0 -0.500003\\
-0.0194706 -0.483336\\
-0.0381665 -0.466669\\
-0.0560909 -0.450002\\
-0.0732465 -0.433336\\
-0.0896362 -0.416669\\
-0.105262 -0.400002\\
-0.120128 -0.383335\\
-0.134235 -0.366669\\
-0.147586 -0.350002\\
-0.160183 -0.333335\\
-0.172029 -0.316668\\
-0.183125 -0.300002\\
-0.193473 -0.283335\\
-0.203075 -0.266668\\
-0.211934 -0.250001\\
-0.220049 -0.233335\\
-0.227424 -0.216668\\
-0.234058 -0.200001\\
-0.239955 -0.183334\\
-0.245114 -0.166668\\
-0.245114 -0.166668\\
-0.263178 -0.150001\\
-0.27916 -0.133334\\
-0.293121 -0.116667\\
-0.305112 -0.100001\\
-0.315181 -0.0833338\\
-0.323366 -0.066667\\
-0.329699 -0.0500003\\
-0.334205 -0.0333335\\
-0.336901 -0.0166668\\
-0.337799 -3.03577e-17\\
-0.336901 0.0166667\\
-0.334205 0.0333335\\
-0.329699 0.0500002\\
-0.323366 0.066667\\
-0.315181 0.0833337\\
-0.305112 0.1\\
-0.293121 0.116667\\
-0.27916 0.133334\\
-0.263178 0.150001\\
-0.245114 0.166667\\
-0.245114 0.166668\\
-0.239955 0.183334\\
-0.234058 0.200001\\
-0.227424 0.216668\\
-0.220049 0.233334\\
-0.211934 0.250001\\
-0.203075 0.266668\\
-0.193473 0.283335\\
-0.183125 0.300001\\
-0.172029 0.316668\\
-0.160183 0.333335\\
-0.147586 0.350002\\
-0.134235 0.366668\\
-0.120128 0.383335\\
-0.105262 0.400002\\
-0.0896362 0.416669\\
-0.0732465 0.433335\\
-0.0560909 0.450002\\
-0.0381665 0.466669\\
-0.0194706 0.483336\\
-3.27418e-17 0.500002\\
};
\addlegendentry{HSDT11A};

\addplot [
color=blue,
mark size=4.3pt,
only marks,
mark=diamond,
mark options={solid,draw=black}
]
table[row sep=crcr]{
-0.0235508 -0.483336\\
-0.0452353 -0.466669\\
-0.0651594 -0.450002\\
-0.0834255 -0.433336\\
-0.100132 -0.416669\\
-0.115375 -0.400002\\
-0.129244 -0.383335\\
-0.141829 -0.366669\\
-0.153214 -0.350002\\
-0.163478 -0.333335\\
-0.1727 -0.316668\\
-0.180952 -0.300002\\
-0.188305 -0.283335\\
-0.194824 -0.266668\\
-0.200573 -0.250001\\
-0.20561 -0.233335\\
-0.20999 -0.216668\\
-0.213766 -0.200001\\
-0.216985 -0.183334\\
-0.219693 -0.166668\\
-0.219693 -0.166668\\
-0.236636 -0.150001\\
-0.251634 -0.133334\\
-0.264745 -0.116667\\
-0.276019 -0.100001\\
-0.285502 -0.0833338\\
-0.29323 -0.066667\\
-0.299236 -0.0500003\\
-0.303544 -0.0333335\\
-0.306173 -0.0166668\\
-0.307135 -3.03577e-17\\
-0.306434 0.0166667\\
-0.304071 0.0333335\\
-0.300038 0.0500002\\
-0.294321 0.066667\\
-0.286898 0.0833337\\
-0.277744 0.1\\
-0.266825 0.116667\\
-0.254101 0.133334\\
-0.239525 0.150001\\
-0.223044 0.166667\\
-0.223044 0.166668\\
-0.218893 0.183334\\
-0.214296 0.200001\\
-0.209214 0.216668\\
-0.203606 0.233334\\
-0.197425 0.250001\\
-0.190623 0.266668\\
-0.183145 0.283335\\
-0.174937 0.300001\\
-0.165937 0.316668\\
-0.156081 0.333335\\
-0.145303 0.350002\\
-0.133531 0.366668\\
-0.12069 0.383335\\
-0.106703 0.400002\\
-0.0914864 0.416669\\
-0.074956 0.433335\\
-0.0570222 0.450002\\
-0.0375925 0.466669\\
-0.0165704 0.483336\\
};
\addlegendentry{HSDT11B};

\addplot [
color=black,
solid,
line width=1.2pt
]
table[row sep=crcr]{
0 -0.500003\\
-0.0151979 -0.483336\\
-0.0298806 -0.466669\\
-0.0440481 -0.450002\\
-0.0577004 -0.433336\\
-0.0708375 -0.416669\\
-0.0834595 -0.400002\\
-0.0955663 -0.383335\\
-0.107158 -0.366669\\
-0.118234 -0.350002\\
-0.128795 -0.333335\\
-0.138842 -0.316668\\
-0.148372 -0.300002\\
-0.157388 -0.283335\\
-0.165889 -0.266668\\
-0.173874 -0.250001\\
-0.181344 -0.233335\\
-0.188299 -0.216668\\
-0.194739 -0.200001\\
-0.200663 -0.183334\\
-0.206073 -0.166668\\
-0.206073 -0.166668\\
-0.224343 -0.150001\\
-0.240689 -0.133334\\
-0.255113 -0.116667\\
-0.267614 -0.100001\\
-0.278191 -0.0833338\\
-0.286845 -0.066667\\
-0.293576 -0.0500003\\
-0.298384 -0.0333335\\
-0.301269 -0.0166668\\
-0.30223 -3.03577e-17\\
-0.301269 0.0166667\\
-0.298384 0.0333335\\
-0.293576 0.0500002\\
-0.286845 0.066667\\
-0.278191 0.0833337\\
-0.267614 0.1\\
-0.255113 0.116667\\
-0.240689 0.133334\\
-0.224343 0.150001\\
-0.206073 0.166667\\
-0.206073 0.166668\\
-0.200663 0.183334\\
-0.194739 0.200001\\
-0.188299 0.216668\\
-0.181344 0.233334\\
-0.173874 0.250001\\
-0.165889 0.266668\\
-0.157388 0.283335\\
-0.148372 0.300001\\
-0.138842 0.316668\\
-0.128795 0.333335\\
-0.118234 0.350002\\
-0.107158 0.366668\\
-0.0955663 0.383335\\
-0.0834595 0.400002\\
-0.0708375 0.416669\\
-0.0577004 0.433335\\
-0.0440481 0.450002\\
-0.0298806 0.466669\\
-0.0151979 0.483336\\
-1.81899e-17 0.500002\\
};
\addlegendentry{TSDT7};

\addplot [
color=blue,
mark size=2.5pt,
only marks,
mark=x,
mark options={solid,draw=black}
]
table[row sep=crcr]{
-0.016664 -0.483336\\
-0.032763 -0.466669\\
-0.0482972 -0.450002\\
-0.0632666 -0.433336\\
-0.077671 -0.416669\\
-0.0915106 -0.400002\\
-0.104785 -0.383335\\
-0.117495 -0.366669\\
-0.12964 -0.350002\\
-0.14122 -0.333335\\
-0.152235 -0.316668\\
-0.162685 -0.300002\\
-0.172571 -0.283335\\
-0.181891 -0.266668\\
-0.190647 -0.250001\\
-0.198838 -0.233335\\
-0.206464 -0.216668\\
-0.213525 -0.200001\\
-0.220021 -0.183334\\
-0.225952 -0.166668\\
-0.225952 -0.166668\\
-0.24431 -0.150001\\
-0.260735 -0.133334\\
-0.275228 -0.116667\\
-0.287789 -0.100001\\
-0.298417 -0.0833338\\
-0.307112 -0.066667\\
-0.313876 -0.0500003\\
-0.318707 -0.0333335\\
-0.321605 -0.0166668\\
-0.322572 -3.03577e-17\\
-0.321605 0.0166667\\
-0.318707 0.0333335\\
-0.313876 0.0500002\\
-0.307112 0.066667\\
-0.298417 0.0833337\\
-0.287789 0.1\\
-0.275228 0.116667\\
-0.260735 0.133334\\
-0.24431 0.150001\\
-0.225952 0.166667\\
-0.225952 0.166668\\
-0.220021 0.183334\\
-0.213525 0.200001\\
-0.206464 0.216668\\
-0.198838 0.233334\\
-0.190647 0.250001\\
-0.181891 0.266668\\
-0.172571 0.283335\\
-0.162685 0.300001\\
-0.152235 0.316668\\
-0.14122 0.333335\\
-0.12964 0.350002\\
-0.117495 0.366668\\
-0.104785 0.383335\\
-0.0915106 0.400002\\
-0.077671 0.416669\\
-0.0632666 0.433335\\
-0.0482972 0.450002\\
-0.032763 0.466669\\
-0.016664 0.483336\\
};
\addlegendentry{FSDT5};

\end{axis}

\begin{axis}[%
width=\figurewidth,
height=\figureheight,
scale only axis,
xmin=0,
xmax=1,
ymin=0,
ymax=1,
hide axis,
axis x line*=bottom,
axis y line*=left
]
\addplot [
color=black,
solid,
line width=0.0pt,
mark=square*,
mark options={solid,fill=black,draw=white},
forget plot
]
table[row sep=crcr]{
0.5 0.988095238095238\\
};
\addplot [
color=black,
solid,
line width=0.0pt,
mark=square*,
mark options={solid,fill=black,draw=white},
forget plot
]
table[row sep=crcr]{
0.991071428571429 0.5\\
};
\addplot [
color=black,
solid,
line width=0.0pt,
mark=square*,
mark options={solid,fill=black,draw=white},
forget plot
]
table[row sep=crcr]{
0.5 0.0119047619047619\\
};
\addplot [
color=black,
solid,
line width=0.0pt,
mark=square*,
mark options={solid,fill=black,draw=white},
forget plot
]
table[row sep=crcr]{
0.00892857142857143 0.5\\
};
\addplot [
color=black,
solid,
line width=0.0pt,
mark=square*,
mark options={solid,fill=black,draw=white},
forget plot
]
table[row sep=crcr]{
0.00892857142857143 0.988095238095238\\
};
\addplot [
color=black,
solid,
line width=0.0pt,
mark=square*,
mark options={solid,fill=black,draw=white},
forget plot
]
table[row sep=crcr]{
0.991071428571429 0.0119047619047619\\
};
\addplot [
color=black,
solid,
line width=0.0pt,
mark=square*,
mark options={solid,fill=black,draw=white},
forget plot
]
table[row sep=crcr]{
0.991071428571429 0.988095238095238\\
};
\addplot [
color=black,
solid,
line width=0.0pt,
mark=square*,
mark options={solid,fill=black,draw=white},
forget plot
]
table[row sep=crcr]{
0.00892857142857143 0.0119047619047619\\
};
\end{axis}
\end{tikzpicture}%
}}
\subfigure[$\overline{\tau}_{xz}=\overline{\tau}_{xz}(-a/2,0,z)$]{\scalebox{0.5}{
%
%
%
%
\begin{tikzpicture}

\begin{axis}[%
width=\figurewidth,
height=\figureheight,
scale only axis,
xmin=-0.2,
xmax=0,
xlabel={$\bar{\tau}_{xz}$},
ymin=-0.5,
ymax=0.5,
ylabel={Normalized thickness, z/h},
legend style={at={(0.715267857142857,0.626785714285716)},anchor=south west,draw=black,fill=white,legend cell align=left}
]
\addplot [
color=blue,
mark size=2.5pt,
only marks,
mark=o,
mark options={solid,draw=black}
]
table[row sep=crcr]{
-0.0149603 -0.483336\\
-0.0291607 -0.466669\\
-0.0426226 -0.450002\\
-0.0553669 -0.433336\\
-0.0674136 -0.416669\\
-0.0787823 -0.400002\\
-0.0894916 -0.383335\\
-0.0995599 -0.366669\\
-0.109005 -0.350002\\
-0.117843 -0.333335\\
-0.12609 -0.316668\\
-0.133763 -0.300002\\
-0.140875 -0.283335\\
-0.147442 -0.266668\\
-0.153476 -0.250001\\
-0.158992 -0.233335\\
-0.164 -0.216668\\
-0.168513 -0.200001\\
-0.172542 -0.183334\\
-0.176097 -0.166668\\
-0.176097 -0.166668\\
-0.158417 -0.150001\\
-0.142778 -0.133334\\
-0.129123 -0.116667\\
-0.117404 -0.100001\\
-0.107578 -0.0833338\\
-0.0996068 -0.066667\\
-0.0934614 -0.0500003\\
-0.0891174 -0.0333335\\
-0.0865568 -0.0166668\\
-0.0857681 -3.03577e-17\\
-0.086746 0.0166667\\
-0.0894915 0.0333335\\
-0.094012 0.0500002\\
-0.100321 0.066667\\
-0.108438 0.0833337\\
-0.11839 0.1\\
-0.13021 0.116667\\
-0.143934 0.133334\\
-0.15961 0.150001\\
-0.177288 0.166667\\
-0.177288 0.166668\\
-0.172796 0.183334\\
-0.167894 0.200001\\
-0.162576 0.216668\\
-0.156837 0.233334\\
-0.150671 0.250001\\
-0.144071 0.266668\\
-0.137031 0.283335\\
-0.129542 0.300001\\
-0.121596 0.316668\\
-0.113183 0.333335\\
-0.104294 0.350002\\
-0.094918 0.366668\\
-0.0850442 0.383335\\
-0.0746608 0.400002\\
-0.0637551 0.416669\\
-0.0523141 0.433335\\
-0.040324 0.450002\\
-0.0277705 0.466669\\
-0.0146384 0.483336\\
};
\addlegendentry{HSDT13};

\addplot [
color=black,
dashed,
line width=1.2pt
]
table[row sep=crcr]{
0 -0.500003\\
-0.014444 -0.483336\\
-0.0283435 -0.466669\\
-0.0416952 -0.450002\\
-0.0544959 -0.433336\\
-0.0667424 -0.416669\\
-0.0784316 -0.400002\\
-0.0895606 -0.383335\\
-0.100127 -0.366669\\
-0.110127 -0.350002\\
-0.119559 -0.333335\\
-0.128421 -0.316668\\
-0.13671 -0.300002\\
-0.144423 -0.283335\\
-0.151559 -0.266668\\
-0.158117 -0.250001\\
-0.164093 -0.233335\\
-0.169486 -0.216668\\
-0.174295 -0.200001\\
-0.178518 -0.183334\\
-0.182154 -0.166668\\
-0.182154 -0.166668\\
-0.164312 -0.150001\\
-0.14852 -0.133334\\
-0.134722 -0.116667\\
-0.122866 -0.100001\\
-0.112908 -0.0833338\\
-0.104812 -0.066667\\
-0.0985464 -0.0500003\\
-0.0940879 -0.0333335\\
-0.0914195 -0.0166668\\
-0.0905312 -3.03577e-17\\
-0.0914195 0.0166667\\
-0.0940879 0.0333335\\
-0.0985464 0.0500002\\
-0.104812 0.066667\\
-0.112908 0.0833337\\
-0.122866 0.1\\
-0.134722 0.116667\\
-0.14852 0.133334\\
-0.164312 0.150001\\
-0.182154 0.166667\\
-0.182154 0.166668\\
-0.178518 0.183334\\
-0.174295 0.200001\\
-0.169486 0.216668\\
-0.164093 0.233334\\
-0.158117 0.250001\\
-0.151559 0.266668\\
-0.144423 0.283335\\
-0.13671 0.300001\\
-0.128421 0.316668\\
-0.119559 0.333335\\
-0.110127 0.350002\\
-0.100127 0.366668\\
-0.0895606 0.383335\\
-0.0784316 0.400002\\
-0.0667424 0.416669\\
-0.0544959 0.433335\\
-0.0416952 0.450002\\
-0.0283435 0.466669\\
-0.014444 0.483336\\
2.54659e-17 0.500002\\
};
\addlegendentry{HSDT11A};

\addplot [
color=blue,
mark size=4.3pt,
only marks,
mark=diamond,
mark options={solid,draw=black}
]
table[row sep=crcr]{
-0.0181372 -0.483336\\
-0.0346758 -0.466669\\
-0.0497141 -0.450002\\
-0.0633469 -0.433336\\
-0.0756657 -0.416669\\
-0.0867588 -0.400002\\
-0.0967109 -0.383335\\
-0.105604 -0.366669\\
-0.113515 -0.350002\\
-0.12052 -0.333335\\
-0.126689 -0.316668\\
-0.132092 -0.300002\\
-0.136793 -0.283335\\
-0.140853 -0.266668\\
-0.14433 -0.250001\\
-0.147279 -0.233335\\
-0.149752 -0.216668\\
-0.151796 -0.200001\\
-0.153456 -0.183334\\
-0.154773 -0.166668\\
-0.154773 -0.166668\\
-0.137711 -0.150001\\
-0.122627 -0.133334\\
-0.109464 -0.116667\\
-0.0981729 -0.100001\\
-0.0887094 -0.0833338\\
-0.0810365 -0.066667\\
-0.0751234 -0.0500003\\
-0.0709454 -0.0333335\\
-0.0684845 -0.0166668\\
-0.0677291 -3.03577e-17\\
-0.0686737 0.0166667\\
-0.0713195 0.0333335\\
-0.0756739 0.0500002\\
-0.0817508 0.066667\\
-0.0895703 0.0833337\\
-0.0991592 0.1\\
-0.11055 0.116667\\
-0.123783 0.133334\\
-0.138904 0.150001\\
-0.155964 0.166667\\
-0.155964 0.166668\\
-0.15371 0.183334\\
-0.151176 0.200001\\
-0.148328 0.216668\\
-0.145124 0.233334\\
-0.141524 0.250001\\
-0.137482 0.266668\\
-0.132949 0.283335\\
-0.127872 0.300001\\
-0.122195 0.316668\\
-0.11586 0.333335\\
-0.108804 0.350002\\
-0.100962 0.366668\\
-0.0922635 0.383335\\
-0.0826373 0.400002\\
-0.0720072 0.416669\\
-0.0602941 0.433335\\
-0.0474155 0.450002\\
-0.0332857 0.466669\\
-0.0178154 0.483336\\
};
\addlegendentry{HSDT11B};

\addplot [
color=black,
solid,
line width=1.2pt
]
table[row sep=crcr]{
0 -0.500003\\
-0.0111411 -0.483336\\
-0.0219046 -0.466669\\
-0.0322904 -0.450002\\
-0.0422985 -0.433336\\
-0.0519289 -0.416669\\
-0.0611817 -0.400002\\
-0.0700569 -0.383335\\
-0.0785543 -0.366669\\
-0.0866741 -0.350002\\
-0.0944163 -0.333335\\
-0.101781 -0.316668\\
-0.108768 -0.300002\\
-0.115377 -0.283335\\
-0.121608 -0.266668\\
-0.127462 -0.250001\\
-0.132938 -0.233335\\
-0.138037 -0.216668\\
-0.142757 -0.200001\\
-0.147101 -0.183334\\
-0.151066 -0.166668\\
-0.151066 -0.166668\\
-0.132627 -0.150001\\
-0.116128 -0.133334\\
-0.101571 -0.116667\\
-0.0889545 -0.100001\\
-0.0782791 -0.0833338\\
-0.0695447 -0.066667\\
-0.0627512 -0.0500003\\
-0.0578988 -0.0333335\\
-0.0549873 -0.0166668\\
-0.0540168 -3.03577e-17\\
-0.0549873 0.0166667\\
-0.0578988 0.0333335\\
-0.0627512 0.0500002\\
-0.0695447 0.066667\\
-0.0782791 0.0833337\\
-0.0889545 0.1\\
-0.101571 0.116667\\
-0.116128 0.133334\\
-0.132627 0.150001\\
-0.151066 0.166667\\
-0.151066 0.166668\\
-0.147101 0.183334\\
-0.142757 0.200001\\
-0.138037 0.216668\\
-0.132938 0.233334\\
-0.127462 0.250001\\
-0.121608 0.266668\\
-0.115377 0.283335\\
-0.108768 0.300001\\
-0.101781 0.316668\\
-0.0944163 0.333335\\
-0.0866741 0.350002\\
-0.0785543 0.366668\\
-0.0700569 0.383335\\
-0.0611817 0.400002\\
-0.0519289 0.416669\\
-0.0422985 0.433335\\
-0.0322904 0.450002\\
-0.0219046 0.466669\\
-0.0111411 0.483336\\
1.89175e-16 0.500002\\
};
\addlegendentry{TSDT7};

\addplot [
color=blue,
mark size=2.5pt,
only marks,
mark=x,
mark options={solid,draw=black}
]
table[row sep=crcr]{
-0.0126521 -0.483336\\
-0.0248753 -0.466669\\
-0.0366696 -0.450002\\
-0.0480351 -0.433336\\
-0.0589716 -0.416669\\
-0.0694793 -0.400002\\
-0.0795581 -0.383335\\
-0.089208 -0.366669\\
-0.098429 -0.350002\\
-0.107221 -0.333335\\
-0.115584 -0.316668\\
-0.123519 -0.300002\\
-0.131024 -0.283335\\
-0.138101 -0.266668\\
-0.144748 -0.250001\\
-0.150967 -0.233335\\
-0.156757 -0.216668\\
-0.162118 -0.200001\\
-0.16705 -0.183334\\
-0.171554 -0.166668\\
-0.171554 -0.166668\\
-0.153067 -0.150001\\
-0.136527 -0.133334\\
-0.121932 -0.116667\\
-0.109283 -0.100001\\
-0.0985805 -0.0833338\\
-0.0898237 -0.066667\\
-0.0830129 -0.0500003\\
-0.078148 -0.0333335\\
-0.0752291 -0.0166668\\
-0.0742561 -3.03577e-17\\
-0.0752291 0.0166667\\
-0.078148 0.0333335\\
-0.0830129 0.0500002\\
-0.0898237 0.066667\\
-0.0985805 0.0833337\\
-0.109283 0.1\\
-0.121932 0.116667\\
-0.136527 0.133334\\
-0.153067 0.150001\\
-0.171554 0.166667\\
-0.171554 0.166668\\
-0.16705 0.183334\\
-0.162118 0.200001\\
-0.156757 0.216668\\
-0.150967 0.233334\\
-0.144748 0.250001\\
-0.138101 0.266668\\
-0.131024 0.283335\\
-0.123519 0.300001\\
-0.115584 0.316668\\
-0.107221 0.333335\\
-0.098429 0.350002\\
-0.089208 0.366668\\
-0.0795581 0.383335\\
-0.0694793 0.400002\\
-0.0589716 0.416669\\
-0.0480351 0.433335\\
-0.0366696 0.450002\\
-0.0248753 0.466669\\
-0.0126521 0.483336\\
};
\addlegendentry{FSDT5};

\end{axis}

\begin{axis}[%
width=\figurewidth,
height=\figureheight,
scale only axis,
xmin=0,
xmax=1,
ymin=0,
ymax=1,
hide axis,
axis x line*=bottom,
axis y line*=left
]
\addplot [
color=black,
solid,
line width=0.0pt,
mark=square*,
mark options={solid,fill=black,draw=white},
forget plot
]
table[row sep=crcr]{
0.5 0.988095238095238\\
};
\addplot [
color=black,
solid,
line width=0.0pt,
mark=square*,
mark options={solid,fill=black,draw=white},
forget plot
]
table[row sep=crcr]{
0.991071428571429 0.5\\
};
\addplot [
color=black,
solid,
line width=0.0pt,
mark=square*,
mark options={solid,fill=black,draw=white},
forget plot
]
table[row sep=crcr]{
0.5 0.0119047619047619\\
};
\addplot [
color=black,
solid,
line width=0.0pt,
mark=square*,
mark options={solid,fill=black,draw=white},
forget plot
]
table[row sep=crcr]{
0.00892857142857143 0.5\\
};
\addplot [
color=black,
solid,
line width=0.0pt,
mark=square*,
mark options={solid,fill=black,draw=white},
forget plot
]
table[row sep=crcr]{
0.00892857142857143 0.988095238095238\\
};
\addplot [
color=black,
solid,
line width=0.0pt,
mark=square*,
mark options={solid,fill=black,draw=white},
forget plot
]
table[row sep=crcr]{
0.991071428571429 0.0119047619047619\\
};
\addplot [
color=black,
solid,
line width=0.0pt,
mark=square*,
mark options={solid,fill=black,draw=white},
forget plot
]
table[row sep=crcr]{
0.991071428571429 0.988095238095238\\
};
\addplot [
color=black,
solid,
line width=0.0pt,
mark=square*,
mark options={solid,fill=black,draw=white},
forget plot
]
table[row sep=crcr]{
0.00892857142857143 0.0119047619047619\\
};
\end{axis}
\end{tikzpicture}%
}}\hspace{60pt}
\subfigure[$\overline{\sigma}_{zz} = \overline{\sigma}_{zz}(0,0,z)$]{\scalebox{0.5}{
%
%
%
%
\begin{tikzpicture}

\begin{axis}[%
width=\figurewidth,
height=\figureheight,
scale only axis,
xmin=-0.4,
xmax=1.2,
xlabel={$\bar{\sigma}_{zz}$},
ymin=-0.5,
ymax=0.5,
ylabel={Normalized thickness, z/h},
legend style={draw=black,fill=white,legend cell align=left},
legend pos = south east
]
\addplot [
color=blue,
mark size=2.5pt,
only marks,
mark=o,
mark options={solid,draw=black}
]
table[row sep=crcr]{
-0.160472 -0.483336\\
-0.126404 -0.466669\\
-0.0931204 -0.450002\\
-0.0605928 -0.433336\\
-0.0287922 -0.416669\\
0.00231027 -0.400002\\
0.0327436 -0.383335\\
0.0625368 -0.366669\\
0.0917186 -0.350002\\
0.120318 -0.333335\\
0.148364 -0.316668\\
0.175886 -0.300002\\
0.202912 -0.283335\\
0.229472 -0.266668\\
0.255594 -0.250001\\
0.281307 -0.233335\\
0.306641 -0.216668\\
0.331624 -0.200001\\
0.356285 -0.183334\\
0.380653 -0.166668\\
0.349753 -0.166668\\
0.360359 -0.150001\\
0.370822 -0.133334\\
0.38116 -0.116667\\
0.391393 -0.100001\\
0.401538 -0.0833338\\
0.411615 -0.066667\\
0.42164 -0.0500003\\
0.431633 -0.0333335\\
0.441612 -0.0166668\\
0.451595 -3.03577e-17\\
0.4616 0.0166667\\
0.471647 0.0333335\\
0.481753 0.0500002\\
0.491936 0.066667\\
0.502215 0.0833337\\
0.512609 0.1\\
0.523136 0.116667\\
0.533813 0.133334\\
0.54466 0.150001\\
0.555694 0.166667\\
0.524202 0.166668\\
0.549104 0.183334\\
0.57435 0.200001\\
0.599969 0.216668\\
0.625989 0.233334\\
0.65244 0.250001\\
0.67935 0.266668\\
0.706749 0.283335\\
0.734666 0.300001\\
0.763128 0.316668\\
0.792167 0.333335\\
0.821809 0.350002\\
0.852084 0.366668\\
0.883022 0.383335\\
0.914651 0.400002\\
0.946999 0.416669\\
0.980097 0.433335\\
1.01397 0.450002\\
1.04866 0.466669\\
1.08417 0.483336\\
};
\addlegendentry{HSDT13};

\addplot [
color=black,
dashed,
line width=1.2pt
]
table[row sep=crcr]{
0 -0.500003\\
0 -0.483336\\
0 -0.466669\\
0 -0.450002\\
0 -0.433336\\
0 -0.416669\\
0 -0.400002\\
0 -0.383335\\
0 -0.366669\\
0 -0.350002\\
0 -0.333335\\
0 -0.316668\\
0 -0.300002\\
0 -0.283335\\
0 -0.266668\\
0 -0.250001\\
0 -0.233335\\
0 -0.216668\\
0 -0.200001\\
0 -0.183334\\
0 -0.166668\\
0 -0.166668\\
0 -0.150001\\
0 -0.133334\\
0 -0.116667\\
0 -0.100001\\
0 -0.0833338\\
0 -0.066667\\
0 -0.0500003\\
0 -0.0333335\\
0 -0.0166668\\
0 -3.03577e-17\\
0 0.0166667\\
0 0.0333335\\
0 0.0500002\\
0 0.066667\\
0 0.0833337\\
0 0.1\\
0 0.116667\\
0 0.133334\\
0 0.150001\\
0 0.166667\\
0 0.166668\\
0 0.183334\\
0 0.200001\\
0 0.216668\\
0 0.233334\\
0 0.250001\\
0 0.266668\\
0 0.283335\\
0 0.300001\\
0 0.316668\\
0 0.333335\\
0 0.350002\\
0 0.366668\\
0 0.383335\\
0 0.400002\\
0 0.416669\\
0 0.433335\\
0 0.450002\\
0 0.466669\\
0 0.483336\\
0 0.500002\\
};
\addlegendentry{HSDT11A};

\addplot [
color=blue,
mark size=4.3pt,
only marks,
mark=diamond,
mark options={solid,draw=black}
]
table[row sep=crcr]{
-0.187869 -0.483336\\
-0.142686 -0.466669\\
-0.0999524 -0.450002\\
-0.0595787 -0.433336\\
-0.0214769 -0.416669\\
0.0144414 -0.400002\\
0.0482645 -0.383335\\
0.0800808 -0.366669\\
0.109979 -0.350002\\
0.138046 -0.333335\\
0.164373 -0.316668\\
0.189045 -0.300002\\
0.212153 -0.283335\\
0.233784 -0.266668\\
0.254027 -0.250001\\
0.272969 -0.233335\\
0.290701 -0.216668\\
0.307309 -0.200001\\
0.322882 -0.183334\\
0.337508 -0.166668\\
0.298767 -0.166668\\
0.316695 -0.150001\\
0.333779 -0.133334\\
0.350114 -0.116667\\
0.365797 -0.100001\\
0.380923 -0.0833338\\
0.395591 -0.066667\\
0.409895 -0.0500003\\
0.423933 -0.0333335\\
0.437801 -0.0166668\\
0.451595 -3.03577e-17\\
0.465411 0.0166667\\
0.479346 0.0333335\\
0.493497 0.0500002\\
0.50796 0.066667\\
0.52283 0.0833337\\
0.538206 0.1\\
0.554182 0.116667\\
0.570856 0.133334\\
0.588323 0.150001\\
0.606681 0.166667\\
0.567346 0.166668\\
0.582507 0.183334\\
0.598665 0.200001\\
0.615909 0.216668\\
0.634327 0.233334\\
0.654007 0.250001\\
0.675038 0.266668\\
0.697509 0.283335\\
0.721507 0.300001\\
0.74712 0.316668\\
0.774438 0.333335\\
0.803549 0.350002\\
0.83454 0.366668\\
0.867501 0.383335\\
0.90252 0.400002\\
0.939684 0.416669\\
0.979083 0.433335\\
1.0208 0.450002\\
1.06494 0.466669\\
1.11157 0.483336\\
};
\addlegendentry{HSDT11B};

\addplot [
color=black,
solid,
line width=1.2pt
]
table[row sep=crcr]{
0 -0.500003\\
0 -0.483336\\
0 -0.466669\\
0 -0.450002\\
0 -0.433336\\
0 -0.416669\\
0 -0.400002\\
0 -0.383335\\
0 -0.366669\\
0 -0.350002\\
0 -0.333335\\
0 -0.316668\\
0 -0.300002\\
0 -0.283335\\
0 -0.266668\\
0 -0.250001\\
0 -0.233335\\
0 -0.216668\\
0 -0.200001\\
0 -0.183334\\
0 -0.166668\\
0 -0.166668\\
0 -0.150001\\
0 -0.133334\\
0 -0.116667\\
0 -0.100001\\
0 -0.0833338\\
0 -0.066667\\
0 -0.0500003\\
0 -0.0333335\\
0 -0.0166668\\
0 -3.03577e-17\\
0 0.0166667\\
0 0.0333335\\
0 0.0500002\\
0 0.066667\\
0 0.0833337\\
0 0.1\\
0 0.116667\\
0 0.133334\\
0 0.150001\\
0 0.166667\\
0 0.166668\\
0 0.183334\\
0 0.200001\\
0 0.216668\\
0 0.233334\\
0 0.250001\\
0 0.266668\\
0 0.283335\\
0 0.300001\\
0 0.316668\\
0 0.333335\\
0 0.350002\\
0 0.366668\\
0 0.383335\\
0 0.400002\\
0 0.416669\\
0 0.433335\\
0 0.450002\\
0 0.466669\\
0 0.483336\\
0 0.500002\\
};
\addlegendentry{TSDT7};

\addplot [
color=blue,
mark size=2.5pt,
only marks,
mark=x,
mark options={solid,draw=black}
]
table[row sep=crcr]{
0 -0.483336\\
0 -0.466669\\
0 -0.450002\\
0 -0.433336\\
0 -0.416669\\
0 -0.400002\\
0 -0.383335\\
0 -0.366669\\
0 -0.350002\\
0 -0.333335\\
0 -0.316668\\
0 -0.300002\\
0 -0.283335\\
0 -0.266668\\
0 -0.250001\\
0 -0.233335\\
0 -0.216668\\
0 -0.200001\\
0 -0.183334\\
0 -0.166668\\
0 -0.166668\\
0 -0.150001\\
0 -0.133334\\
0 -0.116667\\
0 -0.100001\\
0 -0.0833338\\
0 -0.066667\\
0 -0.0500003\\
0 -0.0333335\\
0 -0.0166668\\
0 -3.03577e-17\\
0 0.0166667\\
0 0.0333335\\
0 0.0500002\\
0 0.066667\\
0 0.0833337\\
0 0.1\\
0 0.116667\\
0 0.133334\\
0 0.150001\\
0 0.166667\\
0 0.166668\\
0 0.183334\\
0 0.200001\\
0 0.216668\\
0 0.233334\\
0 0.250001\\
0 0.266668\\
0 0.283335\\
0 0.300001\\
0 0.316668\\
0 0.333335\\
0 0.350002\\
0 0.366668\\
0 0.383335\\
0 0.400002\\
0 0.416669\\
0 0.433335\\
0 0.450002\\
0 0.466669\\
0 0.483336\\
};
\addlegendentry{FSDT5};

\end{axis}

\begin{axis}[%
width=\figurewidth,
height=\figureheight,
scale only axis,
xmin=0,
xmax=1,
ymin=0,
ymax=1,
hide axis,
axis x line*=bottom,
axis y line*=left
]
\addplot [
color=black,
solid,
line width=0.0pt,
mark=square*,
mark options={solid,fill=black,draw=white},
forget plot
]
table[row sep=crcr]{
0.5 0.988095238095238\\
};
\addplot [
color=black,
solid,
line width=0.0pt,
mark=square*,
mark options={solid,fill=black,draw=white},
forget plot
]
table[row sep=crcr]{
0.991071428571429 0.5\\
};
\addplot [
color=black,
solid,
line width=0.0pt,
mark=square*,
mark options={solid,fill=black,draw=white},
forget plot
]
table[row sep=crcr]{
0.5 0.0119047619047619\\
};
\addplot [
color=black,
solid,
line width=0.0pt,
mark=square*,
mark options={solid,fill=black,draw=white},
forget plot
]
table[row sep=crcr]{
0.00892857142857143 0.5\\
};
\addplot [
color=black,
solid,
line width=0.0pt,
mark=square*,
mark options={solid,fill=black,draw=white},
forget plot
]
table[row sep=crcr]{
0.00892857142857143 0.988095238095238\\
};
\addplot [
color=black,
solid,
line width=0.0pt,
mark=square*,
mark options={solid,fill=black,draw=white},
forget plot
]
table[row sep=crcr]{
0.991071428571429 0.0119047619047619\\
};
\addplot [
color=black,
solid,
line width=0.0pt,
mark=square*,
mark options={solid,fill=black,draw=white},
forget plot
]
table[row sep=crcr]{
0.991071428571429 0.988095238095238\\
};
\addplot [
color=black,
solid,
line width=0.0pt,
mark=square*,
mark options={solid,fill=black,draw=white},
forget plot
]
table[row sep=crcr]{
0.00892857142857143 0.0119047619047619\\
};
\end{axis}
\end{tikzpicture}%
}}
\caption{The variation of non-dimensional deflections and stresses through the thickness of a three-layered symmetric curved fibre composite laminate under sinusoidally distributed mechanical load with $a/h=$ 5, $T_o=$ -45$^\circ$, $T_1=$ 45$^\circ$.}
\label{fig:3LayerMechLoad}
\end{figure}

\begin{figure}
\centering
\setlength\figureheight{10cm} 
\setlength\figurewidth{12cm}
\subfigure[$\overline{u} = \overline{u}(-a/2,0,z)$]{\scalebox{0.5}{
%
%
%
%
\begin{tikzpicture}

\begin{axis}[%
width=\figurewidth,
height=\figureheight,
scale only axis,
xmin=-30,
xmax=30,
xlabel={$\bar{u}$},
ymin=-0.5,
ymax=0.5,
ylabel={Normalized thickness, z/h},
legend style={draw=black,fill=white,legend cell align=left}
]
\addplot [
color=blue,
mark size=2.5pt,
only marks,
mark=o,
mark options={solid,draw=black}
]
table[row sep=crcr]{
26.3376 -0.483336\\
24.7897 -0.466669\\
23.304 -0.450002\\
21.8785 -0.433336\\
20.5109 -0.416669\\
19.1989 -0.400002\\
17.9404 -0.383335\\
16.7331 -0.366669\\
15.5747 -0.350002\\
14.4631 -0.333335\\
13.3961 -0.316668\\
12.3713 -0.300002\\
11.3866 -0.283335\\
10.4398 -0.266668\\
9.52856 -0.250001\\
8.65074 -0.233335\\
7.8041 -0.216668\\
6.98639 -0.200001\\
6.1954 -0.183334\\
5.42891 -0.166668\\
5.42891 -0.166668\\
4.82256 -0.150001\\
4.23626 -0.133334\\
3.66776 -0.116667\\
3.11485 -0.100001\\
2.5753 -0.0833338\\
2.04688 -0.066667\\
1.52737 -0.0500003\\
1.01454 -0.0333335\\
0.506155 -0.0166668\\
4.91171e-16 -3.03577e-17\\
-0.506155 0.0166667\\
-1.01454 0.0333335\\
-1.52737 0.0500002\\
-2.04688 0.066667\\
-2.5753 0.0833337\\
-3.11485 0.1\\
-3.66776 0.116667\\
-4.23626 0.133334\\
-4.82256 0.150001\\
-5.42891 0.166667\\
-5.42891 0.166668\\
-6.1954 0.183334\\
-6.98639 0.200001\\
-7.8041 0.216668\\
-8.65074 0.233334\\
-9.52856 0.250001\\
-10.4398 0.266668\\
-11.3866 0.283335\\
-12.3713 0.300001\\
-13.3961 0.316668\\
-14.4631 0.333335\\
-15.5747 0.350002\\
-16.7331 0.366668\\
-17.9404 0.383335\\
-19.1989 0.400002\\
-20.5109 0.416669\\
-21.8785 0.433335\\
-23.304 0.450002\\
-24.7897 0.466669\\
-26.3376 0.483336\\
};
\addlegendentry{HSDT13};

\addplot [
color=black,
dashed,
line width=1.2pt
]
table[row sep=crcr]{
25.2903 -0.500003\\
24.1665 -0.483336\\
23.0652 -0.466669\\
21.9854 -0.450002\\
20.9266 -0.433336\\
19.8878 -0.416669\\
18.8683 -0.400002\\
17.8674 -0.383335\\
16.8842 -0.366669\\
15.918 -0.350002\\
14.9681 -0.333335\\
14.0336 -0.316668\\
13.1138 -0.300002\\
12.2079 -0.283335\\
11.3151 -0.266668\\
10.4346 -0.250001\\
9.56581 -0.233335\\
8.70778 -0.216668\\
7.8598 -0.200001\\
7.02109 -0.183334\\
6.19087 -0.166668\\
6.19087 -0.166668\\
5.54977 -0.150001\\
4.91562 -0.133334\\
4.28765 -0.116667\\
3.66509 -0.100001\\
3.04716 -0.0833338\\
2.43309 -0.066667\\
1.82212 -0.0500003\\
1.21346 -0.0333335\\
0.606342 -0.0166668\\
-2.99313e-16 -3.03577e-17\\
-0.606342 0.0166667\\
-1.21346 0.0333335\\
-1.82212 0.0500002\\
-2.43309 0.066667\\
-3.04716 0.0833337\\
-3.66509 0.1\\
-4.28765 0.116667\\
-4.91562 0.133334\\
-5.54977 0.150001\\
-6.19087 0.166667\\
-6.19087 0.166668\\
-7.02109 0.183334\\
-7.8598 0.200001\\
-8.70778 0.216668\\
-9.56581 0.233334\\
-10.4346 0.250001\\
-11.3151 0.266668\\
-12.2079 0.283335\\
-13.1138 0.300001\\
-14.0336 0.316668\\
-14.9681 0.333335\\
-15.918 0.350002\\
-16.8842 0.366668\\
-17.8674 0.383335\\
-18.8683 0.400002\\
-19.8878 0.416669\\
-20.9266 0.433335\\
-21.9854 0.450002\\
-23.0652 0.466669\\
-24.1665 0.483336\\
-25.2903 0.500002\\
};
\addlegendentry{HSDT11A};

\addplot [
color=blue,
mark size=4.3pt,
only marks,
mark=diamond,
mark options={solid,draw=black}
]
table[row sep=crcr]{
24.9175 -0.483336\\
23.3861 -0.466669\\
21.9255 -0.450002\\
20.5331 -0.433336\\
19.2064 -0.416669\\
17.9429 -0.400002\\
16.74 -0.383335\\
15.5953 -0.366669\\
14.5061 -0.350002\\
13.47 -0.333335\\
12.4844 -0.316668\\
11.5469 -0.300002\\
10.6548 -0.283335\\
9.80569 -0.266668\\
8.99701 -0.250001\\
8.22623 -0.233335\\
7.49082 -0.216668\\
6.78826 -0.200001\\
6.11602 -0.183334\\
5.47158 -0.166668\\
5.47158 -0.166668\\
4.85241 -0.150001\\
4.25598 -0.133334\\
3.67976 -0.116667\\
3.12123 -0.100001\\
2.57786 -0.0833338\\
2.04713 -0.066667\\
1.5265 -0.0500003\\
1.01346 -0.0333335\\
0.505465 -0.0166668\\
2.04845e-16 -3.03577e-17\\
-0.505465 0.0166667\\
-1.01346 0.0333335\\
-1.5265 0.0500002\\
-2.04713 0.066667\\
-2.57786 0.0833337\\
-3.12123 0.1\\
-3.67976 0.116667\\
-4.25598 0.133334\\
-4.85241 0.150001\\
-5.47158 0.166667\\
-5.47158 0.166668\\
-6.11602 0.183334\\
-6.78826 0.200001\\
-7.49082 0.216668\\
-8.22623 0.233334\\
-8.99701 0.250001\\
-9.80569 0.266668\\
-10.6548 0.283335\\
-11.5469 0.300001\\
-12.4844 0.316668\\
-13.47 0.333335\\
-14.5061 0.350002\\
-15.5953 0.366668\\
-16.74 0.383335\\
-17.9429 0.400002\\
-19.2064 0.416669\\
-20.5331 0.433335\\
-21.9255 0.450002\\
-23.3861 0.466669\\
-24.9175 0.483336\\
};
\addlegendentry{HSDT11B};

\addplot [
color=black,
solid,
line width=1.2pt
]
table[row sep=crcr]{
21.9543 -0.500003\\
21.2225 -0.483336\\
20.4907 -0.466669\\
19.7589 -0.450002\\
19.0271 -0.433336\\
18.2953 -0.416669\\
17.5634 -0.400002\\
16.8316 -0.383335\\
16.0998 -0.366669\\
15.368 -0.350002\\
14.6362 -0.333335\\
13.9044 -0.316668\\
13.1726 -0.300002\\
12.4408 -0.283335\\
11.709 -0.266668\\
10.9772 -0.250001\\
10.2453 -0.233335\\
9.51353 -0.216668\\
8.78172 -0.200001\\
8.04991 -0.183334\\
7.3181 -0.166668\\
7.3181 -0.166668\\
6.58629 -0.150001\\
5.85448 -0.133334\\
5.12267 -0.116667\\
4.39086 -0.100001\\
3.65905 -0.0833338\\
2.92724 -0.066667\\
2.19543 -0.0500003\\
1.46362 -0.0333335\\
0.73181 -0.0166668\\
2.61737e-16 -3.03577e-17\\
-0.73181 0.0166667\\
-1.46362 0.0333335\\
-2.19543 0.0500002\\
-2.92724 0.066667\\
-3.65905 0.0833337\\
-4.39086 0.1\\
-5.12267 0.116667\\
-5.85448 0.133334\\
-6.58629 0.150001\\
-7.3181 0.166667\\
-7.3181 0.166668\\
-8.04991 0.183334\\
-8.78172 0.200001\\
-9.51353 0.216668\\
-10.2453 0.233334\\
-10.9772 0.250001\\
-11.709 0.266668\\
-12.4408 0.283335\\
-13.1726 0.300001\\
-13.9044 0.316668\\
-14.6362 0.333335\\
-15.368 0.350002\\
-16.0998 0.366668\\
-16.8316 0.383335\\
-17.5634 0.400002\\
-18.2953 0.416669\\
-19.0271 0.433335\\
-19.7589 0.450002\\
-20.4907 0.466669\\
-21.2225 0.483336\\
-21.9543 0.500002\\
};
\addlegendentry{TSDT7};

\addplot [
color=blue,
mark size=2.5pt,
only marks,
mark=x,
mark options={solid,draw=black}
]
table[row sep=crcr]{
21.3827 -0.483336\\
20.6453 -0.466669\\
19.908 -0.450002\\
19.1707 -0.433336\\
18.4333 -0.416669\\
17.696 -0.400002\\
16.9587 -0.383335\\
16.2213 -0.366669\\
15.484 -0.350002\\
14.7467 -0.333335\\
14.0093 -0.316668\\
13.272 -0.300002\\
12.5347 -0.283335\\
11.7973 -0.266668\\
11.06 -0.250001\\
10.3227 -0.233335\\
9.58533 -0.216668\\
8.848 -0.200001\\
8.11066 -0.183334\\
7.37333 -0.166668\\
7.37333 -0.166668\\
6.636 -0.150001\\
5.89866 -0.133334\\
5.16133 -0.116667\\
4.424 -0.100001\\
3.68666 -0.0833338\\
2.94933 -0.066667\\
2.212 -0.0500003\\
1.47467 -0.0333335\\
0.737333 -0.0166668\\
-2.75252e-16 -3.03577e-17\\
-0.737333 0.0166667\\
-1.47467 0.0333335\\
-2.212 0.0500002\\
-2.94933 0.066667\\
-3.68666 0.0833337\\
-4.424 0.1\\
-5.16133 0.116667\\
-5.89866 0.133334\\
-6.636 0.150001\\
-7.37333 0.166667\\
-7.37333 0.166668\\
-8.11066 0.183334\\
-8.848 0.200001\\
-9.58533 0.216668\\
-10.3227 0.233334\\
-11.06 0.250001\\
-11.7973 0.266668\\
-12.5347 0.283335\\
-13.272 0.300001\\
-14.0093 0.316668\\
-14.7467 0.333335\\
-15.484 0.350002\\
-16.2213 0.366668\\
-16.9587 0.383335\\
-17.696 0.400002\\
-18.4333 0.416669\\
-19.1707 0.433335\\
-19.908 0.450002\\
-20.6453 0.466669\\
-21.3827 0.483336\\
};
\addlegendentry{FSDT5};

\end{axis}
\end{tikzpicture}%
}}\hspace{60pt}
\subfigure[$\overline{w} = \overline{w}(0,0,z)$]{\scalebox{0.5}{
%
%
%
%
\begin{tikzpicture}

\begin{axis}[%
width=\figurewidth,
height=\figureheight,
scale only axis,
xmin=20,
xmax=35,
xlabel={$\bar{w}$},
ymin=-0.5,
ymax=0.5,
ylabel={Normalized thickness, z/h},
legend style={at={(1.0,0.5)},draw=black,fill=white,legend cell align=left}
]
\addplot [
color=blue,
mark size=2.5pt,
only marks,
mark=o,
mark options={solid,draw=black}
]
table[row sep=crcr]{
33.5092 -0.483336\\
32.6805 -0.466669\\
31.881 -0.450002\\
31.1105 -0.433336\\
30.3691 -0.416669\\
29.6568 -0.400002\\
28.9735 -0.383335\\
28.3193 -0.366669\\
27.6942 -0.350002\\
27.0982 -0.333335\\
26.5312 -0.316668\\
25.9934 -0.300002\\
25.4845 -0.283335\\
25.0048 -0.266668\\
24.5541 -0.250001\\
24.1326 -0.233335\\
23.7401 -0.216668\\
23.3766 -0.200001\\
23.0423 -0.183334\\
22.737 -0.166668\\
22.737 -0.166668\\
22.4608 -0.150001\\
22.2136 -0.133334\\
21.9956 -0.116667\\
21.8066 -0.100001\\
21.6467 -0.0833338\\
21.5158 -0.066667\\
21.4141 -0.0500003\\
21.3414 -0.0333335\\
21.2978 -0.0166668\\
21.2832 -3.03577e-17\\
21.2978 0.0166667\\
21.3414 0.0333335\\
21.4141 0.0500002\\
21.5158 0.066667\\
21.6467 0.0833337\\
21.8066 0.1\\
21.9956 0.116667\\
22.2136 0.133334\\
22.4608 0.150001\\
22.737 0.166667\\
22.737 0.166668\\
23.0423 0.183334\\
23.3766 0.200001\\
23.7401 0.216668\\
24.1326 0.233334\\
24.5541 0.250001\\
25.0048 0.266668\\
25.4845 0.283335\\
25.9934 0.300001\\
26.5312 0.316668\\
27.0982 0.333335\\
27.6942 0.350002\\
28.3193 0.366668\\
28.9735 0.383335\\
29.6568 0.400002\\
30.3691 0.416669\\
31.1105 0.433335\\
31.881 0.450002\\
32.6805 0.466669\\
33.5092 0.483336\\
};
\addlegendentry{HSDT13};

\addplot [
color=black,
dashed,
line width=1.2pt
]
table[row sep=crcr]{
23.9682 -0.500003\\
23.9682 -0.483336\\
23.9682 -0.466669\\
23.9682 -0.450002\\
23.9682 -0.433336\\
23.9682 -0.416669\\
23.9682 -0.400002\\
23.9682 -0.383335\\
23.9682 -0.366669\\
23.9682 -0.350002\\
23.9682 -0.333335\\
23.9682 -0.316668\\
23.9682 -0.300002\\
23.9682 -0.283335\\
23.9682 -0.266668\\
23.9682 -0.250001\\
23.9682 -0.233335\\
23.9682 -0.216668\\
23.9682 -0.200001\\
23.9682 -0.183334\\
23.9682 -0.166668\\
23.9682 -0.166668\\
23.9682 -0.150001\\
23.9682 -0.133334\\
23.9682 -0.116667\\
23.9682 -0.100001\\
23.9682 -0.0833338\\
23.9682 -0.066667\\
23.9682 -0.0500003\\
23.9682 -0.0333335\\
23.9682 -0.0166668\\
23.9682 -3.03577e-17\\
23.9682 0.0166667\\
23.9682 0.0333335\\
23.9682 0.0500002\\
23.9682 0.066667\\
23.9682 0.0833337\\
23.9682 0.1\\
23.9682 0.116667\\
23.9682 0.133334\\
23.9682 0.150001\\
23.9682 0.166667\\
23.9682 0.166668\\
23.9682 0.183334\\
23.9682 0.200001\\
23.9682 0.216668\\
23.9682 0.233334\\
23.9682 0.250001\\
23.9682 0.266668\\
23.9682 0.283335\\
23.9682 0.300001\\
23.9682 0.316668\\
23.9682 0.333335\\
23.9682 0.350002\\
23.9682 0.366668\\
23.9682 0.383335\\
23.9682 0.400002\\
23.9682 0.416669\\
23.9682 0.433335\\
23.9682 0.450002\\
23.9682 0.466669\\
23.9682 0.483336\\
23.9682 0.500002\\
};
\addlegendentry{HSDT11A};

\addplot [
color=blue,
mark size=4.3pt,
only marks,
mark=diamond,
mark options={solid,draw=black}
]
table[row sep=crcr]{
32.8765 -0.483336\\
32.0417 -0.466669\\
31.2363 -0.450002\\
30.4601 -0.433336\\
29.7132 -0.416669\\
28.9957 -0.400002\\
28.3074 -0.383335\\
27.6484 -0.366669\\
27.0187 -0.350002\\
26.4182 -0.333335\\
25.8471 -0.316668\\
25.3053 -0.300002\\
24.7927 -0.283335\\
24.3094 -0.266668\\
23.8554 -0.250001\\
23.4308 -0.233335\\
23.0354 -0.216668\\
22.6692 -0.200001\\
22.3324 -0.183334\\
22.0249 -0.166668\\
22.0249 -0.166668\\
21.7466 -0.150001\\
21.4977 -0.133334\\
21.278 -0.116667\\
21.0876 -0.100001\\
20.9266 -0.0833338\\
20.7948 -0.066667\\
20.6922 -0.0500003\\
20.619 -0.0333335\\
20.5751 -0.0166668\\
20.5604 -3.03577e-17\\
20.5751 0.0166667\\
20.619 0.0333335\\
20.6922 0.0500002\\
20.7948 0.066667\\
20.9266 0.0833337\\
21.0876 0.1\\
21.278 0.116667\\
21.4977 0.133334\\
21.7466 0.150001\\
22.0249 0.166667\\
22.0249 0.166668\\
22.3324 0.183334\\
22.6692 0.200001\\
23.0354 0.216668\\
23.4308 0.233334\\
23.8554 0.250001\\
24.3094 0.266668\\
24.7927 0.283335\\
25.3053 0.300001\\
25.8471 0.316668\\
26.4182 0.333335\\
27.0187 0.350002\\
27.6484 0.366668\\
28.3074 0.383335\\
28.9957 0.400002\\
29.7132 0.416669\\
30.4601 0.433335\\
31.2363 0.450002\\
32.0417 0.466669\\
32.8765 0.483336\\
};
\addlegendentry{HSDT11B};

\addplot [
color=black,
solid,
line width=1.2pt
]
table[row sep=crcr]{
21.6789 -0.500003\\
21.6789 -0.483336\\
21.6789 -0.466669\\
21.6789 -0.450002\\
21.6789 -0.433336\\
21.6789 -0.416669\\
21.6789 -0.400002\\
21.6789 -0.383335\\
21.6789 -0.366669\\
21.6789 -0.350002\\
21.6789 -0.333335\\
21.6789 -0.316668\\
21.6789 -0.300002\\
21.6789 -0.283335\\
21.6789 -0.266668\\
21.6789 -0.250001\\
21.6789 -0.233335\\
21.6789 -0.216668\\
21.6789 -0.200001\\
21.6789 -0.183334\\
21.6789 -0.166668\\
21.6789 -0.166668\\
21.6789 -0.150001\\
21.6789 -0.133334\\
21.6789 -0.116667\\
21.6789 -0.100001\\
21.6789 -0.0833338\\
21.6789 -0.066667\\
21.6789 -0.0500003\\
21.6789 -0.0333335\\
21.6789 -0.0166668\\
21.6789 -3.03577e-17\\
21.6789 0.0166667\\
21.6789 0.0333335\\
21.6789 0.0500002\\
21.6789 0.066667\\
21.6789 0.0833337\\
21.6789 0.1\\
21.6789 0.116667\\
21.6789 0.133334\\
21.6789 0.150001\\
21.6789 0.166667\\
21.6789 0.166668\\
21.6789 0.183334\\
21.6789 0.200001\\
21.6789 0.216668\\
21.6789 0.233334\\
21.6789 0.250001\\
21.6789 0.266668\\
21.6789 0.283335\\
21.6789 0.300001\\
21.6789 0.316668\\
21.6789 0.333335\\
21.6789 0.350002\\
21.6789 0.366668\\
21.6789 0.383335\\
21.6789 0.400002\\
21.6789 0.416669\\
21.6789 0.433335\\
21.6789 0.450002\\
21.6789 0.466669\\
21.6789 0.483336\\
21.6789 0.500002\\
};
\addlegendentry{TSDT7};

\addplot [
color=blue,
mark size=2.5pt,
only marks,
mark=x,
mark options={solid,draw=black}
]
table[row sep=crcr]{
22.5125 -0.483336\\
22.5125 -0.466669\\
22.5125 -0.450002\\
22.5125 -0.433336\\
22.5125 -0.416669\\
22.5125 -0.400002\\
22.5125 -0.383335\\
22.5125 -0.366669\\
22.5125 -0.350002\\
22.5125 -0.333335\\
22.5125 -0.316668\\
22.5125 -0.300002\\
22.5125 -0.283335\\
22.5125 -0.266668\\
22.5125 -0.250001\\
22.5125 -0.233335\\
22.5125 -0.216668\\
22.5125 -0.200001\\
22.5125 -0.183334\\
22.5125 -0.166668\\
22.5125 -0.166668\\
22.5125 -0.150001\\
22.5125 -0.133334\\
22.5125 -0.116667\\
22.5125 -0.100001\\
22.5125 -0.0833338\\
22.5125 -0.066667\\
22.5125 -0.0500003\\
22.5125 -0.0333335\\
22.5125 -0.0166668\\
22.5125 -3.03577e-17\\
22.5125 0.0166667\\
22.5125 0.0333335\\
22.5125 0.0500002\\
22.5125 0.066667\\
22.5125 0.0833337\\
22.5125 0.1\\
22.5125 0.116667\\
22.5125 0.133334\\
22.5125 0.150001\\
22.5125 0.166667\\
22.5125 0.166668\\
22.5125 0.183334\\
22.5125 0.200001\\
22.5125 0.216668\\
22.5125 0.233334\\
22.5125 0.250001\\
22.5125 0.266668\\
22.5125 0.283335\\
22.5125 0.300001\\
22.5125 0.316668\\
22.5125 0.333335\\
22.5125 0.350002\\
22.5125 0.366668\\
22.5125 0.383335\\
22.5125 0.400002\\
22.5125 0.416669\\
22.5125 0.433335\\
22.5125 0.450002\\
22.5125 0.466669\\
22.5125 0.483336\\
};
\addlegendentry{FSDT5};

\end{axis}
\end{tikzpicture}%
}}
\subfigure[$\overline{\sigma}_{xx} = \overline{\sigma}_{xx}(0,0,z)$]{\scalebox{0.5}{
%
%
%
%
\begin{tikzpicture}

\begin{axis}[%
width=\figurewidth,
height=\figureheight,
scale only axis,
xmin=-1500,
xmax=1500,
xlabel={$\bar{\sigma}_{xx}$},
ymin=-0.5,
ymax=0.5,
ylabel={Normalized thickness, z/h},
legend style={draw=black,fill=white,legend cell align=left}
]
\addplot [
color=blue,
mark size=2.5pt,
only marks,
mark=o,
mark options={solid,draw=black}
]
table[row sep=crcr]{
40.7495 -0.483336\\
82.0619 -0.466669\\
117.911 -0.450002\\
148.492 -0.433336\\
174.001 -0.416669\\
194.631 -0.400002\\
210.579 -0.383335\\
222.04 -0.366669\\
229.207 -0.350002\\
232.278 -0.333335\\
231.446 -0.316668\\
226.907 -0.300002\\
218.856 -0.283335\\
207.489 -0.266668\\
192.999 -0.250001\\
175.583 -0.233335\\
155.435 -0.216668\\
132.751 -0.200001\\
107.725 -0.183334\\
80.5536 -0.166668\\
-699.527 -0.166668\\
-638.801 -0.150001\\
-575.161 -0.133334\\
-508.931 -0.116667\\
-440.435 -0.100001\\
-369.997 -0.0833338\\
-297.94 -0.066667\\
-224.588 -0.0500003\\
-150.265 -0.0333335\\
-75.2943 -0.0166668\\
-6.63601e-14 -3.03577e-17\\
75.2943 0.0166667\\
150.265 0.0333335\\
224.588 0.0500002\\
297.94 0.066667\\
369.997 0.0833337\\
440.435 0.1\\
508.931 0.116667\\
575.161 0.133334\\
638.801 0.150001\\
699.527 0.166667\\
-80.5536 0.166668\\
-107.725 0.183334\\
-132.751 0.200001\\
-155.435 0.216668\\
-175.583 0.233334\\
-192.999 0.250001\\
-207.489 0.266668\\
-218.856 0.283335\\
-226.907 0.300001\\
-231.446 0.316668\\
-232.278 0.333335\\
-229.207 0.350002\\
-222.04 0.366668\\
-210.579 0.383335\\
-194.631 0.400002\\
-174.001 0.416669\\
-148.492 0.433335\\
-117.911 0.450002\\
-82.0619 0.466669\\
-40.7495 0.483336\\
};
\addlegendentry{HSDT13};

\addplot [
color=black,
dashed,
line width=1.2pt
]
table[row sep=crcr]{
90.9772 -0.500003\\
119.975 -0.483336\\
144.756 -0.466669\\
165.464 -0.450002\\
182.245 -0.433336\\
195.244 -0.416669\\
204.608 -0.400002\\
210.481 -0.383335\\
213.008 -0.366669\\
212.336 -0.350002\\
208.61 -0.333335\\
201.975 -0.316668\\
192.576 -0.300002\\
180.56 -0.283335\\
166.071 -0.266668\\
149.255 -0.250001\\
130.257 -0.233335\\
109.223 -0.216668\\
86.2982 -0.200001\\
61.6282 -0.183334\\
35.3583 -0.166668\\
-740.194 -0.166668\\
-677.568 -0.150001\\
-611.344 -0.133334\\
-541.922 -0.116667\\
-469.702 -0.100001\\
-395.083 -0.0833338\\
-318.465 -0.066667\\
-240.248 -0.0500003\\
-160.831 -0.0333335\\
-80.6156 -0.0166668\\
1.3355e-13 -3.03577e-17\\
80.6156 0.0166667\\
160.831 0.0333335\\
240.248 0.0500002\\
318.465 0.066667\\
395.083 0.0833337\\
469.702 0.1\\
541.922 0.116667\\
611.344 0.133334\\
677.568 0.150001\\
740.194 0.166667\\
-35.3583 0.166668\\
-61.6282 0.183334\\
-86.2982 0.200001\\
-109.223 0.216668\\
-130.257 0.233334\\
-149.255 0.250001\\
-166.071 0.266668\\
-180.56 0.283335\\
-192.576 0.300001\\
-201.975 0.316668\\
-208.61 0.333335\\
-212.336 0.350002\\
-213.008 0.366668\\
-210.481 0.383335\\
-204.608 0.400002\\
-195.244 0.416669\\
-182.245 0.433335\\
-165.464 0.450002\\
-144.756 0.466669\\
-119.975 0.483336\\
-90.9772 0.500002\\
};
\addlegendentry{HSDT11A};

\addplot [
color=blue,
mark size=4.3pt,
only marks,
mark=diamond,
mark options={solid,draw=black}
]
table[row sep=crcr]{
46.7514 -0.483336\\
76.0948 -0.466669\\
102.18 -0.450002\\
125.123 -0.433336\\
145.04 -0.416669\\
162.047 -0.400002\\
176.262 -0.383335\\
187.8 -0.366669\\
196.778 -0.350002\\
203.312 -0.333335\\
207.519 -0.316668\\
209.514 -0.300002\\
209.415 -0.283335\\
207.338 -0.266668\\
203.398 -0.250001\\
197.713 -0.233335\\
190.398 -0.216668\\
181.571 -0.200001\\
171.347 -0.183334\\
159.843 -0.166668\\
-1000.8 -0.166668\\
-889.785 -0.150001\\
-782.223 -0.133334\\
-677.73 -0.116667\\
-575.924 -0.100001\\
-476.419 -0.0833338\\
-378.833 -0.066667\\
-282.782 -0.0500003\\
-187.882 -0.0333335\\
-93.7491 -0.0166668\\
5.3029e-15 -3.03577e-17\\
93.7491 0.0166667\\
187.882 0.0333335\\
282.782 0.0500002\\
378.833 0.066667\\
476.419 0.0833337\\
575.924 0.1\\
677.73 0.116667\\
782.223 0.133334\\
889.785 0.150001\\
1000.8 0.166667\\
-159.843 0.166668\\
-171.347 0.183334\\
-181.571 0.200001\\
-190.398 0.216668\\
-197.713 0.233334\\
-203.398 0.250001\\
-207.338 0.266668\\
-209.415 0.283335\\
-209.514 0.300001\\
-207.519 0.316668\\
-203.312 0.333335\\
-196.778 0.350002\\
-187.8 0.366668\\
-176.262 0.383335\\
-162.047 0.400002\\
-145.04 0.416669\\
-125.123 0.433335\\
-102.18 0.450002\\
-76.0948 0.466669\\
-46.7514 0.483336\\
};
\addlegendentry{HSDT11B};

\addplot [
color=black,
solid,
line width=1.2pt
]
table[row sep=crcr]{
203.089 -0.500003\\
196.32 -0.483336\\
189.55 -0.466669\\
182.78 -0.450002\\
176.011 -0.433336\\
169.241 -0.416669\\
162.471 -0.400002\\
155.702 -0.383335\\
148.932 -0.366669\\
142.162 -0.350002\\
135.393 -0.333335\\
128.623 -0.316668\\
121.854 -0.300002\\
115.084 -0.283335\\
108.314 -0.266668\\
101.545 -0.250001\\
94.775 -0.233335\\
88.0054 -0.216668\\
81.2357 -0.200001\\
74.4661 -0.183334\\
67.6964 -0.166668\\
-1156.12 -0.166668\\
-1040.51 -0.150001\\
-924.898 -0.133334\\
-809.286 -0.116667\\
-693.674 -0.100001\\
-578.061 -0.0833338\\
-462.449 -0.066667\\
-346.837 -0.0500003\\
-231.225 -0.0333335\\
-115.612 -0.0166668\\
-1.74776e-14 -3.03577e-17\\
115.612 0.0166667\\
231.225 0.0333335\\
346.837 0.0500002\\
462.449 0.066667\\
578.061 0.0833337\\
693.674 0.1\\
809.286 0.116667\\
924.898 0.133334\\
1040.51 0.150001\\
1156.12 0.166667\\
-67.6964 0.166668\\
-74.4661 0.183334\\
-81.2357 0.200001\\
-88.0054 0.216668\\
-94.775 0.233334\\
-101.545 0.250001\\
-108.314 0.266668\\
-115.084 0.283335\\
-121.854 0.300001\\
-128.623 0.316668\\
-135.393 0.333335\\
-142.162 0.350002\\
-148.932 0.366668\\
-155.702 0.383335\\
-162.471 0.400002\\
-169.241 0.416669\\
-176.011 0.433335\\
-182.78 0.450002\\
-189.55 0.466669\\
-196.32 0.483336\\
-203.089 0.500002\\
};
\addlegendentry{TSDT7};

\addplot [
color=blue,
mark size=2.5pt,
only marks,
mark=x,
mark options={solid,draw=black}
]
table[row sep=crcr]{
216.499 -0.483336\\
209.034 -0.466669\\
201.568 -0.450002\\
194.103 -0.433336\\
186.637 -0.416669\\
179.172 -0.400002\\
171.706 -0.383335\\
164.241 -0.366669\\
156.775 -0.350002\\
149.31 -0.333335\\
141.844 -0.316668\\
134.379 -0.300002\\
126.913 -0.283335\\
119.448 -0.266668\\
111.982 -0.250001\\
104.517 -0.233335\\
97.0513 -0.216668\\
89.5858 -0.200001\\
82.1204 -0.183334\\
74.6549 -0.166668\\
-1205.59 -0.166668\\
-1085.03 -0.150001\\
-964.469 -0.133334\\
-843.91 -0.116667\\
-723.352 -0.100001\\
-602.793 -0.0833338\\
-482.234 -0.066667\\
-361.676 -0.0500003\\
-241.117 -0.0333335\\
-120.559 -0.0166668\\
-1.18756e-13 -3.03577e-17\\
120.559 0.0166667\\
241.117 0.0333335\\
361.676 0.0500002\\
482.234 0.066667\\
602.793 0.0833337\\
723.352 0.1\\
843.91 0.116667\\
964.469 0.133334\\
1085.03 0.150001\\
1205.59 0.166667\\
-74.6549 0.166668\\
-82.1204 0.183334\\
-89.5858 0.200001\\
-97.0513 0.216668\\
-104.517 0.233334\\
-111.982 0.250001\\
-119.448 0.266668\\
-126.913 0.283335\\
-134.379 0.300001\\
-141.844 0.316668\\
-149.31 0.333335\\
-156.775 0.350002\\
-164.241 0.366668\\
-171.706 0.383335\\
-179.172 0.400002\\
-186.637 0.416669\\
-194.103 0.433335\\
-201.568 0.450002\\
-209.034 0.466669\\
-216.499 0.483336\\
};
\addlegendentry{FSDT5};

\end{axis}

\begin{axis}[%
width=\figurewidth,
height=\figureheight,
scale only axis,
xmin=0,
xmax=1,
ymin=0,
ymax=1,
hide axis,
axis x line*=bottom,
axis y line*=left
]
\addplot [
color=black,
solid,
line width=0.0pt,
mark=square*,
mark options={solid,fill=black,draw=white},
forget plot
]
table[row sep=crcr]{
0.5 0.988095238095238\\
};
\addplot [
color=black,
solid,
line width=0.0pt,
mark=square*,
mark options={solid,fill=black,draw=white},
forget plot
]
table[row sep=crcr]{
0.991071428571429 0.5\\
};
\addplot [
color=black,
solid,
line width=0.0pt,
mark=square*,
mark options={solid,fill=black,draw=white},
forget plot
]
table[row sep=crcr]{
0.5 0.0119047619047619\\
};
\addplot [
color=black,
solid,
line width=0.0pt,
mark=square*,
mark options={solid,fill=black,draw=white},
forget plot
]
table[row sep=crcr]{
0.00892857142857143 0.5\\
};
\addplot [
color=black,
solid,
line width=0.0pt,
mark=square*,
mark options={solid,fill=black,draw=white},
forget plot
]
table[row sep=crcr]{
0.00892857142857143 0.988095238095238\\
};
\addplot [
color=black,
solid,
line width=0.0pt,
mark=square*,
mark options={solid,fill=black,draw=white},
forget plot
]
table[row sep=crcr]{
0.991071428571429 0.0119047619047619\\
};
\addplot [
color=black,
solid,
line width=0.0pt,
mark=square*,
mark options={solid,fill=black,draw=white},
forget plot
]
table[row sep=crcr]{
0.991071428571429 0.988095238095238\\
};
\addplot [
color=black,
solid,
line width=0.0pt,
mark=square*,
mark options={solid,fill=black,draw=white},
forget plot
]
table[row sep=crcr]{
0.00892857142857143 0.0119047619047619\\
};
\end{axis}
\end{tikzpicture}%
}}\hspace{60pt}
\subfigure[$\overline{\tau}_{yz} = \overline{\tau}_{yz}(-a/2,0,z)$]{\scalebox{0.5}{
%
%
%
%
\begin{tikzpicture}

\begin{axis}[%
width=\figurewidth,
height=\figureheight,
scale only axis,
xmin=-250,
xmax=0,
xlabel={$\bar{\tau}_{yz}$},
ymin=-0.5,
ymax=0.5,
ylabel={Normalized thickness, z/h},
legend style={at={(1.0,0.5)},draw=black,fill=white,legend cell align=left}
]
\addplot [
color=blue,
mark size=2.5pt,
only marks,
mark=o,
mark options={solid,draw=black}
]
table[row sep=crcr]{
-12.132 -0.483336\\
-24.7918 -0.466669\\
-37.774 -0.450002\\
-50.8802 -0.433336\\
-63.9192 -0.416669\\
-76.7072 -0.400002\\
-89.0675 -0.383335\\
-100.831 -0.366669\\
-111.834 -0.350002\\
-121.923 -0.333335\\
-130.95 -0.316668\\
-138.774 -0.300002\\
-145.261 -0.283335\\
-150.286 -0.266668\\
-153.729 -0.250001\\
-155.479 -0.233335\\
-155.431 -0.216668\\
-153.487 -0.200001\\
-149.558 -0.183334\\
-143.56 -0.166668\\
-143.56 -0.166668\\
-153.712 -0.150001\\
-162.877 -0.133334\\
-171.028 -0.116667\\
-178.141 -0.100001\\
-184.195 -0.0833338\\
-189.172 -0.066667\\
-193.058 -0.0500003\\
-195.842 -0.0333335\\
-197.516 -0.0166668\\
-198.074 -3.03577e-17\\
-197.516 0.0166667\\
-195.842 0.0333335\\
-193.058 0.0500002\\
-189.172 0.066667\\
-184.195 0.0833337\\
-178.141 0.1\\
-171.028 0.116667\\
-162.877 0.133334\\
-153.712 0.150001\\
-143.56 0.166667\\
-143.56 0.166668\\
-149.558 0.183334\\
-153.487 0.200001\\
-155.431 0.216668\\
-155.479 0.233334\\
-153.729 0.250001\\
-150.286 0.266668\\
-145.261 0.283335\\
-138.774 0.300001\\
-130.95 0.316668\\
-121.923 0.333335\\
-111.834 0.350002\\
-100.831 0.366668\\
-89.0675 0.383335\\
-76.7072 0.400002\\
-63.9192 0.416669\\
-50.8802 0.433335\\
-37.774 0.450002\\
-24.7918 0.466669\\
-12.132 0.483336\\
};
\addlegendentry{HSDT13};

\addplot [
color=black,
dashed,
line width=1.2pt
]
table[row sep=crcr]{
0 -0.500003\\
-11.3693 -0.483336\\
-23.498 -0.466669\\
-36.1497 -0.450002\\
-49.0963 -0.433336\\
-62.1177 -0.416669\\
-75.0026 -0.400002\\
-87.5475 -0.383335\\
-99.5576 -0.366669\\
-110.846 -0.350002\\
-121.235 -0.333335\\
-130.553 -0.316668\\
-138.64 -0.300002\\
-145.341 -0.283335\\
-150.511 -0.266668\\
-154.015 -0.250001\\
-155.722 -0.233335\\
-155.512 -0.216668\\
-153.275 -0.200001\\
-148.905 -0.183334\\
-142.308 -0.166668\\
-142.308 -0.166668\\
-152.766 -0.150001\\
-162.253 -0.133334\\
-170.726 -0.116667\\
-178.145 -0.100001\\
-184.479 -0.0833338\\
-189.699 -0.066667\\
-193.782 -0.0500003\\
-196.712 -0.0333335\\
-198.475 -0.0166668\\
-199.063 -3.03577e-17\\
-198.475 0.0166667\\
-196.712 0.0333335\\
-193.782 0.0500002\\
-189.699 0.066667\\
-184.479 0.0833337\\
-178.145 0.1\\
-170.726 0.116667\\
-162.253 0.133334\\
-152.766 0.150001\\
-142.308 0.166667\\
-142.308 0.166668\\
-148.905 0.183334\\
-153.275 0.200001\\
-155.512 0.216668\\
-155.722 0.233334\\
-154.015 0.250001\\
-150.511 0.266668\\
-145.341 0.283335\\
-138.64 0.300001\\
-130.553 0.316668\\
-121.235 0.333335\\
-110.846 0.350002\\
-99.5576 0.366668\\
-87.5475 0.383335\\
-75.0026 0.400002\\
-62.1177 0.416669\\
-49.0963 0.433335\\
-36.1497 0.450002\\
-23.498 0.466669\\
-11.3693 0.483336\\
-1.29148e-13 0.500002\\
};
\addlegendentry{HSDT11A};

\addplot [
color=blue,
mark size=4.3pt,
only marks,
mark=diamond,
mark options={solid,draw=black}
]
table[row sep=crcr]{
-17.4217 -0.483336\\
-33.0096 -0.466669\\
-46.8924 -0.450002\\
-59.1941 -0.433336\\
-70.0343 -0.416669\\
-79.5281 -0.400002\\
-87.786 -0.383335\\
-94.9142 -0.366669\\
-101.014 -0.350002\\
-106.183 -0.333335\\
-110.512 -0.316668\\
-114.092 -0.300002\\
-117.003 -0.283335\\
-119.327 -0.266668\\
-121.136 -0.250001\\
-122.502 -0.233335\\
-123.489 -0.216668\\
-124.159 -0.200001\\
-124.568 -0.183334\\
-124.767 -0.166668\\
-124.767 -0.166668\\
-138.211 -0.150001\\
-150.052 -0.133334\\
-160.353 -0.116667\\
-169.17 -0.100001\\
-176.549 -0.0833338\\
-182.532 -0.066667\\
-187.151 -0.0500003\\
-190.432 -0.0333335\\
-192.393 -0.0166668\\
-193.046 -3.03577e-17\\
-192.393 0.0166667\\
-190.432 0.0333335\\
-187.151 0.0500002\\
-182.532 0.066667\\
-176.549 0.0833337\\
-169.17 0.1\\
-160.353 0.116667\\
-150.052 0.133334\\
-138.211 0.150001\\
-124.767 0.166667\\
-124.767 0.166668\\
-124.568 0.183334\\
-124.159 0.200001\\
-123.489 0.216668\\
-122.502 0.233334\\
-121.136 0.250001\\
-119.327 0.266668\\
-117.003 0.283335\\
-114.092 0.300001\\
-110.512 0.316668\\
-106.183 0.333335\\
-101.014 0.350002\\
-94.9142 0.366668\\
-87.786 0.383335\\
-79.5281 0.400002\\
-70.0343 0.416669\\
-59.1941 0.433335\\
-46.8924 0.450002\\
-33.0096 0.466669\\
-17.4217 0.483336\\
};
\addlegendentry{HSDT11B};

\addplot [
color=black,
solid,
line width=1.2pt
]
table[row sep=crcr]{
0 -0.500003\\
-8.39878 -0.483336\\
-16.5128 -0.466669\\
-24.3422 -0.450002\\
-31.8869 -0.433336\\
-39.1468 -0.416669\\
-46.1221 -0.400002\\
-52.8126 -0.383335\\
-59.2185 -0.366669\\
-65.3396 -0.350002\\
-71.1761 -0.333335\\
-76.7278 -0.316668\\
-81.9948 -0.300002\\
-86.9772 -0.283335\\
-91.6748 -0.266668\\
-96.0877 -0.250001\\
-100.216 -0.233335\\
-104.059 -0.216668\\
-107.618 -0.200001\\
-110.892 -0.183334\\
-113.882 -0.166668\\
-113.882 -0.166668\\
-130.327 -0.150001\\
-145.041 -0.133334\\
-158.025 -0.116667\\
-169.277 -0.100001\\
-178.798 -0.0833338\\
-186.588 -0.066667\\
-192.647 -0.0500003\\
-196.974 -0.0333335\\
-199.571 -0.0166668\\
-200.437 -3.03577e-17\\
-199.571 0.0166667\\
-196.974 0.0333335\\
-192.647 0.0500002\\
-186.588 0.066667\\
-178.798 0.0833337\\
-169.277 0.1\\
-158.025 0.116667\\
-145.041 0.133334\\
-130.327 0.150001\\
-113.882 0.166667\\
-113.882 0.166668\\
-110.892 0.183334\\
-107.618 0.200001\\
-104.059 0.216668\\
-100.216 0.233334\\
-96.0877 0.250001\\
-91.6748 0.266668\\
-86.9772 0.283335\\
-81.9948 0.300001\\
-76.7278 0.316668\\
-71.1761 0.333335\\
-65.3396 0.350002\\
-59.2185 0.366668\\
-52.8126 0.383335\\
-46.1221 0.400002\\
-39.1468 0.416669\\
-31.8869 0.433335\\
-24.3422 0.450002\\
-16.5128 0.466669\\
-8.39878 0.483336\\
-5.09317e-14 0.500002\\
};
\addlegendentry{TSDT7};

\addplot [
color=blue,
mark size=2.5pt,
only marks,
mark=x,
mark options={solid,draw=black}
]
table[row sep=crcr]{
-8.89217 -0.483336\\
-17.4829 -0.466669\\
-25.7722 -0.450002\\
-33.7601 -0.433336\\
-41.4466 -0.416669\\
-48.8316 -0.400002\\
-55.9152 -0.383335\\
-62.6974 -0.366669\\
-69.1781 -0.350002\\
-75.3574 -0.333335\\
-81.2353 -0.316668\\
-86.8117 -0.300002\\
-92.0868 -0.283335\\
-97.0603 -0.266668\\
-101.733 -0.250001\\
-106.103 -0.233335\\
-110.173 -0.216668\\
-113.94 -0.200001\\
-117.407 -0.183334\\
-120.572 -0.166668\\
-120.572 -0.166668\\
-137.349 -0.150001\\
-152.361 -0.133334\\
-165.606 -0.116667\\
-177.085 -0.100001\\
-186.798 -0.0833338\\
-194.746 -0.066667\\
-200.927 -0.0500003\\
-205.342 -0.0333335\\
-207.991 -0.0166668\\
-208.874 -3.03577e-17\\
-207.991 0.0166667\\
-205.342 0.0333335\\
-200.927 0.0500002\\
-194.746 0.066667\\
-186.798 0.0833337\\
-177.085 0.1\\
-165.606 0.116667\\
-152.361 0.133334\\
-137.349 0.150001\\
-120.572 0.166667\\
-120.572 0.166668\\
-117.407 0.183334\\
-113.94 0.200001\\
-110.173 0.216668\\
-106.103 0.233334\\
-101.733 0.250001\\
-97.0603 0.266668\\
-92.0868 0.283335\\
-86.8117 0.300001\\
-81.2353 0.316668\\
-75.3574 0.333335\\
-69.1781 0.350002\\
-62.6974 0.366668\\
-55.9152 0.383335\\
-48.8316 0.400002\\
-41.4466 0.416669\\
-33.7601 0.433335\\
-25.7722 0.450002\\
-17.4829 0.466669\\
-8.89217 0.483336\\
};
\addlegendentry{FSDT5};

\end{axis}

\begin{axis}[%
width=\figurewidth,
height=\figureheight,
scale only axis,
xmin=0,
xmax=1,
ymin=0,
ymax=1,
hide axis,
axis x line*=bottom,
axis y line*=left
]
\addplot [
color=black,
solid,
line width=0.0pt,
mark=square*,
mark options={solid,fill=black,draw=white},
forget plot
]
table[row sep=crcr]{
0.5 0.988095238095238\\
};
\addplot [
color=black,
solid,
line width=0.0pt,
mark=square*,
mark options={solid,fill=black,draw=white},
forget plot
]
table[row sep=crcr]{
0.991071428571429 0.5\\
};
\addplot [
color=black,
solid,
line width=0.0pt,
mark=square*,
mark options={solid,fill=black,draw=white},
forget plot
]
table[row sep=crcr]{
0.5 0.0119047619047619\\
};
\addplot [
color=black,
solid,
line width=0.0pt,
mark=square*,
mark options={solid,fill=black,draw=white},
forget plot
]
table[row sep=crcr]{
0.00892857142857143 0.5\\
};
\addplot [
color=black,
solid,
line width=0.0pt,
mark=square*,
mark options={solid,fill=black,draw=white},
forget plot
]
table[row sep=crcr]{
0.00892857142857143 0.988095238095238\\
};
\addplot [
color=black,
solid,
line width=0.0pt,
mark=square*,
mark options={solid,fill=black,draw=white},
forget plot
]
table[row sep=crcr]{
0.991071428571429 0.0119047619047619\\
};
\addplot [
color=black,
solid,
line width=0.0pt,
mark=square*,
mark options={solid,fill=black,draw=white},
forget plot
]
table[row sep=crcr]{
0.991071428571429 0.988095238095238\\
};
\addplot [
color=black,
solid,
line width=0.0pt,
mark=square*,
mark options={solid,fill=black,draw=white},
forget plot
]
table[row sep=crcr]{
0.00892857142857143 0.0119047619047619\\
};
\end{axis}
\end{tikzpicture}%
}}
\subfigure[$\overline{\tau}_{xz}=\overline{\tau}_{xz}(-a/2,0,z)$]{\scalebox{0.5}{
%
%
%
%
\begin{tikzpicture}

\begin{axis}[%
width=\figurewidth,
height=\figureheight,
scale only axis,
xmin=-250,
xmax=0,
xlabel={$\bar{\tau}_{xz}$},
ymin=-0.5,
ymax=0.5,
ylabel={Normalized thickness, z/h},
legend style={at={(1.0,0.5)},draw=black,fill=white,legend cell align=left}
]
\addplot [
color=blue,
mark size=2.5pt,
only marks,
mark=o,
mark options={solid,draw=black}
]
table[row sep=crcr]{
-19.1643 -0.483336\\
-38.6101 -0.466669\\
-58.1392 -0.450002\\
-77.5606 -0.433336\\
-96.6901 -0.416669\\
-115.35 -0.400002\\
-133.371 -0.383335\\
-150.59 -0.366669\\
-166.849 -0.350002\\
-182 -0.333335\\
-195.9 -0.316668\\
-208.414 -0.300002\\
-219.412 -0.283335\\
-228.774 -0.266668\\
-236.385 -0.250001\\
-242.137 -0.233335\\
-245.929 -0.216668\\
-247.668 -0.200001\\
-247.266 -0.183334\\
-244.644 -0.166668\\
-244.644 -0.166668\\
-237.897 -0.150001\\
-231.786 -0.133334\\
-226.334 -0.116667\\
-221.564 -0.100001\\
-217.496 -0.0833338\\
-214.145 -0.066667\\
-211.525 -0.0500003\\
-209.646 -0.0333335\\
-208.516 -0.0166668\\
-208.139 -3.03577e-17\\
-208.516 0.0166667\\
-209.646 0.0333335\\
-211.525 0.0500002\\
-214.145 0.066667\\
-217.496 0.0833337\\
-221.564 0.1\\
-226.334 0.116667\\
-231.786 0.133334\\
-237.897 0.150001\\
-244.644 0.166667\\
-244.644 0.166668\\
-247.266 0.183334\\
-247.668 0.200001\\
-245.929 0.216668\\
-242.137 0.233334\\
-236.385 0.250001\\
-228.774 0.266668\\
-219.412 0.283335\\
-208.414 0.300001\\
-195.9 0.316668\\
-182 0.333335\\
-166.849 0.350002\\
-150.59 0.366668\\
-133.371 0.383335\\
-115.35 0.400002\\
-96.6901 0.416669\\
-77.5606 0.433335\\
-58.1392 0.450002\\
-38.6101 0.466669\\
-19.1643 0.483336\\
};
\addlegendentry{HSDT13};

\addplot [
color=black,
dashed,
line width=1.2pt
]
table[row sep=crcr]{
0 -0.500003\\
-18.732 -0.483336\\
-37.9332 -0.466669\\
-57.378 -0.450002\\
-76.8485 -0.433336\\
-96.1351 -0.416669\\
-115.036 -0.400002\\
-133.356 -0.383335\\
-150.911 -0.366669\\
-167.522 -0.350002\\
-183.018 -0.333335\\
-197.237 -0.316668\\
-210.025 -0.300002\\
-221.236 -0.283335\\
-230.73 -0.266668\\
-238.378 -0.250001\\
-244.056 -0.233335\\
-247.65 -0.216668\\
-249.052 -0.200001\\
-248.164 -0.183334\\
-244.895 -0.166668\\
-244.895 -0.166668\\
-237.828 -0.150001\\
-231.381 -0.133334\\
-225.596 -0.116667\\
-220.51 -0.100001\\
-216.152 -0.0833338\\
-212.551 -0.066667\\
-209.728 -0.0500003\\
-207.699 -0.0333335\\
-206.477 -0.0166668\\
-206.069 -3.03577e-17\\
-206.477 0.0166667\\
-207.699 0.0333335\\
-209.728 0.0500002\\
-212.551 0.066667\\
-216.152 0.0833337\\
-220.51 0.1\\
-225.596 0.116667\\
-231.381 0.133334\\
-237.828 0.150001\\
-244.895 0.166667\\
-244.895 0.166668\\
-248.164 0.183334\\
-249.052 0.200001\\
-247.65 0.216668\\
-244.056 0.233334\\
-238.378 0.250001\\
-230.73 0.266668\\
-221.236 0.283335\\
-210.025 0.300001\\
-197.237 0.316668\\
-183.018 0.333335\\
-167.522 0.350002\\
-150.911 0.366668\\
-133.356 0.383335\\
-115.036 0.400002\\
-96.1351 0.416669\\
-76.8485 0.433335\\
-57.378 0.450002\\
-37.9332 0.466669\\
-18.732 0.483336\\
-1.34605e-13 0.500002\\
};
\addlegendentry{HSDT11A};

\addplot [
color=blue,
mark size=4.3pt,
only marks,
mark=diamond,
mark options={solid,draw=black}
]
table[row sep=crcr]{
-24.219 -0.483336\\
-46.5387 -0.466669\\
-67.0708 -0.450002\\
-85.9228 -0.433336\\
-103.198 -0.416669\\
-118.998 -0.400002\\
-133.416 -0.383335\\
-146.546 -0.366669\\
-158.475 -0.350002\\
-169.288 -0.333335\\
-179.064 -0.316668\\
-187.881 -0.300002\\
-195.81 -0.283335\\
-202.92 -0.266668\\
-209.275 -0.250001\\
-214.937 -0.233335\\
-219.962 -0.216668\\
-224.402 -0.200001\\
-228.307 -0.183334\\
-231.722 -0.166668\\
-231.722 -0.166668\\
-221.938 -0.150001\\
-213.372 -0.133334\\
-205.96 -0.116667\\
-199.649 -0.100001\\
-194.389 -0.0833338\\
-190.14 -0.066667\\
-186.87 -0.0500003\\
-184.553 -0.0333335\\
-183.17 -0.0166668\\
-182.71 -3.03577e-17\\
-183.17 0.0166667\\
-184.553 0.0333335\\
-186.87 0.0500002\\
-190.14 0.066667\\
-194.389 0.0833337\\
-199.649 0.1\\
-205.96 0.116667\\
-213.372 0.133334\\
-221.938 0.150001\\
-231.722 0.166667\\
-231.722 0.166668\\
-228.307 0.183334\\
-224.402 0.200001\\
-219.962 0.216668\\
-214.937 0.233334\\
-209.275 0.250001\\
-202.92 0.266668\\
-195.81 0.283335\\
-187.881 0.300001\\
-179.064 0.316668\\
-169.288 0.333335\\
-158.475 0.350002\\
-146.546 0.366668\\
-133.416 0.383335\\
-118.998 0.400002\\
-103.198 0.416669\\
-85.9228 0.433335\\
-67.0708 0.450002\\
-46.5387 0.466669\\
-24.219 0.483336\\
};
\addlegendentry{HSDT11B};

\addplot [
color=black,
solid,
line width=1.2pt
]
table[row sep=crcr]{
0 -0.500003\\
-16.7895 -0.483336\\
-33.0098 -0.466669\\
-48.661 -0.450002\\
-63.7431 -0.433336\\
-78.256 -0.416669\\
-92.1998 -0.400002\\
-105.574 -0.383335\\
-118.38 -0.366669\\
-130.616 -0.350002\\
-142.284 -0.333335\\
-153.382 -0.316668\\
-163.911 -0.300002\\
-173.871 -0.283335\\
-183.261 -0.266668\\
-192.083 -0.250001\\
-200.335 -0.233335\\
-208.019 -0.216668\\
-215.133 -0.200001\\
-221.678 -0.183334\\
-227.654 -0.166668\\
-227.654 -0.166668\\
-214.877 -0.150001\\
-203.445 -0.133334\\
-193.358 -0.116667\\
-184.615 -0.100001\\
-177.218 -0.0833338\\
-171.166 -0.066667\\
-166.459 -0.0500003\\
-163.096 -0.0333335\\
-161.079 -0.0166668\\
-160.406 -3.03577e-17\\
-161.079 0.0166667\\
-163.096 0.0333335\\
-166.459 0.0500002\\
-171.166 0.066667\\
-177.218 0.0833337\\
-184.615 0.1\\
-193.358 0.116667\\
-203.445 0.133334\\
-214.877 0.150001\\
-227.654 0.166667\\
-227.654 0.166668\\
-221.678 0.183334\\
-215.133 0.200001\\
-208.019 0.216668\\
-200.335 0.233334\\
-192.083 0.250001\\
-183.261 0.266668\\
-173.871 0.283335\\
-163.911 0.300001\\
-153.382 0.316668\\
-142.284 0.333335\\
-130.616 0.350002\\
-118.38 0.366668\\
-105.574 0.383335\\
-92.1998 0.400002\\
-78.256 0.416669\\
-63.7431 0.433335\\
-48.661 0.450002\\
-33.0098 0.466669\\
-16.7895 0.483336\\
1.45519e-14 0.500002\\
};
\addlegendentry{TSDT7};

\addplot [
color=blue,
mark size=2.5pt,
only marks,
mark=x,
mark options={solid,draw=black}
]
table[row sep=crcr]{
-17.2783 -0.483336\\
-33.9708 -0.466669\\
-50.0777 -0.450002\\
-65.5989 -0.433336\\
-80.5343 -0.416669\\
-94.8841 -0.400002\\
-108.648 -0.383335\\
-121.826 -0.366669\\
-134.419 -0.350002\\
-146.426 -0.333335\\
-157.847 -0.316668\\
-168.683 -0.300002\\
-178.933 -0.283335\\
-188.597 -0.266668\\
-197.675 -0.250001\\
-206.168 -0.233335\\
-214.075 -0.216668\\
-221.396 -0.200001\\
-228.132 -0.183334\\
-234.282 -0.166668\\
-234.282 -0.166668\\
-221.192 -0.150001\\
-209.48 -0.133334\\
-199.146 -0.116667\\
-190.189 -0.100001\\
-182.611 -0.0833338\\
-176.41 -0.066667\\
-171.588 -0.0500003\\
-168.143 -0.0333335\\
-166.076 -0.0166668\\
-165.387 -3.03577e-17\\
-166.076 0.0166667\\
-168.143 0.0333335\\
-171.588 0.0500002\\
-176.41 0.066667\\
-182.611 0.0833337\\
-190.189 0.1\\
-199.146 0.116667\\
-209.48 0.133334\\
-221.192 0.150001\\
-234.282 0.166667\\
-234.282 0.166668\\
-228.132 0.183334\\
-221.396 0.200001\\
-214.075 0.216668\\
-206.168 0.233334\\
-197.675 0.250001\\
-188.597 0.266668\\
-178.933 0.283335\\
-168.683 0.300001\\
-157.847 0.316668\\
-146.426 0.333335\\
-134.419 0.350002\\
-121.826 0.366668\\
-108.648 0.383335\\
-94.8841 0.400002\\
-80.5343 0.416669\\
-65.5989 0.433335\\
-50.0777 0.450002\\
-33.9708 0.466669\\
-17.2783 0.483336\\
};
\addlegendentry{FSDT5};

\end{axis}
\end{tikzpicture}%
}}\hspace{60pt}
\subfigure[$\overline{\sigma}_{zz} = \overline{\sigma}_{zz}(0,0,z)$]{\scalebox{0.5}{
%
%
%
%
\begin{tikzpicture}

\begin{axis}[%
width=\figurewidth,
height=\figureheight,
scale only axis,
xmin=-25,
xmax=25,
xlabel={$\bar{\sigma}_{zz}$},
ymin=-0.5,
ymax=0.5,
ylabel={Normalized thickness, z/h},
legend style={draw=black,fill=white,legend cell align=left},
legend pos = south east
]
\addplot [
color=blue,
mark size=2.5pt,
only marks,
mark=o,
mark options={solid,draw=black}
]
table[row sep=crcr]{
-14.224 -0.483336\\
-13.0919 -0.466669\\
-11.9227 -0.450002\\
-10.7177 -0.433336\\
-9.47835 -0.416669\\
-8.20584 -0.400002\\
-6.90155 -0.383335\\
-5.5668 -0.366669\\
-4.20291 -0.350002\\
-2.81121 -0.333335\\
-1.39303 -0.316668\\
0.050322 -0.300002\\
1.51751 -0.283335\\
3.00721 -0.266668\\
4.5181 -0.250001\\
6.04886 -0.233335\\
7.59816 -0.216668\\
9.16467 -0.200001\\
10.7471 -0.183334\\
12.3441 -0.166668\\
8.1725 -0.166668\\
7.23843 -0.150001\\
6.34125 -0.133334\\
5.47687 -0.116667\\
4.64117 -0.100001\\
3.83007 -0.0833338\\
3.03946 -0.066667\\
2.26525 -0.0500003\\
1.50334 -0.0333335\\
0.749618 -0.0166668\\
1.5749e-14 -3.03577e-17\\
-0.749618 0.0166667\\
-1.50334 0.0333335\\
-2.26525 0.0500002\\
-3.03946 0.066667\\
-3.83007 0.0833337\\
-4.64117 0.1\\
-5.47687 0.116667\\
-6.34125 0.133334\\
-7.23843 0.150001\\
-8.1725 0.166667\\
-12.3441 0.166668\\
-10.7471 0.183334\\
-9.16467 0.200001\\
-7.59816 0.216668\\
-6.04886 0.233334\\
-4.5181 0.250001\\
-3.00721 0.266668\\
-1.51751 0.283335\\
-0.050322 0.300001\\
1.39303 0.316668\\
2.81121 0.333335\\
4.20291 0.350002\\
5.5668 0.366668\\
6.90155 0.383335\\
8.20584 0.400002\\
9.47835 0.416669\\
10.7177 0.433335\\
11.9227 0.450002\\
13.0919 0.466669\\
14.224 0.483336\\
};
\addlegendentry{HSDT13};

\addplot [
color=black,
dashed,
line width=1.2pt
]
table[row sep=crcr]{
0 -0.500003\\
0 -0.483336\\
0 -0.466669\\
0 -0.450002\\
0 -0.433336\\
0 -0.416669\\
0 -0.400002\\
0 -0.383335\\
0 -0.366669\\
0 -0.350002\\
0 -0.333335\\
0 -0.316668\\
0 -0.300002\\
0 -0.283335\\
0 -0.266668\\
0 -0.250001\\
0 -0.233335\\
0 -0.216668\\
0 -0.200001\\
0 -0.183334\\
0 -0.166668\\
0 -0.166668\\
0 -0.150001\\
0 -0.133334\\
0 -0.116667\\
0 -0.100001\\
0 -0.0833338\\
0 -0.066667\\
0 -0.0500003\\
0 -0.0333335\\
0 -0.0166668\\
0 -3.03577e-17\\
0 0.0166667\\
0 0.0333335\\
0 0.0500002\\
0 0.066667\\
0 0.0833337\\
0 0.1\\
0 0.116667\\
0 0.133334\\
0 0.150001\\
0 0.166667\\
0 0.166668\\
0 0.183334\\
0 0.200001\\
0 0.216668\\
0 0.233334\\
0 0.250001\\
0 0.266668\\
0 0.283335\\
0 0.300001\\
0 0.316668\\
0 0.333335\\
0 0.350002\\
0 0.366668\\
0 0.383335\\
0 0.400002\\
0 0.416669\\
0 0.433335\\
0 0.450002\\
0 0.466669\\
0 0.483336\\
0 0.500002\\
};
\addlegendentry{HSDT11A};

\addplot [
color=blue,
mark size=4.3pt,
only marks,
mark=diamond,
mark options={solid,draw=black}
]
table[row sep=crcr]{
-19.4045 -0.483336\\
-16.1102 -0.466669\\
-13.0923 -0.450002\\
-10.3409 -0.433336\\
-7.84601 -0.416669\\
-5.59787 -0.400002\\
-3.58658 -0.383335\\
-1.80227 -0.366669\\
-0.235085 -0.350002\\
1.12486 -0.333335\\
2.28742 -0.316668\\
3.26248 -0.300002\\
4.05989 -0.283335\\
4.68954 -0.266668\\
5.16128 -0.250001\\
5.48499 -0.233335\\
5.67053 -0.216668\\
5.72778 -0.200001\\
5.66661 -0.183334\\
5.49687 -0.166668\\
-0.709779 -0.166668\\
-0.316796 -0.150001\\
-0.0254992 -0.133334\\
0.175411 -0.116667\\
0.297231 -0.100001\\
0.351262 -0.0833338\\
0.3488 -0.066667\\
0.301144 -0.0500003\\
0.219594 -0.0333335\\
0.115446 -0.0166668\\
1.44897e-14 -3.03577e-17\\
-0.115446 0.0166667\\
-0.219594 0.0333335\\
-0.301144 0.0500002\\
-0.3488 0.066667\\
-0.351262 0.0833337\\
-0.297231 0.1\\
-0.175411 0.116667\\
0.0254992 0.133334\\
0.316796 0.150001\\
0.709779 0.166667\\
-5.49687 0.166668\\
-5.66661 0.183334\\
-5.72778 0.200001\\
-5.67053 0.216668\\
-5.48499 0.233334\\
-5.16128 0.250001\\
-4.68954 0.266668\\
-4.05989 0.283335\\
-3.26248 0.300001\\
-2.28742 0.316668\\
-1.12486 0.333335\\
0.235085 0.350002\\
1.80227 0.366668\\
3.58658 0.383335\\
5.59787 0.400002\\
7.84601 0.416669\\
10.3409 0.433335\\
13.0923 0.450002\\
16.1102 0.466669\\
19.4045 0.483336\\
};
\addlegendentry{HSDT11B};

\addplot [
color=black,
solid,
line width=1.2pt
]
table[row sep=crcr]{
0 -0.500003\\
0 -0.483336\\
0 -0.466669\\
0 -0.450002\\
0 -0.433336\\
0 -0.416669\\
0 -0.400002\\
0 -0.383335\\
0 -0.366669\\
0 -0.350002\\
0 -0.333335\\
0 -0.316668\\
0 -0.300002\\
0 -0.283335\\
0 -0.266668\\
0 -0.250001\\
0 -0.233335\\
0 -0.216668\\
0 -0.200001\\
0 -0.183334\\
0 -0.166668\\
0 -0.166668\\
0 -0.150001\\
0 -0.133334\\
0 -0.116667\\
0 -0.100001\\
0 -0.0833338\\
0 -0.066667\\
0 -0.0500003\\
0 -0.0333335\\
0 -0.0166668\\
0 -3.03577e-17\\
0 0.0166667\\
0 0.0333335\\
0 0.0500002\\
0 0.066667\\
0 0.0833337\\
0 0.1\\
0 0.116667\\
0 0.133334\\
0 0.150001\\
0 0.166667\\
0 0.166668\\
0 0.183334\\
0 0.200001\\
0 0.216668\\
0 0.233334\\
0 0.250001\\
0 0.266668\\
0 0.283335\\
0 0.300001\\
0 0.316668\\
0 0.333335\\
0 0.350002\\
0 0.366668\\
0 0.383335\\
0 0.400002\\
0 0.416669\\
0 0.433335\\
0 0.450002\\
0 0.466669\\
0 0.483336\\
0 0.500002\\
};
\addlegendentry{TSDT7};

\addplot [
color=blue,
mark size=2.5pt,
only marks,
mark=x,
mark options={solid,draw=black}
]
table[row sep=crcr]{
0 -0.483336\\
0 -0.466669\\
0 -0.450002\\
0 -0.433336\\
0 -0.416669\\
0 -0.400002\\
0 -0.383335\\
0 -0.366669\\
0 -0.350002\\
0 -0.333335\\
0 -0.316668\\
0 -0.300002\\
0 -0.283335\\
0 -0.266668\\
0 -0.250001\\
0 -0.233335\\
0 -0.216668\\
0 -0.200001\\
0 -0.183334\\
0 -0.166668\\
0 -0.166668\\
0 -0.150001\\
0 -0.133334\\
0 -0.116667\\
0 -0.100001\\
0 -0.0833338\\
0 -0.066667\\
0 -0.0500003\\
0 -0.0333335\\
0 -0.0166668\\
0 -3.03577e-17\\
0 0.0166667\\
0 0.0333335\\
0 0.0500002\\
0 0.066667\\
0 0.0833337\\
0 0.1\\
0 0.116667\\
0 0.133334\\
0 0.150001\\
0 0.166667\\
0 0.166668\\
0 0.183334\\
0 0.200001\\
0 0.216668\\
0 0.233334\\
0 0.250001\\
0 0.266668\\
0 0.283335\\
0 0.300001\\
0 0.316668\\
0 0.333335\\
0 0.350002\\
0 0.366668\\
0 0.383335\\
0 0.400002\\
0 0.416669\\
0 0.433335\\
0 0.450002\\
0 0.466669\\
0 0.483336\\
};
\addlegendentry{FSDT5};

\end{axis}

\begin{axis}[%
width=\figurewidth,
height=\figureheight,
scale only axis,
xmin=0,
xmax=1,
ymin=0,
ymax=1,
hide axis,
axis x line*=bottom,
axis y line*=left
]
\addplot [
color=black,
solid,
line width=0.0pt,
mark=square*,
mark options={solid,fill=black,draw=white},
forget plot
]
table[row sep=crcr]{
0.5 0.988095238095238\\
};
\addplot [
color=black,
solid,
line width=0.0pt,
mark=square*,
mark options={solid,fill=black,draw=white},
forget plot
]
table[row sep=crcr]{
0.991071428571429 0.5\\
};
\addplot [
color=black,
solid,
line width=0.0pt,
mark=square*,
mark options={solid,fill=black,draw=white},
forget plot
]
table[row sep=crcr]{
0.5 0.0119047619047619\\
};
\addplot [
color=black,
solid,
line width=0.0pt,
mark=square*,
mark options={solid,fill=black,draw=white},
forget plot
]
table[row sep=crcr]{
0.00892857142857143 0.5\\
};
\addplot [
color=black,
solid,
line width=0.0pt,
mark=square*,
mark options={solid,fill=black,draw=white},
forget plot
]
table[row sep=crcr]{
0.00892857142857143 0.988095238095238\\
};
\addplot [
color=black,
solid,
line width=0.0pt,
mark=square*,
mark options={solid,fill=black,draw=white},
forget plot
]
table[row sep=crcr]{
0.991071428571429 0.0119047619047619\\
};
\addplot [
color=black,
solid,
line width=0.0pt,
mark=square*,
mark options={solid,fill=black,draw=white},
forget plot
]
table[row sep=crcr]{
0.991071428571429 0.988095238095238\\
};
\addplot [
color=black,
solid,
line width=0.0pt,
mark=square*,
mark options={solid,fill=black,draw=white},
forget plot
]
table[row sep=crcr]{
0.00892857142857143 0.0119047619047619\\
};
\end{axis}
\end{tikzpicture}%
}}
\caption{The variation of non-dimensional deflections and stresses through the thickness of a three-layered symmetric curved fibre composite laminate under sinusoidally distributed thermal load with $a/h=$ 5, $T_o=$ -45$^\circ$, $T_1=$ 45$^\circ$.}
\label{fig:3LayerThermLoad}
\end{figure}

\begin{figure}
\centering
\setlength\figureheight{10cm} 
\setlength\figurewidth{12cm}
\subfigure[$\overline{u} = \overline{u}(-a/2,0,z)$]{\scalebox{0.5}{
%
%
%
%
\begin{tikzpicture}

\begin{axis}[%
width=\figurewidth,
height=\figureheight,
scale only axis,
xmin=-1.5,
xmax=1,
xlabel={$\bar{u}$},
ymin=-0.5,
ymax=0.5,
ylabel={Normalized thickness, z/h},
legend style={draw=black,fill=white,legend cell align=left}
]
\addplot [
color=blue,
mark size=2.5pt,
only marks,
mark=o,
mark options={solid,draw=black}
]
table[row sep=crcr]{
0.964665 -0.5\\
0.912482 -0.4875\\
0.862453 -0.475\\
0.814521 -0.4625\\
0.76863 -0.45\\
0.724724 -0.4375\\
0.682748 -0.425\\
0.642645 -0.4125\\
0.60436 -0.4\\
0.567837 -0.3875\\
0.53302 -0.375\\
0.499853 -0.3625\\
0.46828 -0.35\\
0.438245 -0.3375\\
0.409692 -0.325\\
0.382566 -0.3125\\
0.35681 -0.3\\
0.332369 -0.2875\\
0.309187 -0.275\\
0.287207 -0.2625\\
0.266374 -0.25\\
0.266374 -0.25\\
0.246455 -0.2375\\
0.227571 -0.225\\
0.209666 -0.2125\\
0.192684 -0.2\\
0.17657 -0.1875\\
0.161267 -0.175\\
0.14672 -0.1625\\
0.132873 -0.15\\
0.119669 -0.1375\\
0.107054 -0.125\\
0.0949708 -0.1125\\
0.0833637 -0.1\\
0.0721769 -0.0875\\
0.0613546 -0.075\\
0.0508407 -0.0625\\
0.0405795 -0.05\\
0.030515 -0.0375\\
0.0205914 -0.025\\
0.0107526 -0.0125\\
0.00094291 3.90313e-17\\
0.00094291 0\\
-0.00871573 0.0125\\
-0.0184571 0.025\\
-0.0283372 0.0375\\
-0.0384117 0.05\\
-0.0487368 0.0625\\
-0.0593681 0.075\\
-0.0703617 0.0875\\
-0.0817734 0.1\\
-0.0936591 0.1125\\
-0.106075 0.125\\
-0.119076 0.1375\\
-0.132719 0.15\\
-0.14706 0.1625\\
-0.162154 0.175\\
-0.178058 0.1875\\
-0.194827 0.2\\
-0.212517 0.2125\\
-0.231184 0.225\\
-0.250884 0.2375\\
-0.271673 0.25\\
-0.271673 0.25\\
-0.293785 0.2625\\
-0.317097 0.275\\
-0.341666 0.2875\\
-0.367547 0.3\\
-0.394797 0.3125\\
-0.423471 0.325\\
-0.453625 0.3375\\
-0.485315 0.35\\
-0.518597 0.3625\\
-0.553527 0.375\\
-0.59016 0.3875\\
-0.628553 0.4\\
-0.668762 0.4125\\
-0.710842 0.425\\
-0.754849 0.4375\\
-0.80084 0.45\\
-0.848869 0.4625\\
-0.898994 0.475\\
-0.951269 0.4875\\
-1.00575 0.5\\
};
\addlegendentry{HSDT13};

\addplot [
color=black,
dashed,
line width=1.2pt
]
table[row sep=crcr]{
0.998304 -0.5\\
0.944622 -0.4875\\
0.893122 -0.475\\
0.843748 -0.4625\\
0.796443 -0.45\\
0.751152 -0.4375\\
0.707819 -0.425\\
0.666388 -0.4125\\
0.626802 -0.4\\
0.589007 -0.3875\\
0.552946 -0.375\\
0.518563 -0.3625\\
0.485802 -0.35\\
0.454607 -0.3375\\
0.424923 -0.325\\
0.396693 -0.3125\\
0.369862 -0.3\\
0.344373 -0.2875\\
0.32017 -0.275\\
0.297198 -0.2625\\
0.275401 -0.25\\
0.275401 -0.25\\
0.254723 -0.2375\\
0.235107 -0.225\\
0.216499 -0.2125\\
0.198841 -0.2\\
0.182078 -0.1875\\
0.166155 -0.175\\
0.151014 -0.1625\\
0.136601 -0.15\\
0.122859 -0.1375\\
0.109732 -0.125\\
0.0971644 -0.1125\\
0.0851005 -0.1\\
0.073484 -0.0875\\
0.0622591 -0.075\\
0.0513698 -0.0625\\
0.0407602 -0.05\\
0.0303744 -0.0375\\
0.0201564 -0.025\\
0.0100502 -0.0125\\
8.0644e-16 3.90313e-17\\
8.37793e-16 0\\
-0.0100502 0.0125\\
-0.0201564 0.025\\
-0.0303744 0.0375\\
-0.0407602 0.05\\
-0.0513698 0.0625\\
-0.0622591 0.075\\
-0.073484 0.0875\\
-0.0851005 0.1\\
-0.0971644 0.1125\\
-0.109732 0.125\\
-0.122859 0.1375\\
-0.136601 0.15\\
-0.151014 0.1625\\
-0.166155 0.175\\
-0.182078 0.1875\\
-0.198841 0.2\\
-0.216499 0.2125\\
-0.235107 0.225\\
-0.254723 0.2375\\
-0.275401 0.25\\
-0.275401 0.25\\
-0.297198 0.2625\\
-0.32017 0.275\\
-0.344373 0.2875\\
-0.369862 0.3\\
-0.396693 0.3125\\
-0.424923 0.325\\
-0.454607 0.3375\\
-0.485802 0.35\\
-0.518563 0.3625\\
-0.552946 0.375\\
-0.589007 0.3875\\
-0.626802 0.4\\
-0.666388 0.4125\\
-0.707819 0.425\\
-0.751152 0.4375\\
-0.796443 0.45\\
-0.843748 0.4625\\
-0.893122 0.475\\
-0.944622 0.4875\\
-0.998304 0.5\\
};
\addlegendentry{HSDT11A};

\addplot [
color=blue,
mark size=4.3pt,
only marks,
mark=diamond,
mark options={solid,draw=black}
]
table[row sep=crcr]{
0.942288 -0.5\\
0.893308 -0.4875\\
0.846277 -0.475\\
0.801143 -0.4625\\
0.757856 -0.45\\
0.716366 -0.4375\\
0.676622 -0.425\\
0.638573 -0.4125\\
0.602168 -0.4\\
0.567357 -0.3875\\
0.534089 -0.375\\
0.502314 -0.3625\\
0.47198 -0.35\\
0.443036 -0.3375\\
0.415434 -0.325\\
0.389121 -0.3125\\
0.364046 -0.3\\
0.34016 -0.2875\\
0.317412 -0.275\\
0.29575 -0.2625\\
0.275125 -0.25\\
0.275125 -0.25\\
0.255485 -0.2375\\
0.236779 -0.225\\
0.218958 -0.2125\\
0.201971 -0.2\\
0.185766 -0.1875\\
0.170293 -0.175\\
0.155501 -0.1625\\
0.14134 -0.15\\
0.127759 -0.1375\\
0.114708 -0.125\\
0.102135 -0.1125\\
0.0899898 -0.1\\
0.078222 -0.0875\\
0.0667808 -0.075\\
0.0556154 -0.0625\\
0.0446752 -0.05\\
0.0339095 -0.0375\\
0.0232676 -0.025\\
0.0126988 -0.0125\\
0.00215246 3.90313e-17\\
0.00215246 0\\
-0.00842213 0.0125\\
-0.0190757 0.025\\
-0.0298588 0.0375\\
-0.0408223 0.05\\
-0.0520167 0.0625\\
-0.0634929 0.075\\
-0.0753014 0.0875\\
-0.087493 0.1\\
-0.100118 0.1125\\
-0.113228 0.125\\
-0.126873 0.1375\\
-0.141104 0.15\\
-0.155971 0.1625\\
-0.171525 0.175\\
-0.187817 0.1875\\
-0.204898 0.2\\
-0.222818 0.2125\\
-0.241628 0.225\\
-0.261378 0.2375\\
-0.28212 0.25\\
-0.28212 0.25\\
-0.303904 0.2625\\
-0.32678 0.275\\
-0.3508 0.2875\\
-0.376014 0.3\\
-0.402472 0.3125\\
-0.430226 0.325\\
-0.459326 0.3375\\
-0.489823 0.35\\
-0.521768 0.3625\\
-0.55521 0.375\\
-0.590201 0.3875\\
-0.626792 0.4\\
-0.665033 0.4125\\
-0.704975 0.425\\
-0.746669 0.4375\\
-0.790165 0.45\\
-0.835513 0.4625\\
-0.882766 0.475\\
-0.931973 0.4875\\
-0.983184 0.5\\
};
\addlegendentry{HSDT11B};

\addplot [
color=black,
solid,
line width=1.2pt
]
table[row sep=crcr]{
0.737439 -0.5\\
0.719003 -0.4875\\
0.700567 -0.475\\
0.682131 -0.4625\\
0.663695 -0.45\\
0.645259 -0.4375\\
0.626823 -0.425\\
0.608387 -0.4125\\
0.589951 -0.4\\
0.571515 -0.3875\\
0.553079 -0.375\\
0.534643 -0.3625\\
0.516207 -0.35\\
0.497771 -0.3375\\
0.479335 -0.325\\
0.460899 -0.3125\\
0.442463 -0.3\\
0.424028 -0.2875\\
0.405592 -0.275\\
0.387156 -0.2625\\
0.36872 -0.25\\
0.36872 -0.25\\
0.350284 -0.2375\\
0.331848 -0.225\\
0.313412 -0.2125\\
0.294976 -0.2\\
0.27654 -0.1875\\
0.258104 -0.175\\
0.239668 -0.1625\\
0.221232 -0.15\\
0.202796 -0.1375\\
0.18436 -0.125\\
0.165924 -0.1125\\
0.147488 -0.1\\
0.129052 -0.0875\\
0.110616 -0.075\\
0.0921799 -0.0625\\
0.0737439 -0.05\\
0.0553079 -0.0375\\
0.036872 -0.025\\
0.018436 -0.0125\\
-2.14093e-15 3.90313e-17\\
-2.08336e-15 0\\
-0.018436 0.0125\\
-0.036872 0.025\\
-0.0553079 0.0375\\
-0.0737439 0.05\\
-0.0921799 0.0625\\
-0.110616 0.075\\
-0.129052 0.0875\\
-0.147488 0.1\\
-0.165924 0.1125\\
-0.18436 0.125\\
-0.202796 0.1375\\
-0.221232 0.15\\
-0.239668 0.1625\\
-0.258104 0.175\\
-0.27654 0.1875\\
-0.294976 0.2\\
-0.313412 0.2125\\
-0.331848 0.225\\
-0.350284 0.2375\\
-0.36872 0.25\\
-0.36872 0.25\\
-0.387156 0.2625\\
-0.405592 0.275\\
-0.424028 0.2875\\
-0.442463 0.3\\
-0.460899 0.3125\\
-0.479335 0.325\\
-0.497771 0.3375\\
-0.516207 0.35\\
-0.534643 0.3625\\
-0.553079 0.375\\
-0.571515 0.3875\\
-0.589951 0.4\\
-0.608387 0.4125\\
-0.626823 0.425\\
-0.645259 0.4375\\
-0.663695 0.45\\
-0.682131 0.4625\\
-0.700567 0.475\\
-0.719003 0.4875\\
-0.737439 0.5\\
};
\addlegendentry{TSDT7};

\addplot [
color=blue,
mark size=2.5pt,
only marks,
mark=x,
mark options={solid,draw=black}
]
table[row sep=crcr]{
0.564543 -0.5\\
0.55043 -0.4875\\
0.536316 -0.475\\
0.522202 -0.4625\\
0.508089 -0.45\\
0.493975 -0.4375\\
0.479862 -0.425\\
0.465748 -0.4125\\
0.451634 -0.4\\
0.437521 -0.3875\\
0.423407 -0.375\\
0.409294 -0.3625\\
0.39518 -0.35\\
0.381067 -0.3375\\
0.366953 -0.325\\
0.352839 -0.3125\\
0.338726 -0.3\\
0.324612 -0.2875\\
0.310499 -0.275\\
0.296385 -0.2625\\
0.282272 -0.25\\
0.282272 -0.25\\
0.268158 -0.2375\\
0.254044 -0.225\\
0.239931 -0.2125\\
0.225817 -0.2\\
0.211704 -0.1875\\
0.19759 -0.175\\
0.183477 -0.1625\\
0.169363 -0.15\\
0.155249 -0.1375\\
0.141136 -0.125\\
0.127022 -0.1125\\
0.112909 -0.1\\
0.098795 -0.0875\\
0.0846815 -0.075\\
0.0705679 -0.0625\\
0.0564543 -0.05\\
0.0423407 -0.0375\\
0.0282272 -0.025\\
0.0141136 -0.0125\\
3.87731e-16 3.90313e-17\\
4.318e-16 0\\
-0.0141136 0.0125\\
-0.0282272 0.025\\
-0.0423407 0.0375\\
-0.0564543 0.05\\
-0.0705679 0.0625\\
-0.0846815 0.075\\
-0.098795 0.0875\\
-0.112909 0.1\\
-0.127022 0.1125\\
-0.141136 0.125\\
-0.155249 0.1375\\
-0.169363 0.15\\
-0.183477 0.1625\\
-0.19759 0.175\\
-0.211704 0.1875\\
-0.225817 0.2\\
-0.239931 0.2125\\
-0.254044 0.225\\
-0.268158 0.2375\\
-0.282272 0.25\\
-0.282272 0.25\\
-0.296385 0.2625\\
-0.310499 0.275\\
-0.324612 0.2875\\
-0.338726 0.3\\
-0.352839 0.3125\\
-0.366953 0.325\\
-0.381067 0.3375\\
-0.39518 0.35\\
-0.409294 0.3625\\
-0.423407 0.375\\
-0.437521 0.3875\\
-0.451634 0.4\\
-0.465748 0.4125\\
-0.479862 0.425\\
-0.493975 0.4375\\
-0.508089 0.45\\
-0.522202 0.4625\\
-0.536316 0.475\\
-0.55043 0.4875\\
-0.564543 0.5\\
};
\addlegendentry{FSDT5};

\end{axis}
\end{tikzpicture}%
}}\hspace{60pt}
\subfigure[$\overline{w} = \overline{w}(0,0,z)$]{\scalebox{0.5}{
%
%
%
%
\begin{tikzpicture}

\begin{axis}[%
width=\figurewidth,
height=\figureheight,
scale only axis,
xmin=1.05,
xmax=1.35,
xlabel={$\bar{w}$},
ymin=-0.5,
ymax=0.5,
ylabel={Normalized thickness, z/h},
legend style={at={(0.5,1.0)},draw=black,fill=white,legend cell align=left}
]
\addplot [
color=blue,
mark size=2.5pt,
only marks,
mark=o,
mark options={solid,draw=black}
]
table[row sep=crcr]{
1.26878 -0.5\\
1.2698 -0.4875\\
1.27082 -0.475\\
1.27183 -0.4625\\
1.27284 -0.45\\
1.27385 -0.4375\\
1.27485 -0.425\\
1.27585 -0.4125\\
1.27684 -0.4\\
1.27783 -0.3875\\
1.27881 -0.375\\
1.27979 -0.3625\\
1.28077 -0.35\\
1.28174 -0.3375\\
1.28271 -0.325\\
1.28368 -0.3125\\
1.28464 -0.3\\
1.28559 -0.2875\\
1.28655 -0.275\\
1.2875 -0.2625\\
1.28844 -0.25\\
1.28844 -0.25\\
1.28938 -0.2375\\
1.29032 -0.225\\
1.29125 -0.2125\\
1.29218 -0.2\\
1.29311 -0.1875\\
1.29403 -0.175\\
1.29494 -0.1625\\
1.29586 -0.15\\
1.29677 -0.1375\\
1.29767 -0.125\\
1.29857 -0.1125\\
1.29947 -0.1\\
1.30036 -0.0875\\
1.30125 -0.075\\
1.30213 -0.0625\\
1.30302 -0.05\\
1.30389 -0.0375\\
1.30476 -0.025\\
1.30563 -0.0125\\
1.3065 3.90313e-17\\
1.3065 0\\
1.30736 0.0125\\
1.30822 0.025\\
1.30907 0.0375\\
1.30992 0.05\\
1.31076 0.0625\\
1.3116 0.075\\
1.31244 0.0875\\
1.31327 0.1\\
1.3141 0.1125\\
1.31493 0.125\\
1.31575 0.1375\\
1.31656 0.15\\
1.31738 0.1625\\
1.31819 0.175\\
1.31899 0.1875\\
1.31979 0.2\\
1.32059 0.2125\\
1.32138 0.225\\
1.32217 0.2375\\
1.32295 0.25\\
1.32295 0.25\\
1.32374 0.2625\\
1.32451 0.275\\
1.32529 0.2875\\
1.32605 0.3\\
1.32682 0.3125\\
1.32758 0.325\\
1.32834 0.3375\\
1.32909 0.35\\
1.32984 0.3625\\
1.33058 0.375\\
1.33132 0.3875\\
1.33206 0.4\\
1.33279 0.4125\\
1.33352 0.425\\
1.33425 0.4375\\
1.33497 0.45\\
1.33568 0.4625\\
1.3364 0.475\\
1.33711 0.4875\\
1.33781 0.5\\
};
\addlegendentry{HSDT13};

\addplot [
color=black,
dashed,
line width=1.2pt
]
table[row sep=crcr]{
1.31727 -0.5\\
1.31727 -0.4875\\
1.31727 -0.475\\
1.31727 -0.4625\\
1.31727 -0.45\\
1.31727 -0.4375\\
1.31727 -0.425\\
1.31727 -0.4125\\
1.31727 -0.4\\
1.31727 -0.3875\\
1.31727 -0.375\\
1.31727 -0.3625\\
1.31727 -0.35\\
1.31727 -0.3375\\
1.31727 -0.325\\
1.31727 -0.3125\\
1.31727 -0.3\\
1.31727 -0.2875\\
1.31727 -0.275\\
1.31727 -0.2625\\
1.31727 -0.25\\
1.31727 -0.25\\
1.31727 -0.2375\\
1.31727 -0.225\\
1.31727 -0.2125\\
1.31727 -0.2\\
1.31727 -0.1875\\
1.31727 -0.175\\
1.31727 -0.1625\\
1.31727 -0.15\\
1.31727 -0.1375\\
1.31727 -0.125\\
1.31727 -0.1125\\
1.31727 -0.1\\
1.31727 -0.0875\\
1.31727 -0.075\\
1.31727 -0.0625\\
1.31727 -0.05\\
1.31727 -0.0375\\
1.31727 -0.025\\
1.31727 -0.0125\\
1.31727 3.90313e-17\\
1.31727 0\\
1.31727 0.0125\\
1.31727 0.025\\
1.31727 0.0375\\
1.31727 0.05\\
1.31727 0.0625\\
1.31727 0.075\\
1.31727 0.0875\\
1.31727 0.1\\
1.31727 0.1125\\
1.31727 0.125\\
1.31727 0.1375\\
1.31727 0.15\\
1.31727 0.1625\\
1.31727 0.175\\
1.31727 0.1875\\
1.31727 0.2\\
1.31727 0.2125\\
1.31727 0.225\\
1.31727 0.2375\\
1.31727 0.25\\
1.31727 0.25\\
1.31727 0.2625\\
1.31727 0.275\\
1.31727 0.2875\\
1.31727 0.3\\
1.31727 0.3125\\
1.31727 0.325\\
1.31727 0.3375\\
1.31727 0.35\\
1.31727 0.3625\\
1.31727 0.375\\
1.31727 0.3875\\
1.31727 0.4\\
1.31727 0.4125\\
1.31727 0.425\\
1.31727 0.4375\\
1.31727 0.45\\
1.31727 0.4625\\
1.31727 0.475\\
1.31727 0.4875\\
1.31727 0.5\\
};
\addlegendentry{HSDT11A};

\addplot [
color=blue,
mark size=4.3pt,
only marks,
mark=diamond,
mark options={solid,draw=black}
]
table[row sep=crcr]{
1.23016 -0.5\\
1.23117 -0.4875\\
1.23218 -0.475\\
1.23318 -0.4625\\
1.23418 -0.45\\
1.23517 -0.4375\\
1.23617 -0.425\\
1.23715 -0.4125\\
1.23814 -0.4\\
1.23912 -0.3875\\
1.2401 -0.375\\
1.24107 -0.3625\\
1.24204 -0.35\\
1.24301 -0.3375\\
1.24397 -0.325\\
1.24493 -0.3125\\
1.24588 -0.3\\
1.24683 -0.2875\\
1.24778 -0.275\\
1.24872 -0.2625\\
1.24966 -0.25\\
1.24966 -0.25\\
1.2506 -0.2375\\
1.25153 -0.225\\
1.25246 -0.2125\\
1.25338 -0.2\\
1.2543 -0.1875\\
1.25522 -0.175\\
1.25613 -0.1625\\
1.25704 -0.15\\
1.25795 -0.1375\\
1.25885 -0.125\\
1.25975 -0.1125\\
1.26064 -0.1\\
1.26154 -0.0875\\
1.26242 -0.075\\
1.26331 -0.0625\\
1.26419 -0.05\\
1.26506 -0.0375\\
1.26593 -0.025\\
1.2668 -0.0125\\
1.26767 3.90313e-17\\
1.26767 0\\
1.26853 0.0125\\
1.26939 0.025\\
1.27024 0.0375\\
1.27109 0.05\\
1.27194 0.0625\\
1.27278 0.075\\
1.27362 0.0875\\
1.27445 0.1\\
1.27528 0.1125\\
1.27611 0.125\\
1.27694 0.1375\\
1.27776 0.15\\
1.27857 0.1625\\
1.27938 0.175\\
1.28019 0.1875\\
1.281 0.2\\
1.2818 0.2125\\
1.2826 0.225\\
1.28339 0.2375\\
1.28418 0.25\\
1.28418 0.25\\
1.28497 0.2625\\
1.28575 0.275\\
1.28653 0.2875\\
1.28731 0.3\\
1.28808 0.3125\\
1.28884 0.325\\
1.28961 0.3375\\
1.29037 0.35\\
1.29113 0.3625\\
1.29188 0.375\\
1.29263 0.3875\\
1.29337 0.4\\
1.29411 0.4125\\
1.29485 0.425\\
1.29559 0.4375\\
1.29632 0.45\\
1.29704 0.4625\\
1.29777 0.475\\
1.29849 0.4875\\
1.2992 0.5\\
};
\addlegendentry{HSDT11B};

\addplot [
color=black,
solid,
line width=1.2pt
]
table[row sep=crcr]{
1.1039 -0.5\\
1.1039 -0.4875\\
1.1039 -0.475\\
1.1039 -0.4625\\
1.1039 -0.45\\
1.1039 -0.4375\\
1.1039 -0.425\\
1.1039 -0.4125\\
1.1039 -0.4\\
1.1039 -0.3875\\
1.1039 -0.375\\
1.1039 -0.3625\\
1.1039 -0.35\\
1.1039 -0.3375\\
1.1039 -0.325\\
1.1039 -0.3125\\
1.1039 -0.3\\
1.1039 -0.2875\\
1.1039 -0.275\\
1.1039 -0.2625\\
1.1039 -0.25\\
1.1039 -0.25\\
1.1039 -0.2375\\
1.1039 -0.225\\
1.1039 -0.2125\\
1.1039 -0.2\\
1.1039 -0.1875\\
1.1039 -0.175\\
1.1039 -0.1625\\
1.1039 -0.15\\
1.1039 -0.1375\\
1.1039 -0.125\\
1.1039 -0.1125\\
1.1039 -0.1\\
1.1039 -0.0875\\
1.1039 -0.075\\
1.1039 -0.0625\\
1.1039 -0.05\\
1.1039 -0.0375\\
1.1039 -0.025\\
1.1039 -0.0125\\
1.1039 3.90313e-17\\
1.1039 0\\
1.1039 0.0125\\
1.1039 0.025\\
1.1039 0.0375\\
1.1039 0.05\\
1.1039 0.0625\\
1.1039 0.075\\
1.1039 0.0875\\
1.1039 0.1\\
1.1039 0.1125\\
1.1039 0.125\\
1.1039 0.1375\\
1.1039 0.15\\
1.1039 0.1625\\
1.1039 0.175\\
1.1039 0.1875\\
1.1039 0.2\\
1.1039 0.2125\\
1.1039 0.225\\
1.1039 0.2375\\
1.1039 0.25\\
1.1039 0.25\\
1.1039 0.2625\\
1.1039 0.275\\
1.1039 0.2875\\
1.1039 0.3\\
1.1039 0.3125\\
1.1039 0.325\\
1.1039 0.3375\\
1.1039 0.35\\
1.1039 0.3625\\
1.1039 0.375\\
1.1039 0.3875\\
1.1039 0.4\\
1.1039 0.4125\\
1.1039 0.425\\
1.1039 0.4375\\
1.1039 0.45\\
1.1039 0.4625\\
1.1039 0.475\\
1.1039 0.4875\\
1.1039 0.5\\
};
\addlegendentry{TSDT7};

\addplot [
color=blue,
mark size=2.5pt,
only marks,
mark=x,
mark options={solid,draw=black}
]
table[row sep=crcr]{
1.10444 -0.5\\
1.10444 -0.4875\\
1.10444 -0.475\\
1.10444 -0.4625\\
1.10444 -0.45\\
1.10444 -0.4375\\
1.10444 -0.425\\
1.10444 -0.4125\\
1.10444 -0.4\\
1.10444 -0.3875\\
1.10444 -0.375\\
1.10444 -0.3625\\
1.10444 -0.35\\
1.10444 -0.3375\\
1.10444 -0.325\\
1.10444 -0.3125\\
1.10444 -0.3\\
1.10444 -0.2875\\
1.10444 -0.275\\
1.10444 -0.2625\\
1.10444 -0.25\\
1.10444 -0.25\\
1.10444 -0.2375\\
1.10444 -0.225\\
1.10444 -0.2125\\
1.10444 -0.2\\
1.10444 -0.1875\\
1.10444 -0.175\\
1.10444 -0.1625\\
1.10444 -0.15\\
1.10444 -0.1375\\
1.10444 -0.125\\
1.10444 -0.1125\\
1.10444 -0.1\\
1.10444 -0.0875\\
1.10444 -0.075\\
1.10444 -0.0625\\
1.10444 -0.05\\
1.10444 -0.0375\\
1.10444 -0.025\\
1.10444 -0.0125\\
1.10444 3.90313e-17\\
1.10444 0\\
1.10444 0.0125\\
1.10444 0.025\\
1.10444 0.0375\\
1.10444 0.05\\
1.10444 0.0625\\
1.10444 0.075\\
1.10444 0.0875\\
1.10444 0.1\\
1.10444 0.1125\\
1.10444 0.125\\
1.10444 0.1375\\
1.10444 0.15\\
1.10444 0.1625\\
1.10444 0.175\\
1.10444 0.1875\\
1.10444 0.2\\
1.10444 0.2125\\
1.10444 0.225\\
1.10444 0.2375\\
1.10444 0.25\\
1.10444 0.25\\
1.10444 0.2625\\
1.10444 0.275\\
1.10444 0.2875\\
1.10444 0.3\\
1.10444 0.3125\\
1.10444 0.325\\
1.10444 0.3375\\
1.10444 0.35\\
1.10444 0.3625\\
1.10444 0.375\\
1.10444 0.3875\\
1.10444 0.4\\
1.10444 0.4125\\
1.10444 0.425\\
1.10444 0.4375\\
1.10444 0.45\\
1.10444 0.4625\\
1.10444 0.475\\
1.10444 0.4875\\
1.10444 0.5\\
};
\addlegendentry{FSDT5};

\end{axis}
\end{tikzpicture}%
}}
\subfigure[$\overline{\sigma}_{xx} = \overline{\sigma}_{xx}(0,0,z)$]{\scalebox{0.5}{
%
%
%
%
\begin{tikzpicture}

\begin{axis}[%
width=\figurewidth,
height=\figureheight,
scale only axis,
xmin=-0.4,
xmax=0.4,
xlabel={$\bar{\sigma}_{xx}$},
ymin=-0.5,
ymax=0.5,
ylabel={Normalized thickness, z/h},
legend style={draw=black,fill=white,legend cell align=left},
legend pos = south east
]
\addplot [
color=blue,
mark size=2.5pt,
only marks,
mark=o,
mark options={solid,draw=black}
]
table[row sep=crcr]{
-0.299098 -0.5\\
-0.282305 -0.4875\\
-0.266445 -0.475\\
-0.251479 -0.4625\\
-0.237368 -0.45\\
-0.224071 -0.4375\\
-0.211549 -0.425\\
-0.199761 -0.4125\\
-0.188668 -0.4\\
-0.178231 -0.3875\\
-0.168409 -0.375\\
-0.159162 -0.3625\\
-0.15045 -0.35\\
-0.142235 -0.3375\\
-0.134476 -0.325\\
-0.127132 -0.3125\\
-0.120166 -0.3\\
-0.113535 -0.2875\\
-0.107202 -0.275\\
-0.101125 -0.2625\\
-0.0952654 -0.25\\
-0.342965 -0.25\\
-0.313639 -0.2375\\
-0.285674 -0.225\\
-0.25903 -0.2125\\
-0.233667 -0.2\\
-0.209544 -0.1875\\
-0.186621 -0.175\\
-0.164857 -0.1625\\
-0.144211 -0.15\\
-0.124644 -0.1375\\
-0.106114 -0.125\\
-0.0885816 -0.1125\\
-0.0720059 -0.1\\
-0.0563466 -0.0875\\
-0.0415629 -0.075\\
-0.0276146 -0.0625\\
-0.0144612 -0.05\\
-0.002062 -0.0375\\
0.0096233 -0.025\\
0.0206352 -0.0125\\
0.0310143 3.90313e-17\\
-0.0363994 0\\
-0.0259 0.0125\\
-0.0147417 0.025\\
-0.00288456 0.0375\\
0.00971119 0.05\\
0.0230854 0.0625\\
0.0372778 0.075\\
0.0523282 0.0875\\
0.0682766 0.1\\
0.0851627 0.1125\\
0.103026 0.125\\
0.121907 0.1375\\
0.141845 0.15\\
0.16288 0.1625\\
0.185052 0.175\\
0.208401 0.1875\\
0.232966 0.2\\
0.258787 0.2125\\
0.285904 0.225\\
0.314358 0.2375\\
0.344187 0.25\\
0.0925783 0.25\\
0.099018 0.2625\\
0.105715 0.275\\
0.112711 0.2875\\
0.120045 0.3\\
0.127759 0.3125\\
0.135892 0.325\\
0.144485 0.3375\\
0.153579 0.35\\
0.163213 0.3625\\
0.17343 0.375\\
0.184268 0.3875\\
0.195769 0.4\\
0.207973 0.4125\\
0.220921 0.425\\
0.234652 0.4375\\
0.249208 0.45\\
0.264629 0.4625\\
0.280956 0.475\\
0.298228 0.4875\\
0.316487 0.5\\
};
\addlegendentry{HSDT13};

\addplot [
color=black,
dashed,
line width=1.2pt
]
table[row sep=crcr]{
-0.306971 -0.5\\
-0.289353 -0.4875\\
-0.272708 -0.475\\
-0.256996 -0.4625\\
-0.242174 -0.45\\
-0.228204 -0.4375\\
-0.215043 -0.425\\
-0.202651 -0.4125\\
-0.190988 -0.4\\
-0.180013 -0.3875\\
-0.169685 -0.375\\
-0.159964 -0.3625\\
-0.150808 -0.35\\
-0.142177 -0.3375\\
-0.13403 -0.325\\
-0.126327 -0.3125\\
-0.119027 -0.3\\
-0.11209 -0.2875\\
-0.105473 -0.275\\
-0.0991377 -0.2625\\
-0.0930422 -0.25\\
-0.347316 -0.25\\
-0.317447 -0.2375\\
-0.288966 -0.225\\
-0.261833 -0.2125\\
-0.236006 -0.2\\
-0.211445 -0.1875\\
-0.18811 -0.175\\
-0.165959 -0.1625\\
-0.144952 -0.15\\
-0.125048 -0.1375\\
-0.106206 -0.125\\
-0.0883861 -0.1125\\
-0.0715472 -0.1\\
-0.0556486 -0.0875\\
-0.0406497 -0.075\\
-0.0265096 -0.0625\\
-0.0131878 -0.05\\
-0.000643406 -0.0375\\
0.0111641 -0.025\\
0.0222756 -0.0125\\
0.0327317 3.90313e-17\\
-0.0327317 0\\
-0.0222756 0.0125\\
-0.0111641 0.025\\
0.000643406 0.0375\\
0.0131878 0.05\\
0.0265096 0.0625\\
0.0406497 0.075\\
0.0556486 0.0875\\
0.0715472 0.1\\
0.0883861 0.1125\\
0.106206 0.125\\
0.125048 0.1375\\
0.144952 0.15\\
0.165959 0.1625\\
0.18811 0.175\\
0.211445 0.1875\\
0.236006 0.2\\
0.261833 0.2125\\
0.288966 0.225\\
0.317447 0.2375\\
0.347316 0.25\\
0.0930422 0.25\\
0.0991377 0.2625\\
0.105473 0.275\\
0.11209 0.2875\\
0.119027 0.3\\
0.126327 0.3125\\
0.13403 0.325\\
0.142177 0.3375\\
0.150808 0.35\\
0.159964 0.3625\\
0.169685 0.375\\
0.180013 0.3875\\
0.190988 0.4\\
0.202651 0.4125\\
0.215043 0.425\\
0.228204 0.4375\\
0.242174 0.45\\
0.256996 0.4625\\
0.272708 0.475\\
0.289353 0.4875\\
0.306971 0.5\\
};
\addlegendentry{HSDT11A};

\addplot [
color=blue,
mark size=4.3pt,
only marks,
mark=diamond,
mark options={solid,draw=black}
]
table[row sep=crcr]{
-0.287688 -0.5\\
-0.27244 -0.4875\\
-0.258112 -0.475\\
-0.244665 -0.4625\\
-0.23206 -0.45\\
-0.220258 -0.4375\\
-0.209221 -0.425\\
-0.19891 -0.4125\\
-0.189286 -0.4\\
-0.180311 -0.3875\\
-0.171945 -0.375\\
-0.16415 -0.3625\\
-0.156888 -0.35\\
-0.150118 -0.3375\\
-0.143804 -0.325\\
-0.137905 -0.3125\\
-0.132384 -0.3\\
-0.127201 -0.2875\\
-0.122318 -0.275\\
-0.117696 -0.2625\\
-0.113296 -0.25\\
-0.323678 -0.25\\
-0.296956 -0.2375\\
-0.271547 -0.225\\
-0.247411 -0.2125\\
-0.224507 -0.2\\
-0.202798 -0.1875\\
-0.182243 -0.175\\
-0.162803 -0.1625\\
-0.144439 -0.15\\
-0.127111 -0.1375\\
-0.11078 -0.125\\
-0.0954064 -0.1125\\
-0.0809512 -0.1\\
-0.0673747 -0.0875\\
-0.0546377 -0.075\\
-0.0427006 -0.0625\\
-0.0315243 -0.05\\
-0.0210692 -0.0375\\
-0.0112959 -0.025\\
-0.00216524 -0.0125\\
0.00636234 3.90313e-17\\
-0.0122973 0\\
-0.00361402 0.0125\\
0.00569937 0.025\\
0.0156817 0.0375\\
0.0263716 0.05\\
0.037808 0.0625\\
0.0500294 0.075\\
0.0630748 0.0875\\
0.0769828 0.1\\
0.0917922 0.1125\\
0.107542 0.125\\
0.12427 0.1375\\
0.142016 0.15\\
0.160818 0.1625\\
0.180716 0.175\\
0.201747 0.1875\\
0.223951 0.2\\
0.247366 0.2125\\
0.272031 0.225\\
0.297985 0.2375\\
0.325267 0.25\\
0.110904 0.25\\
0.115863 0.2625\\
0.121085 0.275\\
0.12661 0.2875\\
0.132477 0.3\\
0.138725 0.3125\\
0.145394 0.325\\
0.152523 0.3375\\
0.160152 0.35\\
0.168319 0.3625\\
0.177065 0.375\\
0.186429 0.3875\\
0.196449 0.4\\
0.207166 0.4125\\
0.218619 0.425\\
0.230847 0.4375\\
0.24389 0.45\\
0.257787 0.4625\\
0.272577 0.475\\
0.288299 0.4875\\
0.304994 0.5\\
};
\addlegendentry{HSDT11B};

\addplot [
color=black,
solid,
line width=1.2pt
]
table[row sep=crcr]{
-0.174137 -0.5\\
-0.178786 -0.4875\\
-0.18299 -0.475\\
-0.186747 -0.4625\\
-0.190059 -0.45\\
-0.192926 -0.4375\\
-0.195347 -0.425\\
-0.197322 -0.4125\\
-0.198852 -0.4\\
-0.199936 -0.3875\\
-0.200574 -0.375\\
-0.200767 -0.3625\\
-0.200514 -0.35\\
-0.199815 -0.3375\\
-0.198671 -0.325\\
-0.197081 -0.3125\\
-0.195045 -0.3\\
-0.192564 -0.2875\\
-0.189637 -0.275\\
-0.186264 -0.2625\\
-0.182446 -0.25\\
-0.335712 -0.25\\
-0.314068 -0.2375\\
-0.29287 -0.225\\
-0.272117 -0.2125\\
-0.25181 -0.2\\
-0.231948 -0.1875\\
-0.212532 -0.175\\
-0.193562 -0.1625\\
-0.175038 -0.15\\
-0.156959 -0.1375\\
-0.139325 -0.125\\
-0.122138 -0.1125\\
-0.105396 -0.1\\
-0.0890996 -0.0875\\
-0.0732489 -0.075\\
-0.0578439 -0.0625\\
-0.0428846 -0.05\\
-0.0283708 -0.0375\\
-0.0143028 -0.025\\
-0.000680339 -0.0125\\
0.0124964 3.90313e-17\\
-0.0124964 0\\
0.000680339 0.0125\\
0.0143028 0.025\\
0.0283708 0.0375\\
0.0428846 0.05\\
0.0578439 0.0625\\
0.0732489 0.075\\
0.0890996 0.0875\\
0.105396 0.1\\
0.122138 0.1125\\
0.139325 0.125\\
0.156959 0.1375\\
0.175038 0.15\\
0.193562 0.1625\\
0.212532 0.175\\
0.231948 0.1875\\
0.25181 0.2\\
0.272117 0.2125\\
0.29287 0.225\\
0.314068 0.2375\\
0.335712 0.25\\
0.182446 0.25\\
0.186264 0.2625\\
0.189637 0.275\\
0.192564 0.2875\\
0.195045 0.3\\
0.197081 0.3125\\
0.198671 0.325\\
0.199815 0.3375\\
0.200514 0.35\\
0.200767 0.3625\\
0.200574 0.375\\
0.199936 0.3875\\
0.198852 0.4\\
0.197322 0.4125\\
0.195347 0.425\\
0.192926 0.4375\\
0.190059 0.45\\
0.186747 0.4625\\
0.18299 0.475\\
0.178786 0.4875\\
0.174137 0.5\\
};
\addlegendentry{TSDT7};

\addplot [
color=blue,
mark size=2.5pt,
only marks,
mark=x,
mark options={solid,draw=black}
]
table[row sep=crcr]{
-0.279382 -0.5\\
-0.270677 -0.4875\\
-0.261972 -0.475\\
-0.253267 -0.4625\\
-0.244562 -0.45\\
-0.235857 -0.4375\\
-0.227152 -0.425\\
-0.218448 -0.4125\\
-0.209743 -0.4\\
-0.201038 -0.3875\\
-0.192333 -0.375\\
-0.183628 -0.3625\\
-0.174923 -0.35\\
-0.166218 -0.3375\\
-0.157514 -0.325\\
-0.148809 -0.3125\\
-0.140104 -0.3\\
-0.131399 -0.2875\\
-0.122694 -0.275\\
-0.113989 -0.2625\\
-0.105284 -0.25\\
-0.24291 -0.25\\
-0.234206 -0.2375\\
-0.225501 -0.225\\
-0.216796 -0.2125\\
-0.208091 -0.2\\
-0.199386 -0.1875\\
-0.190681 -0.175\\
-0.181976 -0.1625\\
-0.173271 -0.15\\
-0.164567 -0.1375\\
-0.155862 -0.125\\
-0.147157 -0.1125\\
-0.138452 -0.1\\
-0.129747 -0.0875\\
-0.121042 -0.075\\
-0.112337 -0.0625\\
-0.103633 -0.05\\
-0.0949277 -0.0375\\
-0.0862228 -0.025\\
-0.0775179 -0.0125\\
-0.0688131 3.90313e-17\\
0.0688131 0\\
0.0775179 0.0125\\
0.0862228 0.025\\
0.0949277 0.0375\\
0.103633 0.05\\
0.112337 0.0625\\
0.121042 0.075\\
0.129747 0.0875\\
0.138452 0.1\\
0.147157 0.1125\\
0.155862 0.125\\
0.164567 0.1375\\
0.173271 0.15\\
0.181976 0.1625\\
0.190681 0.175\\
0.199386 0.1875\\
0.208091 0.2\\
0.216796 0.2125\\
0.225501 0.225\\
0.234206 0.2375\\
0.24291 0.25\\
0.105284 0.25\\
0.113989 0.2625\\
0.122694 0.275\\
0.131399 0.2875\\
0.140104 0.3\\
0.148809 0.3125\\
0.157514 0.325\\
0.166218 0.3375\\
0.174923 0.35\\
0.183628 0.3625\\
0.192333 0.375\\
0.201038 0.3875\\
0.209743 0.4\\
0.218448 0.4125\\
0.227152 0.425\\
0.235857 0.4375\\
0.244562 0.45\\
0.253267 0.4625\\
0.261972 0.475\\
0.270677 0.4875\\
0.279382 0.5\\
};
\addlegendentry{FSDT5};

\end{axis}
\end{tikzpicture}%
}}\hspace{60pt}
\subfigure[$\overline{\tau}_{yz} = \overline{\tau}_{yz}(-a/2,0,z)$]{\scalebox{0.5}{
%
%
%
%
\begin{tikzpicture}

\begin{axis}[%
width=\figurewidth,
height=\figureheight,
scale only axis,
xmin=-0.35,
xmax=0.05,
xlabel={$\bar{\tau}_{yz}$},
ymin=-0.5,
ymax=0.5,
ylabel={Normalized thickness, z/h},
legend style={draw=black,fill=white,legend cell align=left}
]
\addplot [
color=blue,
mark size=2.5pt,
only marks,
mark=o,
mark options={solid,draw=black}
]
table[row sep=crcr]{
0 -0.5\\
-0.0141545 -0.4875\\
-0.0272918 -0.475\\
-0.0394605 -0.4625\\
-0.0507097 -0.45\\
-0.0610892 -0.4375\\
-0.070649 -0.425\\
-0.0794398 -0.4125\\
-0.0875128 -0.4\\
-0.0949197 -0.3875\\
-0.101713 -0.375\\
-0.107944 -0.3625\\
-0.113668 -0.35\\
-0.118937 -0.3375\\
-0.123805 -0.325\\
-0.128329 -0.3125\\
-0.132561 -0.3\\
-0.13656 -0.2875\\
-0.140379 -0.275\\
-0.144077 -0.2625\\
-0.14771 -0.25\\
-0.14771 -0.25\\
-0.164142 -0.2375\\
-0.178577 -0.225\\
-0.191085 -0.2125\\
-0.201733 -0.2\\
-0.210588 -0.1875\\
-0.217715 -0.175\\
-0.223177 -0.1625\\
-0.227038 -0.15\\
-0.229358 -0.1375\\
-0.230198 -0.125\\
-0.229617 -0.1125\\
-0.227673 -0.1\\
-0.224423 -0.0875\\
-0.219922 -0.075\\
-0.214224 -0.0625\\
-0.207382 -0.05\\
-0.199449 -0.0375\\
-0.190474 -0.025\\
-0.180508 -0.0125\\
-0.169598 3.90313e-17\\
-0.169598 0\\
-0.155074 0.0125\\
-0.141872 0.025\\
-0.13006 0.0375\\
-0.119708 0.05\\
-0.110885 0.0625\\
-0.103662 0.075\\
-0.0981104 0.0875\\
-0.0943004 0.1\\
-0.0923042 0.1125\\
-0.0921941 0.125\\
-0.0940431 0.1375\\
-0.0979245 0.15\\
-0.103912 0.1625\\
-0.11208 0.175\\
-0.122504 0.1875\\
-0.135258 0.2\\
-0.150419 0.2125\\
-0.168063 0.225\\
-0.188266 0.2375\\
-0.211107 0.25\\
-0.211107 0.25\\
-0.215021 0.2625\\
-0.218747 0.275\\
-0.222307 0.2875\\
-0.22572 0.3\\
-0.229004 0.3125\\
-0.232176 0.325\\
-0.235252 0.3375\\
-0.238246 0.35\\
-0.241171 0.3625\\
-0.244041 0.375\\
-0.246865 0.3875\\
-0.249653 0.4\\
-0.252414 0.4125\\
-0.255155 0.425\\
-0.257882 0.4375\\
-0.2606 0.45\\
-0.263313 0.4625\\
-0.266024 0.475\\
-0.268733 0.4875\\
-0.27144 0.5\\
};
\addlegendentry{HSDT13};

\addplot [
color=black,
dashed,
line width=1.2pt
]
table[row sep=crcr]{
0 -0.5\\
-0.0142128 -0.4875\\
-0.0274656 -0.475\\
-0.0398031 -0.4625\\
-0.0512708 -0.45\\
-0.0619144 -0.4375\\
-0.0717806 -0.425\\
-0.0809166 -0.4125\\
-0.0893702 -0.4\\
-0.0971899 -0.3875\\
-0.104425 -0.375\\
-0.111124 -0.3625\\
-0.117339 -0.35\\
-0.123121 -0.3375\\
-0.12852 -0.325\\
-0.133588 -0.3125\\
-0.13838 -0.3\\
-0.142949 -0.2875\\
-0.147347 -0.275\\
-0.151631 -0.2625\\
-0.155856 -0.25\\
-0.155856 -0.25\\
-0.172555 -0.2375\\
-0.187225 -0.225\\
-0.199933 -0.2125\\
-0.210748 -0.2\\
-0.219736 -0.1875\\
-0.226962 -0.175\\
-0.232491 -0.1625\\
-0.236385 -0.15\\
-0.238706 -0.1375\\
-0.239516 -0.125\\
-0.238873 -0.1125\\
-0.236836 -0.1\\
-0.233462 -0.0875\\
-0.228808 -0.075\\
-0.222928 -0.0625\\
-0.215876 -0.05\\
-0.207705 -0.0375\\
-0.198467 -0.025\\
-0.188211 -0.0125\\
-0.176988 3.90313e-17\\
-0.176988 0\\
-0.162728 0.0125\\
-0.149786 0.025\\
-0.138232 0.0375\\
-0.128135 0.05\\
-0.119567 0.0625\\
-0.112599 0.075\\
-0.107305 0.0875\\
-0.103756 0.1\\
-0.102026 0.1125\\
-0.10219 0.125\\
-0.104323 0.1375\\
-0.108499 0.15\\
-0.114795 0.1625\\
-0.123288 0.175\\
-0.134055 0.1875\\
-0.147174 0.2\\
-0.162724 0.2125\\
-0.180784 0.225\\
-0.201433 0.2375\\
-0.224752 0.25\\
-0.224752 0.25\\
-0.228289 0.2625\\
-0.23165 0.275\\
-0.234857 0.2875\\
-0.237934 0.3\\
-0.240899 0.3125\\
-0.243772 0.325\\
-0.246572 0.3375\\
-0.249316 0.35\\
-0.252019 0.3625\\
-0.254696 0.375\\
-0.257361 0.3875\\
-0.260026 0.4\\
-0.262702 0.4125\\
-0.2654 0.425\\
-0.268129 0.4375\\
-0.270896 0.45\\
-0.273708 0.4625\\
-0.276571 0.475\\
-0.279489 0.4875\\
-0.282465 0.5\\
};
\addlegendentry{HSDT11A};

\addplot [
color=blue,
mark size=4.3pt,
only marks,
mark=diamond,
mark options={solid,draw=black}
]
table[row sep=crcr]{
0 -0.5\\
-0.0188867 -0.4875\\
-0.036266 -0.475\\
-0.0521767 -0.4625\\
-0.0666579 -0.45\\
-0.0797495 -0.4375\\
-0.091492 -0.425\\
-0.101927 -0.4125\\
-0.111095 -0.4\\
-0.11904 -0.3875\\
-0.125804 -0.375\\
-0.131431 -0.3625\\
-0.135965 -0.35\\
-0.139452 -0.3375\\
-0.141937 -0.325\\
-0.143465 -0.3125\\
-0.144085 -0.3\\
-0.143842 -0.2875\\
-0.142787 -0.275\\
-0.140966 -0.2625\\
-0.138431 -0.25\\
-0.138431 -0.25\\
-0.148539 -0.2375\\
-0.157348 -0.225\\
-0.164935 -0.2125\\
-0.171376 -0.2\\
-0.176746 -0.1875\\
-0.181117 -0.175\\
-0.184559 -0.1625\\
-0.18714 -0.15\\
-0.188927 -0.1375\\
-0.189985 -0.125\\
-0.190376 -0.1125\\
-0.190161 -0.1\\
-0.189398 -0.0875\\
-0.188145 -0.075\\
-0.186457 -0.0625\\
-0.184387 -0.05\\
-0.181985 -0.0375\\
-0.179301 -0.025\\
-0.176383 -0.0125\\
-0.173276 3.90313e-17\\
-0.173276 0\\
-0.168249 0.0125\\
-0.163753 0.025\\
-0.159853 0.0375\\
-0.156613 0.05\\
-0.154099 0.0625\\
-0.152375 0.075\\
-0.151509 0.0875\\
-0.151569 0.1\\
-0.15262 0.1125\\
-0.154733 0.125\\
-0.157976 0.1375\\
-0.162419 0.15\\
-0.168131 0.1625\\
-0.175185 0.175\\
-0.183651 0.1875\\
-0.193602 0.2\\
-0.205111 0.2125\\
-0.21825 0.225\\
-0.233095 0.2375\\
-0.24972 0.25\\
-0.24972 0.25\\
-0.249746 0.2625\\
-0.250099 0.275\\
-0.250779 0.2875\\
-0.251786 0.3\\
-0.253117 0.3125\\
-0.25477 0.325\\
-0.256736 0.3375\\
-0.259008 0.35\\
-0.261576 0.3625\\
-0.264427 0.375\\
-0.267549 0.3875\\
-0.270925 0.4\\
-0.274537 0.4125\\
-0.278367 0.425\\
-0.282391 0.4375\\
-0.286587 0.45\\
-0.29093 0.4625\\
-0.295391 0.475\\
-0.299943 0.4875\\
-0.304554 0.5\\
};
\addlegendentry{HSDT11B};

\addplot [
color=black,
solid,
line width=1.2pt
]
table[row sep=crcr]{
0 -0.5\\
-0.012337 -0.4875\\
-0.0234868 -0.475\\
-0.0334905 -0.4625\\
-0.0423894 -0.45\\
-0.0502246 -0.4375\\
-0.0570374 -0.425\\
-0.0628689 -0.4125\\
-0.0677603 -0.4\\
-0.0717529 -0.3875\\
-0.0748878 -0.375\\
-0.0772062 -0.3625\\
-0.0787494 -0.35\\
-0.0795585 -0.3375\\
-0.0796747 -0.325\\
-0.0791392 -0.3125\\
-0.0779932 -0.3\\
-0.076278 -0.2875\\
-0.0740347 -0.275\\
-0.0713045 -0.2625\\
-0.0681286 -0.25\\
-0.0681286 -0.25\\
-0.0774607 -0.2375\\
-0.0859714 -0.225\\
-0.0936865 -0.2125\\
-0.100632 -0.2\\
-0.106833 -0.1875\\
-0.112316 -0.175\\
-0.117107 -0.1625\\
-0.121231 -0.15\\
-0.124715 -0.1375\\
-0.127584 -0.125\\
-0.129863 -0.1125\\
-0.13158 -0.1\\
-0.132759 -0.0875\\
-0.133427 -0.075\\
-0.133608 -0.0625\\
-0.13333 -0.05\\
-0.132618 -0.0375\\
-0.131498 -0.025\\
-0.129995 -0.0125\\
-0.128135 3.90313e-17\\
-0.128135 0\\
-0.125117 0.0125\\
-0.122559 0.025\\
-0.120502 0.0375\\
-0.118989 0.05\\
-0.118061 0.0625\\
-0.117757 0.075\\
-0.118121 0.0875\\
-0.119192 0.1\\
-0.121012 0.1125\\
-0.123622 0.125\\
-0.127064 0.1375\\
-0.131379 0.15\\
-0.136607 0.1625\\
-0.142791 0.175\\
-0.14997 0.1875\\
-0.158187 0.2\\
-0.167483 0.2125\\
-0.177899 0.225\\
-0.189475 0.2375\\
-0.202254 0.25\\
-0.202254 0.25\\
-0.198871 0.2625\\
-0.195699 0.275\\
-0.192765 0.2875\\
-0.190094 0.3\\
-0.187712 0.3125\\
-0.185644 0.325\\
-0.183918 0.3375\\
-0.182558 0.35\\
-0.18159 0.3625\\
-0.18104 0.375\\
-0.180934 0.3875\\
-0.181298 0.4\\
-0.182157 0.4125\\
-0.183538 0.425\\
-0.185466 0.4375\\
-0.187967 0.45\\
-0.191067 0.4625\\
-0.194792 0.475\\
-0.199167 0.4875\\
-0.204218 0.5\\
};
\addlegendentry{TSDT7};

\addplot [
color=blue,
mark size=2.5pt,
only marks,
mark=x,
mark options={solid,draw=black}
]
table[row sep=crcr]{
0 -0.5\\
0.0031699 -0.4875\\
0.0061328 -0.475\\
0.00888871 -0.4625\\
0.0114376 -0.45\\
0.0137796 -0.4375\\
0.0159145 -0.425\\
0.0178425 -0.4125\\
0.0195634 -0.4\\
0.0210774 -0.3875\\
0.0223844 -0.375\\
0.0234844 -0.3625\\
0.0243774 -0.35\\
0.0250634 -0.3375\\
0.0255424 -0.325\\
0.0258144 -0.3125\\
0.0258794 -0.3\\
0.0257375 -0.2875\\
0.0253885 -0.275\\
0.0248326 -0.2625\\
0.0240697 -0.25\\
0.0240697 -0.25\\
0.0186154 -0.2375\\
0.0132834 -0.225\\
0.00807366 -0.2125\\
0.00298612 -0.2\\
-0.00197919 -0.1875\\
-0.00682228 -0.175\\
-0.0115431 -0.1625\\
-0.0161418 -0.15\\
-0.0206182 -0.1375\\
-0.0249723 -0.125\\
-0.0292043 -0.1125\\
-0.033314 -0.1\\
-0.0373015 -0.0875\\
-0.0411667 -0.075\\
-0.0449098 -0.0625\\
-0.0485305 -0.05\\
-0.0520291 -0.0375\\
-0.0554055 -0.025\\
-0.0586596 -0.0125\\
-0.0617914 3.90313e-17\\
-0.0617914 0\\
-0.0669012 0.0125\\
-0.0722179 0.025\\
-0.0777416 0.0375\\
-0.0834723 0.05\\
-0.08941 0.0625\\
-0.0955547 0.075\\
-0.101906 0.0875\\
-0.108465 0.1\\
-0.115231 0.1125\\
-0.122203 0.125\\
-0.129383 0.1375\\
-0.13677 0.15\\
-0.144363 0.1625\\
-0.152164 0.175\\
-0.160171 0.1875\\
-0.168386 0.2\\
-0.176808 0.2125\\
-0.185436 0.225\\
-0.194272 0.2375\\
-0.203314 0.25\\
-0.203314 0.25\\
-0.203879 0.2625\\
-0.204322 0.275\\
-0.204643 0.2875\\
-0.204841 0.3\\
-0.204917 0.3125\\
-0.204871 0.325\\
-0.204703 0.3375\\
-0.204413 0.35\\
-0.204 0.3625\\
-0.203465 0.375\\
-0.202808 0.3875\\
-0.202028 0.4\\
-0.201127 0.4125\\
-0.200103 0.425\\
-0.198956 0.4375\\
-0.197688 0.45\\
-0.196298 0.4625\\
-0.194785 0.475\\
-0.19315 0.4875\\
-0.191392 0.5\\
};
\addlegendentry{FSDT5};

\end{axis}
\end{tikzpicture}%
}}
\subfigure[$\overline{\tau}_{xz}=\overline{\tau}_{xz}(-a/2,0,z)$]{\scalebox{0.5}{
%
%
%
%
\begin{tikzpicture}

\begin{axis}[%
width=\figurewidth,
height=\figureheight,
scale only axis,
xmin=-0.2,
xmax=0.15,
xlabel={$\bar{\tau}_{xz}$},
ymin=-0.5,
ymax=0.5,
ylabel={Normalized thickness, z/h},
legend style={draw=black,fill=white,legend cell align=left}
]
\addplot [
color=blue,
mark size=2.5pt,
only marks,
mark=o,
mark options={solid,draw=black}
]
table[row sep=crcr]{
0 -0.5\\
-0.0106719 -0.4875\\
-0.0204738 -0.475\\
-0.0294495 -0.4625\\
-0.0376431 -0.45\\
-0.0450993 -0.4375\\
-0.0518637 -0.425\\
-0.057982 -0.4125\\
-0.0635008 -0.4\\
-0.0684671 -0.3875\\
-0.0729286 -0.375\\
-0.0769335 -0.3625\\
-0.0805306 -0.35\\
-0.0837691 -0.3375\\
-0.0866991 -0.325\\
-0.089371 -0.3125\\
-0.0918358 -0.3\\
-0.0941453 -0.2875\\
-0.0963515 -0.275\\
-0.0985074 -0.2625\\
-0.100666 -0.25\\
-0.100666 -0.25\\
-0.0840419 -0.2375\\
-0.0693616 -0.225\\
-0.0565593 -0.2125\\
-0.0455704 -0.2\\
-0.0363315 -0.1875\\
-0.0287806 -0.175\\
-0.0228566 -0.1625\\
-0.0185 -0.15\\
-0.0156524 -0.1375\\
-0.0142565 -0.125\\
-0.0142565 -0.1125\\
-0.0155976 -0.1\\
-0.0182263 -0.0875\\
-0.0220904 -0.075\\
-0.027139 -0.0625\\
-0.0333223 -0.05\\
-0.0405918 -0.0375\\
-0.0489001 -0.025\\
-0.0582013 -0.0125\\
-0.0684506 3.90313e-17\\
-0.0684506 0\\
-0.0548279 0.0125\\
-0.0425261 0.025\\
-0.0316114 0.0375\\
-0.0221509 0.05\\
-0.0142123 0.0625\\
-0.00786368 0.075\\
-0.00317383 0.0875\\
-0.000212045 0.1\\
0.000951781 0.1125\\
0.000247196 0.125\\
-0.00239683 0.1375\\
-0.00705188 0.15\\
-0.0137901 0.1625\\
-0.0226843 0.175\\
-0.0338078 0.1875\\
-0.0472344 0.2\\
-0.0630385 0.2125\\
-0.0812953 0.225\\
-0.10208 0.2375\\
-0.125469 0.25\\
-0.125469 0.25\\
-0.123733 0.2625\\
-0.122272 0.275\\
-0.12107 0.2875\\
-0.120113 0.3\\
-0.119386 0.3125\\
-0.118876 0.325\\
-0.118573 0.3375\\
-0.118466 0.35\\
-0.118546 0.3625\\
-0.118806 0.375\\
-0.11924 0.3875\\
-0.119841 0.4\\
-0.120606 0.4125\\
-0.121533 0.425\\
-0.122619 0.4375\\
-0.123865 0.45\\
-0.125272 0.4625\\
-0.126842 0.475\\
-0.128578 0.4875\\
-0.130485 0.5\\
};
\addlegendentry{HSDT13};

\addplot [
color=black,
dashed,
line width=1.2pt
]
table[row sep=crcr]{
0 -0.5\\
-0.0108396 -0.4875\\
-0.0208522 -0.475\\
-0.0300778 -0.4625\\
-0.038557 -0.45\\
-0.0463309 -0.4375\\
-0.0534416 -0.425\\
-0.0599318 -0.4125\\
-0.0658447 -0.4\\
-0.0712243 -0.3875\\
-0.0761155 -0.375\\
-0.0805635 -0.3625\\
-0.0846145 -0.35\\
-0.0883153 -0.3375\\
-0.0917133 -0.325\\
-0.0948567 -0.3125\\
-0.0977943 -0.3\\
-0.100576 -0.2875\\
-0.103251 -0.275\\
-0.105871 -0.2625\\
-0.108488 -0.25\\
-0.108488 -0.25\\
-0.0915569 -0.2375\\
-0.0765973 -0.225\\
-0.0635434 -0.2125\\
-0.0523302 -0.2\\
-0.042894 -0.1875\\
-0.0351722 -0.175\\
-0.0291035 -0.1625\\
-0.0246276 -0.15\\
-0.0216856 -0.1375\\
-0.0202196 -0.125\\
-0.0201731 -0.1125\\
-0.0214905 -0.1\\
-0.0241178 -0.0875\\
-0.0280018 -0.075\\
-0.0330907 -0.0625\\
-0.0393337 -0.05\\
-0.0466815 -0.0375\\
-0.0550857 -0.025\\
-0.0644993 -0.0125\\
-0.0748763 3.90313e-17\\
-0.0748763 0\\
-0.0613756 0.0125\\
-0.0491811 0.025\\
-0.0383602 0.0375\\
-0.0289811 0.05\\
-0.0211126 0.0625\\
-0.0148243 0.075\\
-0.0101864 0.0875\\
-0.00726976 0.1\\
-0.00614596 0.1125\\
-0.00688731 0.125\\
-0.00956677 0.1375\\
-0.014258 0.15\\
-0.0210353 0.1625\\
-0.0299738 0.175\\
-0.0411492 0.1875\\
-0.0546379 0.2\\
-0.070517 0.2125\\
-0.0888642 0.225\\
-0.109758 0.2375\\
-0.133278 0.25\\
-0.133278 0.25\\
-0.131591 0.2625\\
-0.130158 0.275\\
-0.128962 0.2875\\
-0.127983 0.3\\
-0.127208 0.3125\\
-0.126619 0.325\\
-0.126205 0.3375\\
-0.125951 0.35\\
-0.125847 0.3625\\
-0.125883 0.375\\
-0.126049 0.3875\\
-0.126337 0.4\\
-0.126742 0.4125\\
-0.127256 0.425\\
-0.127878 0.4375\\
-0.128602 0.45\\
-0.129427 0.4625\\
-0.130353 0.475\\
-0.13138 0.4875\\
-0.13251 0.5\\
};
\addlegendentry{HSDT11A};

\addplot [
color=blue,
mark size=4.3pt,
only marks,
mark=diamond,
mark options={solid,draw=black}
]
table[row sep=crcr]{
0 -0.5\\
-0.0145631 -0.4875\\
-0.0278812 -0.475\\
-0.0399828 -0.4625\\
-0.0508973 -0.45\\
-0.0606551 -0.4375\\
-0.0692872 -0.425\\
-0.0768255 -0.4125\\
-0.0833029 -0.4\\
-0.088753 -0.3875\\
-0.0932102 -0.375\\
-0.0967097 -0.3625\\
-0.0992878 -0.35\\
-0.100981 -0.3375\\
-0.101828 -0.325\\
-0.101867 -0.3125\\
-0.101137 -0.3\\
-0.099679 -0.2875\\
-0.0975338 -0.275\\
-0.0947434 -0.2625\\
-0.0913507 -0.25\\
-0.0913507 -0.25\\
-0.0808889 -0.2375\\
-0.0717296 -0.225\\
-0.0637953 -0.2125\\
-0.0570107 -0.2\\
-0.0513024 -0.1875\\
-0.0465992 -0.175\\
-0.0428317 -0.1625\\
-0.0399328 -0.15\\
-0.0378374 -0.1375\\
-0.0364825 -0.125\\
-0.035807 -0.1125\\
-0.0357521 -0.1\\
-0.0362607 -0.0875\\
-0.0372782 -0.075\\
-0.0387518 -0.0625\\
-0.0406306 -0.05\\
-0.0428662 -0.0375\\
-0.0454119 -0.025\\
-0.0482232 -0.0125\\
-0.0512576 3.90313e-17\\
-0.0512576 0\\
-0.0465933 0.0125\\
-0.0424809 0.025\\
-0.0389824 0.0375\\
-0.0361608 0.05\\
-0.0340798 0.0625\\
-0.0328041 0.075\\
-0.032399 0.0875\\
-0.032931 0.1\\
-0.0344671 0.1125\\
-0.0370752 0.125\\
-0.0408242 0.1375\\
-0.0457836 0.15\\
-0.0520241 0.1625\\
-0.0596168 0.175\\
-0.0686339 0.1875\\
-0.0791485 0.2\\
-0.0912343 0.2125\\
-0.104966 0.225\\
-0.120419 0.2375\\
-0.13767 0.25\\
-0.13767 0.25\\
-0.139225 0.2625\\
-0.140591 0.275\\
-0.141771 0.2875\\
-0.142772 0.3\\
-0.143604 0.3125\\
-0.144277 0.325\\
-0.144805 0.3375\\
-0.145203 0.35\\
-0.145488 0.3625\\
-0.145679 0.375\\
-0.145798 0.3875\\
-0.145868 0.4\\
-0.145914 0.4125\\
-0.145965 0.425\\
-0.146049 0.4375\\
-0.146199 0.45\\
-0.146448 0.4625\\
-0.146831 0.475\\
-0.147387 0.4875\\
-0.148156 0.5\\
};
\addlegendentry{HSDT11B};

\addplot [
color=black,
solid,
line width=1.2pt
]
table[row sep=crcr]{
0 -0.5\\
-0.0102856 -0.4875\\
-0.0194586 -0.475\\
-0.0275592 -0.4625\\
-0.0346275 -0.45\\
-0.0407035 -0.4375\\
-0.0458273 -0.425\\
-0.050039 -0.4125\\
-0.0533787 -0.4\\
-0.0558865 -0.3875\\
-0.0576024 -0.375\\
-0.0585665 -0.3625\\
-0.058819 -0.35\\
-0.0583999 -0.3375\\
-0.0573493 -0.325\\
-0.0557073 -0.3125\\
-0.053514 -0.3\\
-0.0508094 -0.2875\\
-0.0476336 -0.275\\
-0.0440268 -0.2625\\
-0.040029 -0.25\\
-0.040029 -0.25\\
-0.0302443 -0.2375\\
-0.0212811 -0.225\\
-0.0131154 -0.2125\\
-0.0057232 -0.2\\
0.000919475 -0.1875\\
0.00683662 -0.175\\
0.0120522 -0.1625\\
0.0165903 -0.15\\
0.0204748 -0.1375\\
0.0237298 -0.125\\
0.0263791 -0.1125\\
0.0284469 -0.1\\
0.0299571 -0.0875\\
0.0309336 -0.075\\
0.0314006 -0.0625\\
0.0313819 -0.05\\
0.0309016 -0.0375\\
0.0299837 -0.025\\
0.0286521 -0.0125\\
0.0269308 3.90313e-17\\
0.0269308 0\\
0.0298798 0.0125\\
0.0323379 0.025\\
0.034265 0.0375\\
0.0356211 0.05\\
0.0363661 0.0625\\
0.0364598 0.075\\
0.0358623 0.0875\\
0.0345333 0.1\\
0.0324329 0.1125\\
0.0295209 0.125\\
0.0257573 0.1375\\
0.0211019 0.15\\
0.0155147 0.1625\\
0.00895561 0.175\\
0.00138453 0.1875\\
-0.00723861 0.2\\
-0.0169539 0.2125\\
-0.0278015 0.225\\
-0.0398213 0.2375\\
-0.0530536 0.25\\
-0.0530536 0.25\\
-0.0574163 0.2625\\
-0.0616408 0.275\\
-0.0657032 0.2875\\
-0.0695796 0.3\\
-0.0732458 0.3125\\
-0.076678 0.325\\
-0.0798522 0.3375\\
-0.0827443 0.35\\
-0.0853303 0.3625\\
-0.0875863 0.375\\
-0.0894883 0.3875\\
-0.0910123 0.4\\
-0.0921343 0.4125\\
-0.0928303 0.425\\
-0.0930763 0.4375\\
-0.0928484 0.45\\
-0.0921225 0.4625\\
-0.0908746 0.475\\
-0.0890808 0.4875\\
-0.0867171 0.5\\
};
\addlegendentry{TSDT7};

\addplot [
color=blue,
mark size=2.5pt,
only marks,
mark=x,
mark options={solid,draw=black}
]
table[row sep=crcr]{
0 -0.5\\
0.00398665 -0.4875\\
0.00774961 -0.475\\
0.0112889 -0.4625\\
0.0146044 -0.45\\
0.0176963 -0.4375\\
0.0205645 -0.425\\
0.023209 -0.4125\\
0.0256298 -0.4\\
0.0278269 -0.3875\\
0.0298003 -0.375\\
0.03155 -0.3625\\
0.033076 -0.35\\
0.0343783 -0.3375\\
0.035457 -0.325\\
0.0363119 -0.3125\\
0.0369432 -0.3\\
0.0373507 -0.2875\\
0.0375346 -0.275\\
0.0374947 -0.2625\\
0.0372312 -0.25\\
0.0372312 -0.25\\
0.0428338 -0.2375\\
0.048294 -0.225\\
0.0536119 -0.2125\\
0.0587875 -0.2\\
0.0638208 -0.1875\\
0.0687117 -0.175\\
0.0734603 -0.1625\\
0.0780666 -0.15\\
0.0825305 -0.1375\\
0.0868521 -0.125\\
0.0910314 -0.1125\\
0.0950683 -0.1\\
0.098963 -0.0875\\
0.102715 -0.075\\
0.106325 -0.0625\\
0.109793 -0.05\\
0.113118 -0.0375\\
0.116301 -0.025\\
0.119342 -0.0125\\
0.12224 3.90313e-17\\
0.12224 0\\
0.117279 0.0125\\
0.112094 0.025\\
0.106686 0.0375\\
0.101053 0.05\\
0.0951975 0.0625\\
0.089118 0.075\\
0.0828147 0.0875\\
0.0762877 0.1\\
0.0695371 0.1125\\
0.0625627 0.125\\
0.0553647 0.1375\\
0.0479429 0.15\\
0.0402975 0.1625\\
0.0324284 0.175\\
0.0243356 0.1875\\
0.0160191 0.2\\
0.00747885 0.2125\\
-0.00128505 0.225\\
-0.0102726 0.2375\\
-0.0194839 0.25\\
-0.0194839 0.25\\
-0.0195746 0.2625\\
-0.0198076 0.275\\
-0.0201829 0.2875\\
-0.0207005 0.3\\
-0.0213605 0.3125\\
-0.0221628 0.325\\
-0.0231074 0.3375\\
-0.0241943 0.35\\
-0.0254236 0.3625\\
-0.0267953 0.375\\
-0.0283092 0.3875\\
-0.0299655 0.4\\
-0.0317641 0.4125\\
-0.033705 0.425\\
-0.0357883 0.4375\\
-0.0380139 0.45\\
-0.0403819 0.4625\\
-0.0428921 0.475\\
-0.0455447 0.4875\\
-0.0483396 0.5\\
};
\addlegendentry{FSDT5};

\end{axis}
\end{tikzpicture}%
}}\hspace{60pt}
\subfigure[$\overline{\sigma}_{zz} = \overline{\sigma}_{zz}(0,0,z)$]{\scalebox{0.5}{
%
%
%
%
\begin{tikzpicture}

\begin{axis}[%
width=\figurewidth,
height=\figureheight,
scale only axis,
xmin=-0.4,
xmax=1.2,
xlabel={$\bar{\sigma}_{zz}$},
ymin=-0.5,
ymax=0.5,
ylabel={Normalized thickness, z/h},
legend style={draw=black,fill=white,legend cell align=left},
legend pos = south east
]
\addplot [
color=blue,
mark size=2.5pt,
only marks,
mark=o,
mark options={solid,draw=black}
]
table[row sep=crcr]{
-0.230297 -0.5\\
-0.192549 -0.4875\\
-0.15641 -0.475\\
-0.121836 -0.4625\\
-0.0887842 -0.45\\
-0.05721 -0.4375\\
-0.0270699 -0.425\\
0.00167967 -0.4125\\
0.0290825 -0.4\\
0.0551824 -0.3875\\
0.080023 -0.375\\
0.103648 -0.3625\\
0.126101 -0.35\\
0.147426 -0.3375\\
0.167667 -0.325\\
0.186866 -0.3125\\
0.205069 -0.3\\
0.222319 -0.2875\\
0.238659 -0.275\\
0.254133 -0.2625\\
0.268785 -0.25\\
0.23567 -0.25\\
0.25274 -0.2375\\
0.268911 -0.225\\
0.284228 -0.2125\\
0.298734 -0.2\\
0.312474 -0.1875\\
0.32549 -0.175\\
0.337827 -0.1625\\
0.349528 -0.15\\
0.360638 -0.1375\\
0.371199 -0.125\\
0.381257 -0.1125\\
0.390854 -0.1\\
0.400035 -0.0875\\
0.408843 -0.075\\
0.417322 -0.0625\\
0.425516 -0.05\\
0.433469 -0.0375\\
0.441224 -0.025\\
0.448826 -0.0125\\
0.456317 3.90313e-17\\
0.447305 0\\
0.45479 0.0125\\
0.462415 0.025\\
0.470223 0.0375\\
0.478259 0.05\\
0.486565 0.0625\\
0.495186 0.075\\
0.504165 0.0875\\
0.513545 0.1\\
0.523372 0.1125\\
0.533687 0.125\\
0.544536 0.1375\\
0.555962 0.15\\
0.568008 0.1625\\
0.580718 0.175\\
0.594136 0.1875\\
0.608306 0.2\\
0.623271 0.2125\\
0.639075 0.225\\
0.655762 0.2375\\
0.673376 0.25\\
0.639738 0.25\\
0.655041 0.2625\\
0.671198 0.275\\
0.688253 0.2875\\
0.706249 0.3\\
0.72523 0.3125\\
0.74524 0.325\\
0.766323 0.3375\\
0.788523 0.35\\
0.811883 0.3625\\
0.836447 0.375\\
0.86226 0.3875\\
0.889364 0.4\\
0.917804 0.4125\\
0.947624 0.425\\
0.978866 0.4375\\
1.01158 0.45\\
1.0458 0.4625\\
1.08157 0.475\\
1.11895 0.4875\\
1.15796 0.5\\
};
\addlegendentry{HSDT13};

\addplot [
color=black,
dashed,
line width=1.2pt
]
table[row sep=crcr]{
0 -0.5\\
0 -0.4875\\
0 -0.475\\
0 -0.4625\\
0 -0.45\\
0 -0.4375\\
0 -0.425\\
0 -0.4125\\
0 -0.4\\
0 -0.3875\\
0 -0.375\\
0 -0.3625\\
0 -0.35\\
0 -0.3375\\
0 -0.325\\
0 -0.3125\\
0 -0.3\\
0 -0.2875\\
0 -0.275\\
0 -0.2625\\
0 -0.25\\
0 -0.25\\
0 -0.2375\\
0 -0.225\\
0 -0.2125\\
0 -0.2\\
0 -0.1875\\
0 -0.175\\
0 -0.1625\\
0 -0.15\\
0 -0.1375\\
0 -0.125\\
0 -0.1125\\
0 -0.1\\
0 -0.0875\\
0 -0.075\\
0 -0.0625\\
0 -0.05\\
0 -0.0375\\
0 -0.025\\
0 -0.0125\\
0 3.90313e-17\\
0 0\\
0 0.0125\\
0 0.025\\
0 0.0375\\
0 0.05\\
0 0.0625\\
0 0.075\\
0 0.0875\\
0 0.1\\
0 0.1125\\
0 0.125\\
0 0.1375\\
0 0.15\\
0 0.1625\\
0 0.175\\
0 0.1875\\
0 0.2\\
0 0.2125\\
0 0.225\\
0 0.2375\\
0 0.25\\
0 0.25\\
0 0.2625\\
0 0.275\\
0 0.2875\\
0 0.3\\
0 0.3125\\
0 0.325\\
0 0.3375\\
0 0.35\\
0 0.3625\\
0 0.375\\
0 0.3875\\
0 0.4\\
0 0.4125\\
0 0.425\\
0 0.4375\\
0 0.45\\
0 0.4625\\
0 0.475\\
0 0.4875\\
0 0.5\\
};
\addlegendentry{HSDT11A};

\addplot [
color=blue,
mark size=4.3pt,
only marks,
mark=diamond,
mark options={solid,draw=black}
]
table[row sep=crcr]{
-0.225119 -0.5\\
-0.188217 -0.4875\\
-0.15288 -0.475\\
-0.119065 -0.4625\\
-0.0867294 -0.45\\
-0.0558312 -0.4375\\
-0.0263277 -0.425\\
0.00182352 -0.4125\\
0.028665 -0.4\\
0.0542393 -0.3875\\
0.0785888 -0.375\\
0.101756 -0.3625\\
0.123784 -0.35\\
0.144714 -0.3375\\
0.164589 -0.325\\
0.183453 -0.3125\\
0.201346 -0.3\\
0.218312 -0.2875\\
0.234394 -0.275\\
0.249633 -0.2625\\
0.264073 -0.25\\
0.235947 -0.25\\
0.252637 -0.2375\\
0.268457 -0.225\\
0.283449 -0.2125\\
0.297655 -0.2\\
0.311118 -0.1875\\
0.32388 -0.175\\
0.335985 -0.1625\\
0.347474 -0.15\\
0.35839 -0.1375\\
0.368776 -0.125\\
0.378675 -0.1125\\
0.388128 -0.1\\
0.39718 -0.0875\\
0.405871 -0.075\\
0.414245 -0.0625\\
0.422344 -0.05\\
0.430211 -0.0375\\
0.437889 -0.025\\
0.445419 -0.0125\\
0.452846 3.90313e-17\\
0.450351 0\\
0.457812 0.0125\\
0.465408 0.025\\
0.473181 0.0375\\
0.481175 0.05\\
0.489431 0.0625\\
0.497992 0.075\\
0.506901 0.0875\\
0.5162 0.1\\
0.525932 0.1125\\
0.536139 0.125\\
0.546863 0.1375\\
0.558147 0.15\\
0.570035 0.1625\\
0.582567 0.175\\
0.595786 0.1875\\
0.609736 0.2\\
0.624459 0.2125\\
0.639996 0.225\\
0.656391 0.2375\\
0.673686 0.25\\
0.645028 0.25\\
0.660099 0.2625\\
0.676002 0.275\\
0.69278 0.2875\\
0.710476 0.3\\
0.729132 0.3125\\
0.74879 0.325\\
0.769494 0.3375\\
0.791286 0.35\\
0.814209 0.3625\\
0.838304 0.375\\
0.863615 0.3875\\
0.890184 0.4\\
0.918054 0.4125\\
0.947268 0.425\\
0.977867 0.4375\\
1.0099 0.45\\
1.04339 0.4625\\
1.07841 0.475\\
1.11498 0.4875\\
1.15314 0.5\\
};
\addlegendentry{HSDT11B};

\addplot [
color=black,
solid,
line width=1.2pt
]
table[row sep=crcr]{
0 -0.5\\
0 -0.4875\\
0 -0.475\\
0 -0.4625\\
0 -0.45\\
0 -0.4375\\
0 -0.425\\
0 -0.4125\\
0 -0.4\\
0 -0.3875\\
0 -0.375\\
0 -0.3625\\
0 -0.35\\
0 -0.3375\\
0 -0.325\\
0 -0.3125\\
0 -0.3\\
0 -0.2875\\
0 -0.275\\
0 -0.2625\\
0 -0.25\\
0 -0.25\\
0 -0.2375\\
0 -0.225\\
0 -0.2125\\
0 -0.2\\
0 -0.1875\\
0 -0.175\\
0 -0.1625\\
0 -0.15\\
0 -0.1375\\
0 -0.125\\
0 -0.1125\\
0 -0.1\\
0 -0.0875\\
0 -0.075\\
0 -0.0625\\
0 -0.05\\
0 -0.0375\\
0 -0.025\\
0 -0.0125\\
0 3.90313e-17\\
0 0\\
0 0.0125\\
0 0.025\\
0 0.0375\\
0 0.05\\
0 0.0625\\
0 0.075\\
0 0.0875\\
0 0.1\\
0 0.1125\\
0 0.125\\
0 0.1375\\
0 0.15\\
0 0.1625\\
0 0.175\\
0 0.1875\\
0 0.2\\
0 0.2125\\
0 0.225\\
0 0.2375\\
0 0.25\\
0 0.25\\
0 0.2625\\
0 0.275\\
0 0.2875\\
0 0.3\\
0 0.3125\\
0 0.325\\
0 0.3375\\
0 0.35\\
0 0.3625\\
0 0.375\\
0 0.3875\\
0 0.4\\
0 0.4125\\
0 0.425\\
0 0.4375\\
0 0.45\\
0 0.4625\\
0 0.475\\
0 0.4875\\
0 0.5\\
};
\addlegendentry{TSDT7};

\addplot [
color=blue,
mark size=2.5pt,
only marks,
mark=x,
mark options={solid,draw=black}
]
table[row sep=crcr]{
0 -0.5\\
0 -0.4875\\
0 -0.475\\
0 -0.4625\\
0 -0.45\\
0 -0.4375\\
0 -0.425\\
0 -0.4125\\
0 -0.4\\
0 -0.3875\\
0 -0.375\\
0 -0.3625\\
0 -0.35\\
0 -0.3375\\
0 -0.325\\
0 -0.3125\\
0 -0.3\\
0 -0.2875\\
0 -0.275\\
0 -0.2625\\
0 -0.25\\
0 -0.25\\
0 -0.2375\\
0 -0.225\\
0 -0.2125\\
0 -0.2\\
0 -0.1875\\
0 -0.175\\
0 -0.1625\\
0 -0.15\\
0 -0.1375\\
0 -0.125\\
0 -0.1125\\
0 -0.1\\
0 -0.0875\\
0 -0.075\\
0 -0.0625\\
0 -0.05\\
0 -0.0375\\
0 -0.025\\
0 -0.0125\\
0 3.90313e-17\\
0 0\\
0 0.0125\\
0 0.025\\
0 0.0375\\
0 0.05\\
0 0.0625\\
0 0.075\\
0 0.0875\\
0 0.1\\
0 0.1125\\
0 0.125\\
0 0.1375\\
0 0.15\\
0 0.1625\\
0 0.175\\
0 0.1875\\
0 0.2\\
0 0.2125\\
0 0.225\\
0 0.2375\\
0 0.25\\
0 0.25\\
0 0.2625\\
0 0.275\\
0 0.2875\\
0 0.3\\
0 0.3125\\
0 0.325\\
0 0.3375\\
0 0.35\\
0 0.3625\\
0 0.375\\
0 0.3875\\
0 0.4\\
0 0.4125\\
0 0.425\\
0 0.4375\\
0 0.45\\
0 0.4625\\
0 0.475\\
0 0.4875\\
0 0.5\\
};
\addlegendentry{FSDT5};

\end{axis}
\end{tikzpicture}%
}}
\caption{The variation of non-dimensional deflections and stresses through the thickness of a four-layered anti-symmetric curved fibre composite laminate under sinusoidally distributed mechanical load with $a/h=$ 5, $T_o=$ -45$^\circ$, $T_1=$ 45$^\circ$.}
\label{fig:4LayerMechLoad}
\end{figure}

\begin{figure}
\centering
\setlength\figureheight{10cm} 
\setlength\figurewidth{12cm}
\subfigure[$\overline{u} = \overline{u}(-a/2,0,z)$]{\scalebox{0.5}{
%
%
%
%
\begin{tikzpicture}

\begin{axis}[%
width=\figurewidth,
height=\figureheight,
scale only axis,
xmin=-20,
xmax=20,
xlabel={$\bar{u}$},
ymin=-0.5,
ymax=0.5,
ylabel={Normalized thickness, z/h},
legend style={draw=black,fill=white,legend cell align=left}
]
\addplot [
color=blue,
mark size=2.5pt,
only marks,
mark=o,
mark options={solid,draw=black}
]
table[row sep=crcr]{
18.9563 -0.5\\
17.8202 -0.4875\\
16.7344 -0.475\\
15.6976 -0.4625\\
14.7085 -0.45\\
13.7659 -0.4375\\
12.8684 -0.425\\
12.0147 -0.4125\\
11.2035 -0.4\\
10.4337 -0.3875\\
9.7038 -0.375\\
9.0126 -0.3625\\
8.3588 -0.35\\
7.74111 -0.3375\\
7.15823 -0.325\\
6.60889 -0.3125\\
6.09178 -0.3\\
5.60562 -0.2875\\
5.14912 -0.275\\
4.72099 -0.2625\\
4.31994 -0.25\\
4.31994 -0.25\\
3.94469 -0.2375\\
3.59393 -0.225\\
3.26639 -0.2125\\
2.96077 -0.2\\
2.67578 -0.1875\\
2.41013 -0.175\\
2.16254 -0.1625\\
1.93171 -0.15\\
1.71636 -0.1375\\
1.5152 -0.125\\
1.32692 -0.1125\\
1.15026 -0.1\\
0.983908 -0.0875\\
0.826586 -0.075\\
0.677 -0.0625\\
0.533863 -0.05\\
0.395884 -0.0375\\
0.261773 -0.025\\
0.130242 -0.0125\\
-8.311e-14 3.90313e-17\\
-8.27039e-14 0\\
-0.130242 0.0125\\
-0.261773 0.025\\
-0.395884 0.0375\\
-0.533863 0.05\\
-0.677 0.0625\\
-0.826586 0.075\\
-0.983908 0.0875\\
-1.15026 0.1\\
-1.32692 0.1125\\
-1.5152 0.125\\
-1.71636 0.1375\\
-1.93171 0.15\\
-2.16254 0.1625\\
-2.41013 0.175\\
-2.67578 0.1875\\
-2.96077 0.2\\
-3.26639 0.2125\\
-3.59393 0.225\\
-3.94469 0.2375\\
-4.31994 0.25\\
-4.31994 0.25\\
-4.72099 0.2625\\
-5.14912 0.275\\
-5.60562 0.2875\\
-6.09178 0.3\\
-6.60889 0.3125\\
-7.15823 0.325\\
-7.74111 0.3375\\
-8.3588 0.35\\
-9.0126 0.3625\\
-9.7038 0.375\\
-10.4337 0.3875\\
-11.2035 0.4\\
-12.0147 0.4125\\
-12.8684 0.425\\
-13.7659 0.4375\\
-14.7085 0.45\\
-15.6976 0.4625\\
-16.7344 0.475\\
-17.8202 0.4875\\
-18.9563 0.5\\
};
\addlegendentry{HSDT13};

\addplot [
color=black,
dashed,
line width=1.2pt
]
table[row sep=crcr]{
16.0594 -0.5\\
15.2931 -0.4875\\
14.5545 -0.475\\
13.8429 -0.4625\\
13.1576 -0.45\\
12.4979 -0.4375\\
11.8631 -0.425\\
11.2524 -0.4125\\
10.6651 -0.4\\
10.1006 -0.3875\\
9.55805 -0.375\\
9.03686 -0.3625\\
8.53627 -0.35\\
8.05557 -0.3375\\
7.59405 -0.325\\
7.15101 -0.3125\\
6.72572 -0.3\\
6.31749 -0.2875\\
5.9256 -0.275\\
5.54933 -0.2625\\
5.18798 -0.25\\
5.18798 -0.25\\
4.84085 -0.2375\\
4.50721 -0.225\\
4.18636 -0.2125\\
3.87758 -0.2\\
3.58017 -0.1875\\
3.29342 -0.175\\
3.01662 -0.1625\\
2.74905 -0.15\\
2.49001 -0.1375\\
2.23878 -0.125\\
1.99465 -0.1125\\
1.75692 -0.1\\
1.52487 -0.0875\\
1.2978 -0.075\\
1.07499 -0.0625\\
0.855727 -0.05\\
0.639308 -0.0375\\
0.425022 -0.025\\
0.212156 -0.0125\\
1.80678e-14 3.90313e-17\\
1.87299e-14 0\\
-0.212156 0.0125\\
-0.425022 0.025\\
-0.639308 0.0375\\
-0.855727 0.05\\
-1.07499 0.0625\\
-1.2978 0.075\\
-1.52487 0.0875\\
-1.75692 0.1\\
-1.99465 0.1125\\
-2.23878 0.125\\
-2.49001 0.1375\\
-2.74905 0.15\\
-3.01662 0.1625\\
-3.29342 0.175\\
-3.58017 0.1875\\
-3.87758 0.2\\
-4.18636 0.2125\\
-4.50721 0.225\\
-4.84085 0.2375\\
-5.18798 0.25\\
-5.18798 0.25\\
-5.54933 0.2625\\
-5.9256 0.275\\
-6.31749 0.2875\\
-6.72572 0.3\\
-7.15101 0.3125\\
-7.59405 0.325\\
-8.05557 0.3375\\
-8.53627 0.35\\
-9.03686 0.3625\\
-9.55805 0.375\\
-10.1006 0.3875\\
-10.6651 0.4\\
-11.2524 0.4125\\
-11.8631 0.425\\
-12.4979 0.4375\\
-13.1576 0.45\\
-13.8429 0.4625\\
-14.5545 0.475\\
-15.2931 0.4875\\
-16.0594 0.5\\
};
\addlegendentry{HSDT11A};

\addplot [
color=blue,
mark size=4.3pt,
only marks,
mark=diamond,
mark options={solid,draw=black}
]
table[row sep=crcr]{
19.0149 -0.5\\
17.8962 -0.4875\\
16.8264 -0.475\\
15.8042 -0.4625\\
14.8283 -0.45\\
13.8976 -0.4375\\
13.0107 -0.425\\
12.1664 -0.4125\\
11.3634 -0.4\\
10.6005 -0.3875\\
9.87647 -0.375\\
9.19001 -0.3625\\
8.53988 -0.35\\
7.92482 -0.3375\\
7.34359 -0.325\\
6.79493 -0.3125\\
6.27759 -0.3\\
5.79032 -0.2875\\
5.33186 -0.275\\
4.90096 -0.2625\\
4.49637 -0.25\\
4.49637 -0.25\\
4.11683 -0.2375\\
3.7611 -0.225\\
3.42792 -0.2125\\
3.11603 -0.2\\
2.82419 -0.1875\\
2.55114 -0.175\\
2.29563 -0.1625\\
2.0564 -0.15\\
1.83221 -0.1375\\
1.6218 -0.125\\
1.42392 -0.1125\\
1.23731 -0.1\\
1.06072 -0.0875\\
0.892903 -0.075\\
0.732602 -0.0625\\
0.578565 -0.05\\
0.429539 -0.0375\\
0.284271 -0.025\\
0.141509 -0.0125\\
7.72582e-14 3.90313e-17\\
7.76994e-14 0\\
-0.141509 0.0125\\
-0.284271 0.025\\
-0.429539 0.0375\\
-0.578565 0.05\\
-0.732602 0.0625\\
-0.892903 0.075\\
-1.06072 0.0875\\
-1.23731 0.1\\
-1.42392 0.1125\\
-1.6218 0.125\\
-1.83221 0.1375\\
-2.0564 0.15\\
-2.29563 0.1625\\
-2.55114 0.175\\
-2.82419 0.1875\\
-3.11603 0.2\\
-3.42792 0.2125\\
-3.7611 0.225\\
-4.11683 0.2375\\
-4.49637 0.25\\
-4.49637 0.25\\
-4.90096 0.2625\\
-5.33186 0.275\\
-5.79032 0.2875\\
-6.27759 0.3\\
-6.79493 0.3125\\
-7.34359 0.325\\
-7.92482 0.3375\\
-8.53988 0.35\\
-9.19001 0.3625\\
-9.87647 0.375\\
-10.6005 0.3875\\
-11.3634 0.4\\
-12.1664 0.4125\\
-13.0107 0.425\\
-13.8976 0.4375\\
-14.8283 0.45\\
-15.8042 0.4625\\
-16.8264 0.475\\
-17.8962 0.4875\\
-19.0149 0.5\\
};
\addlegendentry{HSDT11B};

\addplot [
color=black,
solid,
line width=1.2pt
]
table[row sep=crcr]{
13.9784 -0.5\\
13.6289 -0.4875\\
13.2794 -0.475\\
12.93 -0.4625\\
12.5805 -0.45\\
12.2311 -0.4375\\
11.8816 -0.425\\
11.5321 -0.4125\\
11.1827 -0.4\\
10.8332 -0.3875\\
10.4838 -0.375\\
10.1343 -0.3625\\
9.78485 -0.35\\
9.43539 -0.3375\\
9.08593 -0.325\\
8.73647 -0.3125\\
8.38701 -0.3\\
8.03755 -0.2875\\
7.68809 -0.275\\
7.33863 -0.2625\\
6.98918 -0.25\\
6.98918 -0.25\\
6.63972 -0.2375\\
6.29026 -0.225\\
5.9408 -0.2125\\
5.59134 -0.2\\
5.24188 -0.1875\\
4.89242 -0.175\\
4.54296 -0.1625\\
4.19351 -0.15\\
3.84405 -0.1375\\
3.49459 -0.125\\
3.14513 -0.1125\\
2.79567 -0.1\\
2.44621 -0.0875\\
2.09675 -0.075\\
1.74729 -0.0625\\
1.39784 -0.05\\
1.04838 -0.0375\\
0.698918 -0.025\\
0.349459 -0.0125\\
-7.49629e-14 3.90313e-17\\
-7.38717e-14 0\\
-0.349459 0.0125\\
-0.698918 0.025\\
-1.04838 0.0375\\
-1.39784 0.05\\
-1.74729 0.0625\\
-2.09675 0.075\\
-2.44621 0.0875\\
-2.79567 0.1\\
-3.14513 0.1125\\
-3.49459 0.125\\
-3.84405 0.1375\\
-4.19351 0.15\\
-4.54296 0.1625\\
-4.89242 0.175\\
-5.24188 0.1875\\
-5.59134 0.2\\
-5.9408 0.2125\\
-6.29026 0.225\\
-6.63972 0.2375\\
-6.98918 0.25\\
-6.98918 0.25\\
-7.33863 0.2625\\
-7.68809 0.275\\
-8.03755 0.2875\\
-8.38701 0.3\\
-8.73647 0.3125\\
-9.08593 0.325\\
-9.43539 0.3375\\
-9.78485 0.35\\
-10.1343 0.3625\\
-10.4838 0.375\\
-10.8332 0.3875\\
-11.1827 0.4\\
-11.5321 0.4125\\
-11.8816 0.425\\
-12.2311 0.4375\\
-12.5805 0.45\\
-12.93 0.4625\\
-13.2794 0.475\\
-13.6289 0.4875\\
-13.9784 0.5\\
};
\addlegendentry{TSDT7};

\addplot [
color=blue,
mark size=2.5pt,
only marks,
mark=x,
mark options={solid,draw=black}
]
table[row sep=crcr]{
7.07914 -0.5\\
6.90216 -0.4875\\
6.72518 -0.475\\
6.5482 -0.4625\\
6.37123 -0.45\\
6.19425 -0.4375\\
6.01727 -0.425\\
5.84029 -0.4125\\
5.66331 -0.4\\
5.48633 -0.3875\\
5.30936 -0.375\\
5.13238 -0.3625\\
4.9554 -0.35\\
4.77842 -0.3375\\
4.60144 -0.325\\
4.42446 -0.3125\\
4.24748 -0.3\\
4.07051 -0.2875\\
3.89353 -0.275\\
3.71655 -0.2625\\
3.53957 -0.25\\
3.53957 -0.25\\
3.36259 -0.2375\\
3.18561 -0.225\\
3.00863 -0.2125\\
2.83166 -0.2\\
2.65468 -0.1875\\
2.4777 -0.175\\
2.30072 -0.1625\\
2.12374 -0.15\\
1.94676 -0.1375\\
1.76979 -0.125\\
1.59281 -0.1125\\
1.41583 -0.1\\
1.23885 -0.0875\\
1.06187 -0.075\\
0.884893 -0.0625\\
0.707914 -0.05\\
0.530936 -0.0375\\
0.353957 -0.025\\
0.176979 -0.0125\\
4.2148e-15 3.90313e-17\\
4.76742e-15 0\\
-0.176979 0.0125\\
-0.353957 0.025\\
-0.530936 0.0375\\
-0.707914 0.05\\
-0.884893 0.0625\\
-1.06187 0.075\\
-1.23885 0.0875\\
-1.41583 0.1\\
-1.59281 0.1125\\
-1.76979 0.125\\
-1.94676 0.1375\\
-2.12374 0.15\\
-2.30072 0.1625\\
-2.4777 0.175\\
-2.65468 0.1875\\
-2.83166 0.2\\
-3.00863 0.2125\\
-3.18561 0.225\\
-3.36259 0.2375\\
-3.53957 0.25\\
-3.53957 0.25\\
-3.71655 0.2625\\
-3.89353 0.275\\
-4.07051 0.2875\\
-4.24748 0.3\\
-4.42446 0.3125\\
-4.60144 0.325\\
-4.77842 0.3375\\
-4.9554 0.35\\
-5.13238 0.3625\\
-5.30936 0.375\\
-5.48633 0.3875\\
-5.66331 0.4\\
-5.84029 0.4125\\
-6.01727 0.425\\
-6.19425 0.4375\\
-6.37123 0.45\\
-6.5482 0.4625\\
-6.72518 0.475\\
-6.90216 0.4875\\
-7.07914 0.5\\
};
\addlegendentry{FSDT5};

\end{axis}
\end{tikzpicture}%
}}\hspace{60pt}
\subfigure[$\overline{w} = \overline{w}(0,0,z)$]{\scalebox{0.5}{
%
%
%
%
\begin{tikzpicture}

\begin{axis}[%
width=\figurewidth,
height=\figureheight,
scale only axis,
xmin=10,
xmax=30,
xlabel={$\bar{w}$},
ymin=-0.5,
ymax=0.5,
ylabel={Normalized thickness, z/h},
legend style={at={(1.0,0.5)},draw=black,fill=white,legend cell align=left}
]
\addplot [
color=blue,
mark size=2.5pt,
only marks,
mark=o,
mark options={solid,draw=black}
]
table[row sep=crcr]{
28.9814 -0.5\\
28.3202 -0.4875\\
27.6758 -0.475\\
27.0481 -0.4625\\
26.4371 -0.45\\
25.8428 -0.4375\\
25.2653 -0.425\\
24.7046 -0.4125\\
24.1606 -0.4\\
23.6333 -0.3875\\
23.1227 -0.375\\
22.6289 -0.3625\\
22.1519 -0.35\\
21.6915 -0.3375\\
21.248 -0.325\\
20.8211 -0.3125\\
20.411 -0.3\\
20.0176 -0.2875\\
19.641 -0.275\\
19.2811 -0.2625\\
18.938 -0.25\\
18.938 -0.25\\
18.6115 -0.2375\\
18.3019 -0.225\\
18.0089 -0.2125\\
17.7327 -0.2\\
17.4733 -0.1875\\
17.2306 -0.175\\
17.0046 -0.1625\\
16.7954 -0.15\\
16.6029 -0.1375\\
16.4271 -0.125\\
16.2681 -0.1125\\
16.1258 -0.1\\
16.0002 -0.0875\\
15.8914 -0.075\\
15.7994 -0.0625\\
15.724 -0.05\\
15.6655 -0.0375\\
15.6236 -0.025\\
15.5985 -0.0125\\
15.5901 3.90313e-17\\
15.5901 0\\
15.5985 0.0125\\
15.6236 0.025\\
15.6655 0.0375\\
15.724 0.05\\
15.7994 0.0625\\
15.8914 0.075\\
16.0002 0.0875\\
16.1258 0.1\\
16.2681 0.1125\\
16.4271 0.125\\
16.6029 0.1375\\
16.7954 0.15\\
17.0046 0.1625\\
17.2306 0.175\\
17.4733 0.1875\\
17.7327 0.2\\
18.0089 0.2125\\
18.3019 0.225\\
18.6115 0.2375\\
18.938 0.25\\
18.938 0.25\\
19.2811 0.2625\\
19.641 0.275\\
20.0176 0.2875\\
20.411 0.3\\
20.8211 0.3125\\
21.248 0.325\\
21.6915 0.3375\\
22.1519 0.35\\
22.6289 0.3625\\
23.1227 0.375\\
23.6333 0.3875\\
24.1606 0.4\\
24.7046 0.4125\\
25.2653 0.425\\
25.8428 0.4375\\
26.4371 0.45\\
27.0481 0.4625\\
27.6758 0.475\\
28.3202 0.4875\\
28.9814 0.5\\
};
\addlegendentry{HSDT13};

\addplot [
color=black,
dashed,
line width=1.2pt
]
table[row sep=crcr]{
18.1409 -0.5\\
18.1409 -0.4875\\
18.1409 -0.475\\
18.1409 -0.4625\\
18.1409 -0.45\\
18.1409 -0.4375\\
18.1409 -0.425\\
18.1409 -0.4125\\
18.1409 -0.4\\
18.1409 -0.3875\\
18.1409 -0.375\\
18.1409 -0.3625\\
18.1409 -0.35\\
18.1409 -0.3375\\
18.1409 -0.325\\
18.1409 -0.3125\\
18.1409 -0.3\\
18.1409 -0.2875\\
18.1409 -0.275\\
18.1409 -0.2625\\
18.1409 -0.25\\
18.1409 -0.25\\
18.1409 -0.2375\\
18.1409 -0.225\\
18.1409 -0.2125\\
18.1409 -0.2\\
18.1409 -0.1875\\
18.1409 -0.175\\
18.1409 -0.1625\\
18.1409 -0.15\\
18.1409 -0.1375\\
18.1409 -0.125\\
18.1409 -0.1125\\
18.1409 -0.1\\
18.1409 -0.0875\\
18.1409 -0.075\\
18.1409 -0.0625\\
18.1409 -0.05\\
18.1409 -0.0375\\
18.1409 -0.025\\
18.1409 -0.0125\\
18.1409 3.90313e-17\\
18.1409 0\\
18.1409 0.0125\\
18.1409 0.025\\
18.1409 0.0375\\
18.1409 0.05\\
18.1409 0.0625\\
18.1409 0.075\\
18.1409 0.0875\\
18.1409 0.1\\
18.1409 0.1125\\
18.1409 0.125\\
18.1409 0.1375\\
18.1409 0.15\\
18.1409 0.1625\\
18.1409 0.175\\
18.1409 0.1875\\
18.1409 0.2\\
18.1409 0.2125\\
18.1409 0.225\\
18.1409 0.2375\\
18.1409 0.25\\
18.1409 0.25\\
18.1409 0.2625\\
18.1409 0.275\\
18.1409 0.2875\\
18.1409 0.3\\
18.1409 0.3125\\
18.1409 0.325\\
18.1409 0.3375\\
18.1409 0.35\\
18.1409 0.3625\\
18.1409 0.375\\
18.1409 0.3875\\
18.1409 0.4\\
18.1409 0.4125\\
18.1409 0.425\\
18.1409 0.4375\\
18.1409 0.45\\
18.1409 0.4625\\
18.1409 0.475\\
18.1409 0.4875\\
18.1409 0.5\\
};
\addlegendentry{HSDT11A};

\addplot [
color=blue,
mark size=4.3pt,
only marks,
mark=diamond,
mark options={solid,draw=black}
]
table[row sep=crcr]{
28.8606 -0.5\\
28.1994 -0.4875\\
27.5549 -0.475\\
26.9272 -0.4625\\
26.3163 -0.45\\
25.722 -0.4375\\
25.1445 -0.425\\
24.5837 -0.4125\\
24.0397 -0.4\\
23.5124 -0.3875\\
23.0019 -0.375\\
22.5081 -0.3625\\
22.031 -0.35\\
21.5707 -0.3375\\
21.1271 -0.325\\
20.7003 -0.3125\\
20.2902 -0.3\\
19.8968 -0.2875\\
19.5202 -0.275\\
19.1603 -0.2625\\
18.8171 -0.25\\
18.8171 -0.25\\
18.4907 -0.2375\\
18.181 -0.225\\
17.8881 -0.2125\\
17.6119 -0.2\\
17.3524 -0.1875\\
17.1097 -0.175\\
16.8837 -0.1625\\
16.6745 -0.15\\
16.482 -0.1375\\
16.3062 -0.125\\
16.1472 -0.1125\\
16.0049 -0.1\\
15.8794 -0.0875\\
15.7706 -0.075\\
15.6785 -0.0625\\
15.6032 -0.05\\
15.5446 -0.0375\\
15.5028 -0.025\\
15.4777 -0.0125\\
15.4693 3.90313e-17\\
15.4693 0\\
15.4777 0.0125\\
15.5028 0.025\\
15.5446 0.0375\\
15.6032 0.05\\
15.6785 0.0625\\
15.7706 0.075\\
15.8794 0.0875\\
16.0049 0.1\\
16.1472 0.1125\\
16.3062 0.125\\
16.482 0.1375\\
16.6745 0.15\\
16.8837 0.1625\\
17.1097 0.175\\
17.3524 0.1875\\
17.6119 0.2\\
17.8881 0.2125\\
18.181 0.225\\
18.4907 0.2375\\
18.8171 0.25\\
18.8171 0.25\\
19.1603 0.2625\\
19.5202 0.275\\
19.8968 0.2875\\
20.2902 0.3\\
20.7003 0.3125\\
21.1271 0.325\\
21.5707 0.3375\\
22.031 0.35\\
22.5081 0.3625\\
23.0019 0.375\\
23.5124 0.3875\\
24.0397 0.4\\
24.5837 0.4125\\
25.1445 0.425\\
25.722 0.4375\\
26.3163 0.45\\
26.9272 0.4625\\
27.5549 0.475\\
28.1994 0.4875\\
28.8606 0.5\\
};
\addlegendentry{HSDT11B};

\addplot [
color=black,
solid,
line width=1.2pt
]
table[row sep=crcr]{
15.957 -0.5\\
15.957 -0.4875\\
15.957 -0.475\\
15.957 -0.4625\\
15.957 -0.45\\
15.957 -0.4375\\
15.957 -0.425\\
15.957 -0.4125\\
15.957 -0.4\\
15.957 -0.3875\\
15.957 -0.375\\
15.957 -0.3625\\
15.957 -0.35\\
15.957 -0.3375\\
15.957 -0.325\\
15.957 -0.3125\\
15.957 -0.3\\
15.957 -0.2875\\
15.957 -0.275\\
15.957 -0.2625\\
15.957 -0.25\\
15.957 -0.25\\
15.957 -0.2375\\
15.957 -0.225\\
15.957 -0.2125\\
15.957 -0.2\\
15.957 -0.1875\\
15.957 -0.175\\
15.957 -0.1625\\
15.957 -0.15\\
15.957 -0.1375\\
15.957 -0.125\\
15.957 -0.1125\\
15.957 -0.1\\
15.957 -0.0875\\
15.957 -0.075\\
15.957 -0.0625\\
15.957 -0.05\\
15.957 -0.0375\\
15.957 -0.025\\
15.957 -0.0125\\
15.957 3.90313e-17\\
15.957 0\\
15.957 0.0125\\
15.957 0.025\\
15.957 0.0375\\
15.957 0.05\\
15.957 0.0625\\
15.957 0.075\\
15.957 0.0875\\
15.957 0.1\\
15.957 0.1125\\
15.957 0.125\\
15.957 0.1375\\
15.957 0.15\\
15.957 0.1625\\
15.957 0.175\\
15.957 0.1875\\
15.957 0.2\\
15.957 0.2125\\
15.957 0.225\\
15.957 0.2375\\
15.957 0.25\\
15.957 0.25\\
15.957 0.2625\\
15.957 0.275\\
15.957 0.2875\\
15.957 0.3\\
15.957 0.3125\\
15.957 0.325\\
15.957 0.3375\\
15.957 0.35\\
15.957 0.3625\\
15.957 0.375\\
15.957 0.3875\\
15.957 0.4\\
15.957 0.4125\\
15.957 0.425\\
15.957 0.4375\\
15.957 0.45\\
15.957 0.4625\\
15.957 0.475\\
15.957 0.4875\\
15.957 0.5\\
};
\addlegendentry{TSDT7};

\addplot [
color=blue,
mark size=2.5pt,
only marks,
mark=x,
mark options={solid,draw=black}
]
table[row sep=crcr]{
11.5764 -0.5\\
11.5764 -0.4875\\
11.5764 -0.475\\
11.5764 -0.4625\\
11.5764 -0.45\\
11.5764 -0.4375\\
11.5764 -0.425\\
11.5764 -0.4125\\
11.5764 -0.4\\
11.5764 -0.3875\\
11.5764 -0.375\\
11.5764 -0.3625\\
11.5764 -0.35\\
11.5764 -0.3375\\
11.5764 -0.325\\
11.5764 -0.3125\\
11.5764 -0.3\\
11.5764 -0.2875\\
11.5764 -0.275\\
11.5764 -0.2625\\
11.5764 -0.25\\
11.5764 -0.25\\
11.5764 -0.2375\\
11.5764 -0.225\\
11.5764 -0.2125\\
11.5764 -0.2\\
11.5764 -0.1875\\
11.5764 -0.175\\
11.5764 -0.1625\\
11.5764 -0.15\\
11.5764 -0.1375\\
11.5764 -0.125\\
11.5764 -0.1125\\
11.5764 -0.1\\
11.5764 -0.0875\\
11.5764 -0.075\\
11.5764 -0.0625\\
11.5764 -0.05\\
11.5764 -0.0375\\
11.5764 -0.025\\
11.5764 -0.0125\\
11.5764 3.90313e-17\\
11.5764 0\\
11.5764 0.0125\\
11.5764 0.025\\
11.5764 0.0375\\
11.5764 0.05\\
11.5764 0.0625\\
11.5764 0.075\\
11.5764 0.0875\\
11.5764 0.1\\
11.5764 0.1125\\
11.5764 0.125\\
11.5764 0.1375\\
11.5764 0.15\\
11.5764 0.1625\\
11.5764 0.175\\
11.5764 0.1875\\
11.5764 0.2\\
11.5764 0.2125\\
11.5764 0.225\\
11.5764 0.2375\\
11.5764 0.25\\
11.5764 0.25\\
11.5764 0.2625\\
11.5764 0.275\\
11.5764 0.2875\\
11.5764 0.3\\
11.5764 0.3125\\
11.5764 0.325\\
11.5764 0.3375\\
11.5764 0.35\\
11.5764 0.3625\\
11.5764 0.375\\
11.5764 0.3875\\
11.5764 0.4\\
11.5764 0.4125\\
11.5764 0.425\\
11.5764 0.4375\\
11.5764 0.45\\
11.5764 0.4625\\
11.5764 0.475\\
11.5764 0.4875\\
11.5764 0.5\\
};
\addlegendentry{FSDT5};

\end{axis}
\end{tikzpicture}%
}}
\subfigure[$\overline{\sigma}_{xx} = \overline{\sigma}_{xx}(0,0,z)$]{\scalebox{0.5}{
%
%
%
%
\begin{tikzpicture}

\begin{axis}[%
width=\figurewidth,
height=\figureheight,
scale only axis,
xmin=-800,
xmax=800,
xlabel={$\bar{\sigma}_{xx}$},
ymin=-0.5,
ymax=0.5,
ylabel={Normalized thickness, z/h},
legend style={draw=black,fill=white,legend cell align=left},
legend pos = south east
]
\addplot [
color=blue,
mark size=2.5pt,
only marks,
mark=o,
mark options={solid,draw=black}
]
table[row sep=crcr]{
166.942 -0.5\\
176.195 -0.4875\\
183.176 -0.475\\
188.003 -0.4625\\
190.796 -0.45\\
191.674 -0.4375\\
190.757 -0.425\\
188.163 -0.4125\\
184.012 -0.4\\
178.423 -0.3875\\
171.515 -0.375\\
163.407 -0.3625\\
154.218 -0.35\\
144.069 -0.3375\\
133.076 -0.325\\
121.361 -0.3125\\
109.043 -0.3\\
96.2391 -0.2875\\
83.0702 -0.275\\
69.655 -0.2625\\
56.1127 -0.25\\
-768.098 -0.25\\
-688.991 -0.2375\\
-614.523 -0.225\\
-544.577 -0.2125\\
-479.033 -0.2\\
-417.772 -0.1875\\
-360.674 -0.175\\
-307.621 -0.1625\\
-258.493 -0.15\\
-213.171 -0.1375\\
-171.536 -0.125\\
-133.469 -0.1125\\
-98.8501 -0.1\\
-67.5607 -0.0875\\
-39.4814 -0.075\\
-14.4931 -0.0625\\
7.52349 -0.05\\
26.6874 -0.0375\\
43.118 -0.025\\
56.9343 -0.0125\\
68.2557 3.90313e-17\\
-68.2557 0\\
-56.9343 0.0125\\
-43.118 0.025\\
-26.6874 0.0375\\
-7.52349 0.05\\
14.4931 0.0625\\
39.4814 0.075\\
67.5607 0.0875\\
98.8501 0.1\\
133.469 0.1125\\
171.536 0.125\\
213.171 0.1375\\
258.493 0.15\\
307.621 0.1625\\
360.674 0.175\\
417.772 0.1875\\
479.033 0.2\\
544.577 0.2125\\
614.523 0.225\\
688.991 0.2375\\
768.098 0.25\\
-56.1127 0.25\\
-69.655 0.2625\\
-83.0702 0.275\\
-96.2391 0.2875\\
-109.043 0.3\\
-121.361 0.3125\\
-133.076 0.325\\
-144.069 0.3375\\
-154.218 0.35\\
-163.407 0.3625\\
-171.515 0.375\\
-178.423 0.3875\\
-184.012 0.4\\
-188.163 0.4125\\
-190.757 0.425\\
-191.674 0.4375\\
-190.796 0.45\\
-188.003 0.4625\\
-183.176 0.475\\
-176.195 0.4875\\
-166.942 0.5\\
};
\addlegendentry{HSDT13};

\addplot [
color=black,
dashed,
line width=1.2pt
]
table[row sep=crcr]{
279.092 -0.5\\
270.913 -0.4875\\
261.616 -0.475\\
251.288 -0.4625\\
240.018 -0.45\\
227.895 -0.4375\\
215.005 -0.425\\
201.438 -0.4125\\
187.282 -0.4\\
172.625 -0.3875\\
157.555 -0.375\\
142.161 -0.3625\\
126.531 -0.35\\
110.752 -0.3375\\
94.9139 -0.325\\
79.1041 -0.3125\\
63.411 -0.3\\
47.9229 -0.2875\\
32.728 -0.275\\
17.9145 -0.2625\\
3.57072 -0.25\\
-792.61 -0.25\\
-715.808 -0.2375\\
-643.006 -0.225\\
-574.115 -0.2125\\
-509.046 -0.2\\
-447.712 -0.1875\\
-390.025 -0.175\\
-335.896 -0.1625\\
-285.236 -0.15\\
-237.959 -0.1375\\
-193.974 -0.125\\
-153.195 -0.1125\\
-115.533 -0.1\\
-80.8996 -0.0875\\
-49.2065 -0.075\\
-20.3657 -0.0625\\
5.71118 -0.05\\
29.1123 -0.0375\\
49.926 -0.025\\
68.2404 -0.0125\\
84.1439 3.90313e-17\\
-84.1439 0\\
-68.2404 0.0125\\
-49.926 0.025\\
-29.1123 0.0375\\
-5.71118 0.05\\
20.3657 0.0625\\
49.2065 0.075\\
80.8996 0.0875\\
115.533 0.1\\
153.195 0.1125\\
193.974 0.125\\
237.959 0.1375\\
285.236 0.15\\
335.896 0.1625\\
390.025 0.175\\
447.712 0.1875\\
509.046 0.2\\
574.115 0.2125\\
643.006 0.225\\
715.808 0.2375\\
792.61 0.25\\
-3.57072 0.25\\
-17.9145 0.2625\\
-32.728 0.275\\
-47.9229 0.2875\\
-63.411 0.3\\
-79.1041 0.3125\\
-94.9139 0.325\\
-110.752 0.3375\\
-126.531 0.35\\
-142.161 0.3625\\
-157.555 0.375\\
-172.625 0.3875\\
-187.282 0.4\\
-201.438 0.4125\\
-215.005 0.425\\
-227.895 0.4375\\
-240.018 0.45\\
-251.288 0.4625\\
-261.616 0.475\\
-270.913 0.4875\\
-279.092 0.5\\
};
\addlegendentry{HSDT11A};

\addplot [
color=blue,
mark size=4.3pt,
only marks,
mark=diamond,
mark options={solid,draw=black}
]
table[row sep=crcr]{
169.577 -0.5\\
177.512 -0.4875\\
183.293 -0.475\\
187.035 -0.4625\\
188.853 -0.45\\
188.861 -0.4375\\
187.175 -0.425\\
183.909 -0.4125\\
179.178 -0.4\\
173.098 -0.3875\\
165.782 -0.375\\
157.345 -0.3625\\
147.903 -0.35\\
137.57 -0.3375\\
126.462 -0.325\\
114.691 -0.3125\\
102.375 -0.3\\
89.6271 -0.2875\\
76.5624 -0.275\\
63.2958 -0.2625\\
49.942 -0.25\\
-762.428 -0.25\\
-685.118 -0.2375\\
-612.314 -0.225\\
-543.901 -0.2125\\
-479.763 -0.2\\
-419.787 -0.1875\\
-363.857 -0.175\\
-311.858 -0.1625\\
-263.675 -0.15\\
-219.195 -0.1375\\
-178.301 -0.125\\
-140.88 -0.1125\\
-106.815 -0.1\\
-75.9938 -0.0875\\
-48.2998 -0.075\\
-23.6186 -0.0625\\
-1.83557 -0.05\\
17.1643 -0.0375\\
33.4957 -0.025\\
47.2735 -0.0125\\
58.6125 3.90313e-17\\
-58.6125 0\\
-47.2735 0.0125\\
-33.4957 0.025\\
-17.1643 0.0375\\
1.83557 0.05\\
23.6186 0.0625\\
48.2998 0.075\\
75.9938 0.0875\\
106.815 0.1\\
140.88 0.1125\\
178.301 0.125\\
219.195 0.1375\\
263.675 0.15\\
311.858 0.1625\\
363.857 0.175\\
419.787 0.1875\\
479.763 0.2\\
543.901 0.2125\\
612.314 0.225\\
685.118 0.2375\\
762.428 0.25\\
-49.942 0.25\\
-63.2958 0.2625\\
-76.5624 0.275\\
-89.6271 0.2875\\
-102.375 0.3\\
-114.691 0.3125\\
-126.462 0.325\\
-137.57 0.3375\\
-147.903 0.35\\
-157.345 0.3625\\
-165.782 0.375\\
-173.098 0.3875\\
-179.178 0.4\\
-183.909 0.4125\\
-187.175 0.425\\
-188.861 0.4375\\
-188.853 0.45\\
-187.035 0.4625\\
-183.293 0.475\\
-177.512 0.4875\\
-169.577 0.5\\
};
\addlegendentry{HSDT11B};

\addplot [
color=black,
solid,
line width=1.2pt
]
table[row sep=crcr]{
509.318 -0.5\\
457.005 -0.4875\\
406.628 -0.475\\
358.187 -0.4625\\
311.682 -0.45\\
267.113 -0.4375\\
224.479 -0.425\\
183.781 -0.4125\\
145.019 -0.4\\
108.193 -0.3875\\
73.3022 -0.375\\
40.3475 -0.3625\\
9.32867 -0.35\\
-19.7544 -0.3375\\
-46.9016 -0.325\\
-72.1131 -0.3125\\
-95.3887 -0.3\\
-116.729 -0.2875\\
-136.133 -0.275\\
-153.601 -0.2625\\
-169.133 -0.25\\
-796.929 -0.25\\
-735.029 -0.2375\\
-675.065 -0.225\\
-617.037 -0.2125\\
-560.944 -0.2\\
-506.788 -0.1875\\
-454.567 -0.175\\
-404.282 -0.1625\\
-355.933 -0.15\\
-309.52 -0.1375\\
-265.043 -0.125\\
-222.501 -0.1125\\
-181.895 -0.1\\
-143.225 -0.0875\\
-106.491 -0.075\\
-71.6921 -0.0625\\
-38.8295 -0.05\\
-7.90262 -0.0375\\
21.0884 -0.025\\
48.1437 -0.0125\\
73.2631 3.90313e-17\\
-73.2631 0\\
-48.1437 0.0125\\
-21.0884 0.025\\
7.90262 0.0375\\
38.8295 0.05\\
71.6921 0.0625\\
106.491 0.075\\
143.225 0.0875\\
181.895 0.1\\
222.501 0.1125\\
265.043 0.125\\
309.52 0.1375\\
355.933 0.15\\
404.282 0.1625\\
454.567 0.175\\
506.788 0.1875\\
560.944 0.2\\
617.037 0.2125\\
675.065 0.225\\
735.029 0.2375\\
796.929 0.25\\
169.133 0.25\\
153.601 0.2625\\
136.133 0.275\\
116.729 0.2875\\
95.3887 0.3\\
72.1131 0.3125\\
46.9016 0.325\\
19.7544 0.3375\\
-9.32867 0.35\\
-40.3475 0.3625\\
-73.3022 0.375\\
-108.193 0.3875\\
-145.019 0.4\\
-183.781 0.4125\\
-224.479 0.425\\
-267.113 0.4375\\
-311.682 0.45\\
-358.187 0.4625\\
-406.628 0.475\\
-457.005 0.4875\\
-509.318 0.5\\
};
\addlegendentry{TSDT7};

\addplot [
color=blue,
mark size=2.5pt,
only marks,
mark=x,
mark options={solid,draw=black}
]
table[row sep=crcr]{
25.1599 -0.5\\
31.6623 -0.4875\\
38.1646 -0.475\\
44.667 -0.4625\\
51.1694 -0.45\\
57.6717 -0.4375\\
64.1741 -0.425\\
70.6765 -0.4125\\
77.1788 -0.4\\
83.6812 -0.3875\\
90.1836 -0.375\\
96.6859 -0.3625\\
103.188 -0.35\\
109.691 -0.3375\\
116.193 -0.325\\
122.695 -0.3125\\
129.198 -0.3\\
135.7 -0.2875\\
142.202 -0.275\\
148.705 -0.2625\\
155.207 -0.25\\
-415.302 -0.25\\
-408.799 -0.2375\\
-402.297 -0.225\\
-395.795 -0.2125\\
-389.292 -0.2\\
-382.79 -0.1875\\
-376.288 -0.175\\
-369.785 -0.1625\\
-363.283 -0.15\\
-356.781 -0.1375\\
-350.278 -0.125\\
-343.776 -0.1125\\
-337.273 -0.1\\
-330.771 -0.0875\\
-324.269 -0.075\\
-317.766 -0.0625\\
-311.264 -0.05\\
-304.762 -0.0375\\
-298.259 -0.025\\
-291.757 -0.0125\\
-285.255 3.90313e-17\\
285.255 0\\
291.757 0.0125\\
298.259 0.025\\
304.762 0.0375\\
311.264 0.05\\
317.766 0.0625\\
324.269 0.075\\
330.771 0.0875\\
337.273 0.1\\
343.776 0.1125\\
350.278 0.125\\
356.781 0.1375\\
363.283 0.15\\
369.785 0.1625\\
376.288 0.175\\
382.79 0.1875\\
389.292 0.2\\
395.795 0.2125\\
402.297 0.225\\
408.799 0.2375\\
415.302 0.25\\
-155.207 0.25\\
-148.705 0.2625\\
-142.202 0.275\\
-135.7 0.2875\\
-129.198 0.3\\
-122.695 0.3125\\
-116.193 0.325\\
-109.691 0.3375\\
-103.188 0.35\\
-96.6859 0.3625\\
-90.1836 0.375\\
-83.6812 0.3875\\
-77.1788 0.4\\
-70.6765 0.4125\\
-64.1741 0.425\\
-57.6717 0.4375\\
-51.1694 0.45\\
-44.667 0.4625\\
-38.1646 0.475\\
-31.6623 0.4875\\
-25.1599 0.5\\
};
\addlegendentry{FSDT5};

\end{axis}
\end{tikzpicture}%
}}\hspace{60pt}
\subfigure[$\overline{\tau}_{yz} = \overline{\tau}_{yz}(-a/2,0,z)$]{\scalebox{0.5}{
%
%
%
%
\begin{tikzpicture}

\begin{axis}[%
width=\figurewidth,
height=\figureheight,
scale only axis,
xmin=-200,
xmax=100,
xlabel={$\bar{\tau}_{yz}$},
ymin=-0.5,
ymax=0.5,
ylabel={Normalized thickness, z/h},
legend style={draw=black,fill=white,legend cell align=left}
]
\addplot [
color=blue,
mark size=2.5pt,
only marks,
mark=o,
mark options={solid,draw=black}
]
table[row sep=crcr]{
0 -0.5\\
-10.8343 -0.4875\\
-20.6713 -0.475\\
-29.5355 -0.4625\\
-37.4523 -0.45\\
-44.4473 -0.4375\\
-50.5471 -0.425\\
-55.7785 -0.4125\\
-60.1691 -0.4\\
-63.7471 -0.3875\\
-66.5411 -0.375\\
-68.5804 -0.3625\\
-69.8949 -0.35\\
-70.515 -0.3375\\
-70.4718 -0.325\\
-69.7968 -0.3125\\
-68.5223 -0.3\\
-66.6809 -0.2875\\
-64.3061 -0.275\\
-61.4318 -0.2625\\
-58.0925 -0.25\\
-58.0925 -0.25\\
-68.5684 -0.2375\\
-77.9173 -0.225\\
-86.1883 -0.2125\\
-93.4295 -0.2\\
-99.6883 -0.1875\\
-105.011 -0.175\\
-109.444 -0.1625\\
-113.032 -0.15\\
-115.819 -0.1375\\
-117.849 -0.125\\
-119.163 -0.1125\\
-119.804 -0.1\\
-119.812 -0.0875\\
-119.228 -0.075\\
-118.091 -0.0625\\
-116.439 -0.05\\
-114.31 -0.0375\\
-111.741 -0.025\\
-108.767 -0.0125\\
-105.424 3.90313e-17\\
-105.424 0\\
-101.381 0.0125\\
-97.7779 0.025\\
-94.6637 0.0375\\
-92.0865 0.05\\
-90.0953 0.0625\\
-88.7397 0.075\\
-88.0699 0.0875\\
-88.1365 0.1\\
-88.9909 0.1125\\
-90.6851 0.125\\
-93.2714 0.1375\\
-96.803 0.15\\
-101.333 0.1625\\
-106.917 0.175\\
-113.608 0.1875\\
-121.463 0.2\\
-130.537 0.2125\\
-140.887 0.225\\
-152.569 0.2375\\
-165.643 0.25\\
-165.643 0.25\\
-162.223 0.2625\\
-159.003 0.275\\
-156 0.2875\\
-153.228 0.3\\
-150.702 0.3125\\
-148.437 0.325\\
-146.446 0.3375\\
-144.742 0.35\\
-143.335 0.3625\\
-142.237 0.375\\
-141.458 0.3875\\
-141.007 0.4\\
-140.892 0.4125\\
-141.121 0.425\\
-141.7 0.4375\\
-142.636 0.45\\
-143.934 0.4625\\
-145.597 0.475\\
-147.63 0.4875\\
-150.035 0.5\\
};
\addlegendentry{HSDT13};

\addplot [
color=black,
dashed,
line width=1.2pt
]
table[row sep=crcr]{
0 -0.5\\
-11.3322 -0.4875\\
-21.5288 -0.475\\
-30.6267 -0.4625\\
-38.663 -0.45\\
-45.6752 -0.4375\\
-51.7009 -0.425\\
-56.7779 -0.4125\\
-60.9444 -0.4\\
-64.2387 -0.3875\\
-66.6994 -0.375\\
-68.3653 -0.3625\\
-69.2755 -0.35\\
-69.4693 -0.3375\\
-68.9863 -0.325\\
-67.8662 -0.3125\\
-66.149 -0.3\\
-63.8751 -0.2875\\
-61.085 -0.275\\
-57.8193 -0.2625\\
-54.1191 -0.25\\
-54.1191 -0.25\\
-64.8661 -0.2375\\
-74.4824 -0.225\\
-83.0123 -0.2125\\
-90.4995 -0.2\\
-96.9871 -0.1875\\
-102.518 -0.175\\
-107.133 -0.1625\\
-110.875 -0.15\\
-113.783 -0.1375\\
-115.899 -0.125\\
-117.262 -0.1125\\
-117.91 -0.1\\
-117.883 -0.0875\\
-117.217 -0.075\\
-115.951 -0.0625\\
-114.121 -0.05\\
-111.763 -0.0375\\
-108.912 -0.025\\
-105.603 -0.0125\\
-101.871 3.90313e-17\\
-101.871 0\\
-97.1588 0.0125\\
-92.9741 0.025\\
-89.3638 0.0375\\
-86.3747 0.05\\
-84.0537 0.0625\\
-82.4481 0.075\\
-81.6053 0.0875\\
-81.5731 0.1\\
-82.3994 0.1125\\
-84.1324 0.125\\
-86.8204 0.1375\\
-90.5122 0.15\\
-95.2567 0.1625\\
-101.103 0.175\\
-108.101 0.1875\\
-116.299 0.2\\
-125.748 0.2125\\
-136.498 0.225\\
-148.599 0.2375\\
-162.101 0.25\\
-162.101 0.25\\
-158.521 0.2625\\
-155.124 0.275\\
-151.929 0.2875\\
-148.959 0.3\\
-146.231 0.3125\\
-143.765 0.325\\
-141.58 0.3375\\
-139.693 0.35\\
-138.122 0.3625\\
-136.883 0.375\\
-135.992 0.3875\\
-135.465 0.4\\
-135.317 0.4125\\
-135.561 0.425\\
-136.213 0.4375\\
-137.284 0.45\\
-138.788 0.4625\\
-140.736 0.475\\
-143.14 0.4875\\
-146.01 0.5\\
};
\addlegendentry{HSDT11A};

\addplot [
color=blue,
mark size=4.3pt,
only marks,
mark=diamond,
mark options={solid,draw=black}
]
table[row sep=crcr]{
0 -0.5\\
-10.7326 -0.4875\\
-20.4663 -0.475\\
-29.2272 -0.4625\\
-37.0419 -0.45\\
-43.9373 -0.4375\\
-49.941 -0.425\\
-55.0812 -0.4125\\
-59.3863 -0.4\\
-62.8855 -0.3875\\
-65.6085 -0.375\\
-67.5854 -0.3625\\
-68.8469 -0.35\\
-69.4241 -0.3375\\
-69.3488 -0.325\\
-68.6532 -0.3125\\
-67.3701 -0.3\\
-65.5327 -0.2875\\
-63.1748 -0.275\\
-60.3308 -0.2625\\
-57.0353 -0.25\\
-57.0353 -0.25\\
-67.2019 -0.2375\\
-76.2836 -0.225\\
-84.3294 -0.2125\\
-91.3873 -0.2\\
-97.5045 -0.1875\\
-102.728 -0.175\\
-107.102 -0.1625\\
-110.672 -0.15\\
-113.482 -0.1375\\
-115.575 -0.125\\
-116.994 -0.1125\\
-117.778 -0.1\\
-117.969 -0.0875\\
-117.607 -0.075\\
-116.729 -0.0625\\
-115.375 -0.05\\
-113.581 -0.0375\\
-111.383 -0.025\\
-108.817 -0.0125\\
-105.918 3.90313e-17\\
-105.918 0\\
-102.107 0.0125\\
-98.7417 0.025\\
-95.8696 0.0375\\
-93.5376 0.05\\
-91.7934 0.0625\\
-90.6853 0.075\\
-90.2623 0.0875\\
-90.5736 0.1\\
-91.6691 0.1125\\
-93.5992 0.125\\
-96.4148 0.1375\\
-100.167 0.15\\
-104.909 0.1625\\
-110.692 0.175\\
-117.569 0.1875\\
-125.594 0.2\\
-134.82 0.2125\\
-145.304 0.225\\
-157.098 0.2375\\
-170.26 0.25\\
-170.26 0.25\\
-166.821 0.2625\\
-163.594 0.275\\
-160.593 0.2875\\
-157.831 0.3\\
-155.323 0.3125\\
-153.081 0.325\\
-151.117 0.3375\\
-149.44 0.35\\
-148.062 0.3625\\
-146.991 0.375\\
-146.235 0.3875\\
-145.802 0.4\\
-145.699 0.4125\\
-145.93 0.425\\
-146.501 0.4375\\
-147.416 0.45\\
-148.678 0.4625\\
-150.289 0.475\\
-152.25 0.4875\\
-154.563 0.5\\
};
\addlegendentry{HSDT11B};

\addplot [
color=black,
solid,
line width=1.2pt
]
table[row sep=crcr]{
0 -0.5\\
-8.94429 -0.4875\\
-16.8496 -0.475\\
-23.7533 -0.4625\\
-29.6927 -0.45\\
-34.7053 -0.4375\\
-38.8283 -0.425\\
-42.0993 -0.4125\\
-44.5554 -0.4\\
-46.2342 -0.3875\\
-47.173 -0.375\\
-47.4091 -0.3625\\
-46.9799 -0.35\\
-45.9229 -0.3375\\
-44.2753 -0.325\\
-42.0745 -0.3125\\
-39.358 -0.3\\
-36.163 -0.2875\\
-32.527 -0.275\\
-28.4873 -0.2625\\
-24.0813 -0.25\\
-24.0813 -0.25\\
-33.8048 -0.2375\\
-42.6433 -0.225\\
-50.6233 -0.2125\\
-57.771 -0.2\\
-64.1131 -0.1875\\
-69.6758 -0.175\\
-74.4857 -0.1625\\
-78.5692 -0.15\\
-81.9527 -0.1375\\
-84.6626 -0.125\\
-86.7255 -0.1125\\
-88.1676 -0.1\\
-89.0155 -0.0875\\
-89.2956 -0.075\\
-89.0343 -0.0625\\
-88.2581 -0.05\\
-86.9933 -0.0375\\
-85.2665 -0.025\\
-83.104 -0.0125\\
-80.5323 3.90313e-17\\
-80.5323 0\\
-77.0667 0.0125\\
-74.0569 0.025\\
-71.5404 0.0375\\
-69.5545 0.05\\
-68.1366 0.0625\\
-67.324 0.075\\
-67.1541 0.0875\\
-67.6643 0.1\\
-68.892 0.1125\\
-70.8746 0.125\\
-73.6493 0.1375\\
-77.2536 0.15\\
-81.7249 0.1625\\
-87.1005 0.175\\
-93.4177 0.1875\\
-100.714 0.2\\
-109.027 0.2125\\
-118.393 0.225\\
-128.851 0.2375\\
-140.437 0.25\\
-140.437 0.25\\
-135.379 0.2625\\
-130.492 0.275\\
-125.804 0.2875\\
-121.342 0.3\\
-117.131 0.3125\\
-113.198 0.325\\
-109.569 0.3375\\
-106.272 0.35\\
-103.332 0.3625\\
-100.776 0.375\\
-98.6296 0.3875\\
-96.9202 0.4\\
-95.674 0.4125\\
-94.9173 0.425\\
-94.6766 0.4375\\
-94.9783 0.45\\
-95.8488 0.4625\\
-97.3146 0.475\\
-99.4021 0.4875\\
-102.138 0.5\\
};
\addlegendentry{TSDT7};

\addplot [
color=blue,
mark size=2.5pt,
only marks,
mark=x,
mark options={solid,draw=black}
]
table[row sep=crcr]{
0 -0.5\\
5.13579 -0.4875\\
10.0481 -0.475\\
14.7369 -0.4625\\
19.2021 -0.45\\
23.4439 -0.4375\\
27.4621 -0.425\\
31.2569 -0.4125\\
34.8281 -0.4\\
38.1758 -0.3875\\
41.3 -0.375\\
44.2007 -0.3625\\
46.8779 -0.35\\
49.3316 -0.3375\\
51.5617 -0.325\\
53.5684 -0.3125\\
55.3515 -0.3\\
56.9112 -0.2875\\
58.2473 -0.275\\
59.3599 -0.2625\\
60.249 -0.25\\
60.249 -0.25\\
53.8907 -0.2375\\
47.7325 -0.225\\
41.7744 -0.2125\\
36.0166 -0.2\\
30.4589 -0.1875\\
25.1013 -0.175\\
19.9439 -0.1625\\
14.9867 -0.15\\
10.2297 -0.1375\\
5.67278 -0.125\\
1.31606 -0.1125\\
-2.84051 -0.1\\
-6.79691 -0.0875\\
-10.5531 -0.075\\
-14.1092 -0.0625\\
-17.4651 -0.05\\
-20.6209 -0.0375\\
-23.5765 -0.025\\
-26.3319 -0.0125\\
-28.8872 3.90313e-17\\
-28.8872 0\\
-32.6918 0.0125\\
-36.7199 0.025\\
-40.9715 0.0375\\
-45.4466 0.05\\
-50.1452 0.0625\\
-55.0674 0.075\\
-60.213 0.0875\\
-65.5822 0.1\\
-71.1749 0.1125\\
-76.9911 0.125\\
-83.0308 0.1375\\
-89.294 0.15\\
-95.7807 0.1625\\
-102.491 0.175\\
-109.425 0.1875\\
-116.582 0.2\\
-123.963 0.2125\\
-131.567 0.225\\
-139.395 0.2375\\
-147.446 0.25\\
-147.446 0.25\\
-145.798 0.2625\\
-143.949 0.275\\
-141.901 0.2875\\
-139.652 0.3\\
-137.204 0.3125\\
-134.555 0.325\\
-131.705 0.3375\\
-128.656 0.35\\
-125.407 0.3625\\
-121.957 0.375\\
-118.307 0.3875\\
-114.457 0.4\\
-110.407 0.4125\\
-106.157 0.425\\
-101.706 0.4375\\
-97.0558 0.45\\
-92.205 0.4625\\
-87.1541 0.475\\
-81.903 0.4875\\
-76.4518 0.5\\
};
\addlegendentry{FSDT5};

\end{axis}
\end{tikzpicture}%
}}
\subfigure[$\overline{\tau}_{xz}=\overline{\tau}_{xz}(-a/2,0,z)$]{\scalebox{0.5}{
%
%
%
%
\begin{tikzpicture}

\begin{axis}[%
width=\figurewidth,
height=\figureheight,
scale only axis,
xmin=-180,
xmax=0,
xlabel={$\bar{\tau}_{xz}$},
ymin=-0.5,
ymax=0.5,
ylabel={Normalized thickness, z/h},
legend style={draw=black,fill=white,legend cell align=left}
]
\addplot [
color=blue,
mark size=2.5pt,
only marks,
mark=o,
mark options={solid,draw=black}
]
table[row sep=crcr]{
0 -0.5\\
-16.4736 -0.4875\\
-31.9015 -0.475\\
-46.302 -0.4625\\
-59.6943 -0.45\\
-72.098 -0.4375\\
-83.5337 -0.425\\
-94.0225 -0.4125\\
-103.586 -0.4\\
-112.248 -0.3875\\
-120.03 -0.375\\
-126.957 -0.3625\\
-133.054 -0.35\\
-138.345 -0.3375\\
-142.857 -0.325\\
-146.617 -0.3125\\
-149.652 -0.3\\
-151.99 -0.2875\\
-153.66 -0.275\\
-154.692 -0.2625\\
-155.115 -0.25\\
-155.115 -0.25\\
-148.65 -0.2375\\
-143.085 -0.225\\
-138.371 -0.2125\\
-134.46 -0.2\\
-131.306 -0.1875\\
-128.863 -0.175\\
-127.086 -0.1625\\
-125.931 -0.15\\
-125.355 -0.1375\\
-125.317 -0.125\\
-125.773 -0.1125\\
-126.686 -0.1\\
-128.014 -0.0875\\
-129.72 -0.075\\
-131.766 -0.0625\\
-134.116 -0.05\\
-136.734 -0.0375\\
-139.584 -0.025\\
-142.634 -0.0125\\
-145.85 3.90313e-17\\
-145.85 0\\
-141.901 0.0125\\
-138.185 0.025\\
-134.747 0.0375\\
-131.636 0.05\\
-128.898 0.0625\\
-126.581 0.075\\
-124.736 0.0875\\
-123.411 0.1\\
-122.657 0.1125\\
-122.524 0.125\\
-123.066 0.1375\\
-124.334 0.15\\
-126.382 0.1625\\
-129.263 0.175\\
-133.032 0.1875\\
-137.744 0.2\\
-143.455 0.2125\\
-150.222 0.225\\
-158.102 0.2375\\
-167.154 0.25\\
-167.154 0.25\\
-166.939 0.2625\\
-166.398 0.275\\
-165.519 0.2875\\
-164.292 0.3\\
-162.706 0.3125\\
-160.753 0.325\\
-158.423 0.3375\\
-155.711 0.35\\
-152.61 0.3625\\
-149.115 0.375\\
-145.22 0.3875\\
-140.922 0.4\\
-136.22 0.4125\\
-131.11 0.425\\
-125.593 0.4375\\
-119.669 0.45\\
-113.338 0.4625\\
-106.602 0.475\\
-99.4647 0.4875\\
-91.9296 0.5\\
};
\addlegendentry{HSDT13};

\addplot [
color=black,
dashed,
line width=1.2pt
]
table[row sep=crcr]{
0 -0.5\\
-17.3001 -0.4875\\
-33.3901 -0.475\\
-48.302 -0.4625\\
-62.0679 -0.45\\
-74.7204 -0.4375\\
-86.2922 -0.425\\
-96.8165 -0.4125\\
-106.327 -0.4\\
-114.857 -0.3875\\
-122.441 -0.375\\
-129.114 -0.3625\\
-134.91 -0.35\\
-139.864 -0.3375\\
-144.012 -0.325\\
-147.389 -0.3125\\
-150.033 -0.3\\
-151.978 -0.2875\\
-153.262 -0.275\\
-153.922 -0.2625\\
-153.996 -0.25\\
-153.996 -0.25\\
-147.272 -0.2375\\
-141.447 -0.225\\
-136.478 -0.2125\\
-132.32 -0.2\\
-128.933 -0.1875\\
-126.275 -0.175\\
-124.306 -0.1625\\
-122.984 -0.15\\
-122.272 -0.1375\\
-122.13 -0.125\\
-122.519 -0.1125\\
-123.404 -0.1\\
-124.746 -0.0875\\
-126.511 -0.075\\
-128.662 -0.0625\\
-131.164 -0.05\\
-133.985 -0.0375\\
-137.089 -0.025\\
-140.445 -0.0125\\
-144.02 3.90313e-17\\
-144.02 0\\
-139.424 0.0125\\
-135.148 0.025\\
-131.235 0.0375\\
-127.731 0.05\\
-124.681 0.0625\\
-122.132 0.075\\
-120.13 0.0875\\
-118.72 0.1\\
-117.951 0.1125\\
-117.869 0.125\\
-118.522 0.1375\\
-119.957 0.15\\
-122.223 0.1625\\
-125.368 0.175\\
-129.442 0.1875\\
-134.492 0.2\\
-140.57 0.2125\\
-147.723 0.225\\
-156.004 0.2375\\
-165.462 0.25\\
-165.462 0.25\\
-165.383 0.2625\\
-164.985 0.275\\
-164.25 0.2875\\
-163.161 0.3\\
-161.703 0.3125\\
-159.861 0.325\\
-157.621 0.3375\\
-154.969 0.35\\
-151.892 0.3625\\
-148.378 0.375\\
-144.416 0.3875\\
-139.993 0.4\\
-135.1 0.4125\\
-129.728 0.425\\
-123.867 0.4375\\
-117.509 0.45\\
-110.646 0.4625\\
-103.271 0.475\\
-95.3778 0.4875\\
-86.9609 0.5\\
};
\addlegendentry{HSDT11A};

\addplot [
color=blue,
mark size=4.3pt,
only marks,
mark=diamond,
mark options={solid,draw=black}
]
table[row sep=crcr]{
0 -0.5\\
-16.3363 -0.4875\\
-31.6337 -0.475\\
-45.9115 -0.4625\\
-59.1897 -0.45\\
-71.4889 -0.4375\\
-82.8303 -0.425\\
-93.2359 -0.4125\\
-102.728 -0.4\\
-111.33 -0.3875\\
-119.066 -0.375\\
-125.96 -0.3625\\
-132.037 -0.35\\
-137.323 -0.3375\\
-141.845 -0.325\\
-145.629 -0.3125\\
-148.704 -0.3\\
-151.098 -0.2875\\
-152.839 -0.275\\
-153.957 -0.2625\\
-154.483 -0.25\\
-154.483 -0.25\\
-148.349 -0.2375\\
-143.074 -0.225\\
-138.609 -0.2125\\
-134.908 -0.2\\
-131.922 -0.1875\\
-129.607 -0.175\\
-127.918 -0.1625\\
-126.81 -0.15\\
-126.24 -0.1375\\
-126.168 -0.125\\
-126.551 -0.1125\\
-127.35 -0.1\\
-128.526 -0.0875\\
-130.04 -0.075\\
-131.856 -0.0625\\
-133.937 -0.05\\
-136.248 -0.0375\\
-138.754 -0.025\\
-141.424 -0.0125\\
-144.223 3.90313e-17\\
-144.223 0\\
-140.461 0.0125\\
-136.938 0.025\\
-133.698 0.0375\\
-130.789 0.05\\
-128.255 0.0625\\
-126.144 0.075\\
-124.505 0.0875\\
-123.385 0.1\\
-122.834 0.1125\\
-122.901 0.125\\
-123.637 0.1375\\
-125.093 0.15\\
-127.321 0.1625\\
-130.372 0.175\\
-134.3 0.1875\\
-139.159 0.2\\
-145.002 0.2125\\
-151.885 0.225\\
-159.864 0.2375\\
-168.994 0.25\\
-168.994 0.25\\
-168.764 0.2625\\
-168.197 0.275\\
-167.282 0.2875\\
-166.011 0.3\\
-164.374 0.3125\\
-162.365 0.325\\
-159.976 0.3375\\
-157.204 0.35\\
-154.043 0.3625\\
-150.49 0.375\\
-146.543 0.3875\\
-142.199 0.4\\
-137.459 0.4125\\
-132.323 0.425\\
-126.792 0.4375\\
-120.868 0.45\\
-114.555 0.4625\\
-107.857 0.475\\
-100.779 0.4875\\
-93.3275 0.5\\
};
\addlegendentry{HSDT11B};

\addplot [
color=black,
solid,
line width=1.2pt
]
table[row sep=crcr]{
0 -0.5\\
-15.7641 -0.4875\\
-30.345 -0.475\\
-43.7788 -0.4625\\
-56.1014 -0.45\\
-67.349 -0.4375\\
-77.5576 -0.425\\
-86.7633 -0.4125\\
-95.0022 -0.4\\
-102.31 -0.3875\\
-108.724 -0.375\\
-114.278 -0.3625\\
-119.01 -0.35\\
-122.956 -0.3375\\
-126.151 -0.325\\
-128.631 -0.3125\\
-130.434 -0.3\\
-131.594 -0.2875\\
-132.147 -0.275\\
-132.131 -0.2625\\
-131.581 -0.25\\
-131.581 -0.25\\
-125.809 -0.2375\\
-120.694 -0.225\\
-116.211 -0.2125\\
-112.336 -0.2\\
-109.045 -0.1875\\
-106.311 -0.175\\
-104.111 -0.1625\\
-102.421 -0.15\\
-101.215 -0.1375\\
-100.469 -0.125\\
-100.158 -0.1125\\
-100.257 -0.1\\
-100.743 -0.0875\\
-101.59 -0.075\\
-102.774 -0.0625\\
-104.27 -0.05\\
-106.053 -0.0375\\
-108.099 -0.025\\
-110.384 -0.0125\\
-112.882 3.90313e-17\\
-112.882 0\\
-109.445 0.0125\\
-106.269 0.025\\
-103.387 0.0375\\
-100.837 0.05\\
-98.6542 0.0625\\
-96.8751 0.075\\
-95.5355 0.0875\\
-94.6716 0.1\\
-94.3194 0.1125\\
-94.515 0.125\\
-95.2944 0.1375\\
-96.6938 0.15\\
-98.7491 0.1625\\
-101.496 0.175\\
-104.972 0.1875\\
-109.211 0.2\\
-114.251 0.2125\\
-120.127 0.225\\
-126.876 0.2375\\
-134.533 0.25\\
-134.533 0.25\\
-135.825 0.2625\\
-136.79 0.275\\
-137.401 0.2875\\
-137.635 0.3\\
-137.467 0.3125\\
-136.872 0.325\\
-135.825 0.3375\\
-134.302 0.35\\
-132.279 0.3625\\
-129.73 0.375\\
-126.631 0.3875\\
-122.957 0.4\\
-118.685 0.4125\\
-113.788 0.425\\
-108.242 0.4375\\
-102.024 0.45\\
-95.1076 0.4625\\
-87.4688 0.475\\
-79.083 0.4875\\
-69.9255 0.5\\
};
\addlegendentry{TSDT7};

\addplot [
color=blue,
mark size=2.5pt,
only marks,
mark=x,
mark options={solid,draw=black}
]
table[row sep=crcr]{
0 -0.5\\
-3.20288 -0.4875\\
-6.41497 -0.475\\
-9.63628 -0.4625\\
-12.8668 -0.45\\
-16.1065 -0.4375\\
-19.3554 -0.425\\
-22.6136 -0.4125\\
-25.8809 -0.4\\
-29.1575 -0.3875\\
-32.4433 -0.375\\
-35.7383 -0.3625\\
-39.0425 -0.35\\
-42.3559 -0.3375\\
-45.6785 -0.325\\
-49.0103 -0.3125\\
-52.3514 -0.3\\
-55.7016 -0.2875\\
-59.0611 -0.275\\
-62.4297 -0.2625\\
-65.8076 -0.25\\
-65.8076 -0.25\\
-63.7926 -0.2375\\
-61.7644 -0.225\\
-59.7229 -0.2125\\
-57.6683 -0.2\\
-55.6005 -0.1875\\
-53.5195 -0.175\\
-51.4252 -0.1625\\
-49.3178 -0.15\\
-47.1972 -0.1375\\
-45.0633 -0.125\\
-42.9163 -0.1125\\
-40.7561 -0.1\\
-38.5827 -0.0875\\
-36.396 -0.075\\
-34.1962 -0.0625\\
-31.9832 -0.05\\
-29.757 -0.0375\\
-27.5175 -0.025\\
-25.2649 -0.0125\\
-22.9991 3.90313e-17\\
-22.9991 0\\
-26.5704 0.0125\\
-30.1509 0.025\\
-33.7406 0.0375\\
-37.3395 0.05\\
-40.9477 0.0625\\
-44.565 0.075\\
-48.1916 0.0875\\
-51.8273 0.1\\
-55.4723 0.1125\\
-59.1265 0.125\\
-62.7899 0.1375\\
-66.4625 0.15\\
-70.1444 0.1625\\
-73.8354 0.175\\
-77.5356 0.1875\\
-81.2451 0.2\\
-84.9638 0.2125\\
-88.6916 0.225\\
-92.4287 0.2375\\
-96.175 0.25\\
-96.175 0.25\\
-93.632 0.2625\\
-91.0757 0.275\\
-88.5063 0.2875\\
-85.9237 0.3\\
-83.3278 0.3125\\
-80.7188 0.325\\
-78.0966 0.3375\\
-75.4611 0.35\\
-72.8125 0.3625\\
-70.1507 0.375\\
-67.4756 0.3875\\
-64.7874 0.4\\
-62.0859 0.4125\\
-59.3713 0.425\\
-56.6435 0.4375\\
-53.9024 0.45\\
-51.1482 0.4625\\
-48.3807 0.475\\
-45.6001 0.4875\\
-42.8063 0.5\\
};
\addlegendentry{FSDT5};

\end{axis}
\end{tikzpicture}%
}}\hspace{60pt}
\subfigure[$\overline{\sigma}_{zz} = \overline{\sigma}_{zz}(0,0,z)$]{\scalebox{0.5}{
%
%
%
%
\begin{tikzpicture}

\begin{axis}[%
width=\figurewidth,
height=\figureheight,
scale only axis,
xmin=-25,
xmax=25,
xlabel={$\bar{\sigma}_{zz}$},
ymin=-0.5,
ymax=0.5,
ylabel={Normalized thickness, z/h},
legend style={draw=black,fill=white,legend cell align=left},
legend pos = south east
]
\addplot [
color=blue,
mark size=2.5pt,
only marks,
mark=o,
mark options={solid,draw=black}
]
table[row sep=crcr]{
-21.6977 -0.5\\
-18.8205 -0.4875\\
-16.1272 -0.475\\
-13.613 -0.4625\\
-11.2727 -0.45\\
-9.10128 -0.4375\\
-7.09372 -0.425\\
-5.24496 -0.4125\\
-3.54996 -0.4\\
-2.00368 -0.3875\\
-0.601068 -0.375\\
0.662918 -0.3625\\
1.79332 -0.35\\
2.79519 -0.3375\\
3.67356 -0.325\\
4.43349 -0.3125\\
5.08001 -0.3\\
5.61817 -0.2875\\
6.05301 -0.275\\
6.38958 -0.2625\\
6.63293 -0.25\\
2.22538 -0.25\\
2.87604 -0.2375\\
3.41816 -0.225\\
3.85676 -0.2125\\
4.19691 -0.2\\
4.44364 -0.1875\\
4.60201 -0.175\\
4.67704 -0.1625\\
4.6738 -0.15\\
4.59731 -0.1375\\
4.45264 -0.125\\
4.24481 -0.1125\\
3.97888 -0.1\\
3.65989 -0.0875\\
3.29288 -0.075\\
2.88291 -0.0625\\
2.435 -0.05\\
1.95422 -0.0375\\
1.44559 -0.025\\
0.914172 -0.0125\\
0.365004 3.90313e-17\\
-0.365004 0\\
-0.914172 0.0125\\
-1.44559 0.025\\
-1.95422 0.0375\\
-2.435 0.05\\
-2.88291 0.0625\\
-3.29288 0.075\\
-3.65989 0.0875\\
-3.97888 0.1\\
-4.24481 0.1125\\
-4.45264 0.125\\
-4.59731 0.1375\\
-4.6738 0.15\\
-4.67704 0.1625\\
-4.60201 0.175\\
-4.44364 0.1875\\
-4.19691 0.2\\
-3.85676 0.2125\\
-3.41816 0.225\\
-2.87604 0.2375\\
-2.22538 0.25\\
-6.63293 0.25\\
-6.38958 0.2625\\
-6.05301 0.275\\
-5.61817 0.2875\\
-5.08001 0.3\\
-4.43349 0.3125\\
-3.67356 0.325\\
-2.79519 0.3375\\
-1.79332 0.35\\
-0.662918 0.3625\\
0.601068 0.375\\
2.00368 0.3875\\
3.54996 0.4\\
5.24496 0.4125\\
7.09372 0.425\\
9.10128 0.4375\\
11.2727 0.45\\
13.613 0.4625\\
16.1272 0.475\\
18.8205 0.4875\\
21.6977 0.5\\
};
\addlegendentry{HSDT13};

\addplot [
color=black,
dashed,
line width=1.2pt
]
table[row sep=crcr]{
0 -0.5\\
0 -0.4875\\
0 -0.475\\
0 -0.4625\\
0 -0.45\\
0 -0.4375\\
0 -0.425\\
0 -0.4125\\
0 -0.4\\
0 -0.3875\\
0 -0.375\\
0 -0.3625\\
0 -0.35\\
0 -0.3375\\
0 -0.325\\
0 -0.3125\\
0 -0.3\\
0 -0.2875\\
0 -0.275\\
0 -0.2625\\
0 -0.25\\
0 -0.25\\
0 -0.2375\\
0 -0.225\\
0 -0.2125\\
0 -0.2\\
0 -0.1875\\
0 -0.175\\
0 -0.1625\\
0 -0.15\\
0 -0.1375\\
0 -0.125\\
0 -0.1125\\
0 -0.1\\
0 -0.0875\\
0 -0.075\\
0 -0.0625\\
0 -0.05\\
0 -0.0375\\
0 -0.025\\
0 -0.0125\\
0 3.90313e-17\\
0 0\\
0 0.0125\\
0 0.025\\
0 0.0375\\
0 0.05\\
0 0.0625\\
0 0.075\\
0 0.0875\\
0 0.1\\
0 0.1125\\
0 0.125\\
0 0.1375\\
0 0.15\\
0 0.1625\\
0 0.175\\
0 0.1875\\
0 0.2\\
0 0.2125\\
0 0.225\\
0 0.2375\\
0 0.25\\
0 0.25\\
0 0.2625\\
0 0.275\\
0 0.2875\\
0 0.3\\
0 0.3125\\
0 0.325\\
0 0.3375\\
0 0.35\\
0 0.3625\\
0 0.375\\
0 0.3875\\
0 0.4\\
0 0.4125\\
0 0.425\\
0 0.4375\\
0 0.45\\
0 0.4625\\
0 0.475\\
0 0.4875\\
0 0.5\\
};
\addlegendentry{HSDT11A};

\addplot [
color=blue,
mark size=4.3pt,
only marks,
mark=diamond,
mark options={solid,draw=black}
]
table[row sep=crcr]{
-20.1698 -0.5\\
-17.4231 -0.4875\\
-14.8533 -0.475\\
-12.4555 -0.4625\\
-10.2249 -0.45\\
-8.15662 -0.4375\\
-6.24581 -0.425\\
-4.48762 -0.4125\\
-2.87719 -0.4\\
-1.40967 -0.3875\\
-0.0801978 -0.375\\
1.11608 -0.3625\\
2.18401 -0.35\\
3.12845 -0.3375\\
3.95425 -0.325\\
4.66628 -0.3125\\
5.26938 -0.3\\
5.7684 -0.2875\\
6.16821 -0.275\\
6.47365 -0.2625\\
6.68958 -0.25\\
2.34536 -0.25\\
2.96132 -0.2375\\
3.47262 -0.225\\
3.88412 -0.2125\\
4.20067 -0.2\\
4.42712 -0.1875\\
4.56834 -0.175\\
4.62917 -0.1625\\
4.61446 -0.15\\
4.52908 -0.1375\\
4.37788 -0.125\\
4.1657 -0.1125\\
3.89741 -0.1\\
3.57786 -0.0875\\
3.2119 -0.075\\
2.80439 -0.0625\\
2.36018 -0.05\\
1.88412 -0.0375\\
1.38108 -0.025\\
0.855898 -0.0125\\
0.313436 3.90313e-17\\
-0.313436 0\\
-0.855898 0.0125\\
-1.38108 0.025\\
-1.88412 0.0375\\
-2.36018 0.05\\
-2.80439 0.0625\\
-3.2119 0.075\\
-3.57786 0.0875\\
-3.89741 0.1\\
-4.1657 0.1125\\
-4.37788 0.125\\
-4.52908 0.1375\\
-4.61446 0.15\\
-4.62917 0.1625\\
-4.56834 0.175\\
-4.42712 0.1875\\
-4.20067 0.2\\
-3.88412 0.2125\\
-3.47262 0.225\\
-2.96132 0.2375\\
-2.34536 0.25\\
-6.68958 0.25\\
-6.47365 0.2625\\
-6.16821 0.275\\
-5.7684 0.2875\\
-5.26938 0.3\\
-4.66628 0.3125\\
-3.95425 0.325\\
-3.12845 0.3375\\
-2.18401 0.35\\
-1.11608 0.3625\\
0.0801978 0.375\\
1.40967 0.3875\\
2.87719 0.4\\
4.48762 0.4125\\
6.24581 0.425\\
8.15662 0.4375\\
10.2249 0.45\\
12.4555 0.4625\\
14.8533 0.475\\
17.4231 0.4875\\
20.1698 0.5\\
};
\addlegendentry{HSDT11B};

\addplot [
color=black,
solid,
line width=1.2pt
]
table[row sep=crcr]{
0 -0.5\\
0 -0.4875\\
0 -0.475\\
0 -0.4625\\
0 -0.45\\
0 -0.4375\\
0 -0.425\\
0 -0.4125\\
0 -0.4\\
0 -0.3875\\
0 -0.375\\
0 -0.3625\\
0 -0.35\\
0 -0.3375\\
0 -0.325\\
0 -0.3125\\
0 -0.3\\
0 -0.2875\\
0 -0.275\\
0 -0.2625\\
0 -0.25\\
0 -0.25\\
0 -0.2375\\
0 -0.225\\
0 -0.2125\\
0 -0.2\\
0 -0.1875\\
0 -0.175\\
0 -0.1625\\
0 -0.15\\
0 -0.1375\\
0 -0.125\\
0 -0.1125\\
0 -0.1\\
0 -0.0875\\
0 -0.075\\
0 -0.0625\\
0 -0.05\\
0 -0.0375\\
0 -0.025\\
0 -0.0125\\
0 3.90313e-17\\
0 0\\
0 0.0125\\
0 0.025\\
0 0.0375\\
0 0.05\\
0 0.0625\\
0 0.075\\
0 0.0875\\
0 0.1\\
0 0.1125\\
0 0.125\\
0 0.1375\\
0 0.15\\
0 0.1625\\
0 0.175\\
0 0.1875\\
0 0.2\\
0 0.2125\\
0 0.225\\
0 0.2375\\
0 0.25\\
0 0.25\\
0 0.2625\\
0 0.275\\
0 0.2875\\
0 0.3\\
0 0.3125\\
0 0.325\\
0 0.3375\\
0 0.35\\
0 0.3625\\
0 0.375\\
0 0.3875\\
0 0.4\\
0 0.4125\\
0 0.425\\
0 0.4375\\
0 0.45\\
0 0.4625\\
0 0.475\\
0 0.4875\\
0 0.5\\
};
\addlegendentry{TSDT7};

\addplot [
color=blue,
mark size=2.5pt,
only marks,
mark=x,
mark options={solid,draw=black}
]
table[row sep=crcr]{
0 -0.5\\
0 -0.4875\\
0 -0.475\\
0 -0.4625\\
0 -0.45\\
0 -0.4375\\
0 -0.425\\
0 -0.4125\\
0 -0.4\\
0 -0.3875\\
0 -0.375\\
0 -0.3625\\
0 -0.35\\
0 -0.3375\\
0 -0.325\\
0 -0.3125\\
0 -0.3\\
0 -0.2875\\
0 -0.275\\
0 -0.2625\\
0 -0.25\\
0 -0.25\\
0 -0.2375\\
0 -0.225\\
0 -0.2125\\
0 -0.2\\
0 -0.1875\\
0 -0.175\\
0 -0.1625\\
0 -0.15\\
0 -0.1375\\
0 -0.125\\
0 -0.1125\\
0 -0.1\\
0 -0.0875\\
0 -0.075\\
0 -0.0625\\
0 -0.05\\
0 -0.0375\\
0 -0.025\\
0 -0.0125\\
0 3.90313e-17\\
0 0\\
0 0.0125\\
0 0.025\\
0 0.0375\\
0 0.05\\
0 0.0625\\
0 0.075\\
0 0.0875\\
0 0.1\\
0 0.1125\\
0 0.125\\
0 0.1375\\
0 0.15\\
0 0.1625\\
0 0.175\\
0 0.1875\\
0 0.2\\
0 0.2125\\
0 0.225\\
0 0.2375\\
0 0.25\\
0 0.25\\
0 0.2625\\
0 0.275\\
0 0.2875\\
0 0.3\\
0 0.3125\\
0 0.325\\
0 0.3375\\
0 0.35\\
0 0.3625\\
0 0.375\\
0 0.3875\\
0 0.4\\
0 0.4125\\
0 0.425\\
0 0.4375\\
0 0.45\\
0 0.4625\\
0 0.475\\
0 0.4875\\
0 0.5\\
};
\addlegendentry{FSDT5};

\end{axis}
\end{tikzpicture}%
}}
\caption{The variation of non-dimensional deflections and stresses through the thickness of a four-layered anti-symmetric curved fibre composite laminate under sinusoidally distributed thermal load with $a/h=$ 5, $T_o=$ -45$^\circ$, $T_1=$ 45$^\circ$.}
\label{fig:4LayerThermLoad}
\end{figure}

Similar study is made considering four-layered anti-symmetric laminates and the maximum displacements and stresses evaluated considering both mechanical and thermal loads are presented in Tables \ref{table:4LmechAntiSymm} - \ref{table:4LthermAntiSymm}. The displacement and stress plots through the thickness of the laminate are depicted in \frefs{fig:4LayerMechLoad} - \ref{fig:4LayerThermLoad} for mechanical and thermal loads, respectively. The behaviour of different structural models are qualitatively similar to those of symmetric case. However, from \frefs{fig:3LayerMechLoad} - \ref{fig:4LayerThermLoad}, it is seen that the variation of stresses through the thickness predicted by lower-order theories are very much quantitatively different from those of symmetric case.

\begin{table}[htpb]
\centering
\caption{Maximum non-dimensionalized deflections and stresses for a four-layered anti-symmetric curvilinear fibre composite panels $( \langle -45^\circ,45^\circ \rangle, \langle 45^\circ,-45^\circ \rangle, \langle -45^\circ,45^\circ \rangle, \langle 45^\circ,-45^\circ \rangle)$ under sinusoidally distributed mechanical load.}
\begin{tabular}{crrrrr}
\hline
 $a/h$ & Structural & $\overline{u}(-a/2,0,h/2)$ & $\overline{w}(0,0,h/2)$ & $\overline{\sigma}_{xx}(0,0,h/2)$ & $\overline{\tau}_{xz}(-a/2,0,0)$ \\
\hline
\multirow{5}{*}{5} & HSDT13 & -1.006 & 1.338 & 0.316 & -0.068 \\
& HSDT11A & -0.998 & 1.317 & 0.307 & -0.075 \\
& HSDT11B & -0.983 & 1.299 & 0.305 & -0.051 \\
& TSDT7 & -0.737 & 1.104 & 0.174 & 0.027 \\
& FSDT5& -0.565 & 1.104 & 0.279 & 0.122 \\
\cline{2-6}
\multirow{5}{*}{10} & HSDT13 & -0.727 & 0.667 & 0.286 & 0.013 \\
& HSDT11A & -0.728 & 0.671 & 0.285 & 0.010 \\
& HSDT11B & -0.720 & 0.647 & 0.277 & 0.032 \\
& TSDT7 & -0.653 & 0.600 & 0.237 & 0.068 \\
& FSDT5& -0.578 & 0.574 & 0.276 & 0.171 \\
\cline{2-6}
\multirow{5}{*}{50} & HSDT13 & -0.561 & 0.398 & 0.250 & 0.164 \\
& HSDT11A & -0.561 & 0.399 & 0.250 & 0.162 \\
& HSDT11B & -0.560 & 0.397 & 0.250 & 0.168 \\
& TSDT7 & -0.557 & 0.394 & 0.249 & 0.165 \\
& FSDT5& -0.531 & 0.381 & 0.257 & 0.311 \\
\hline
\end{tabular}
\label{table:4LmechAntiSymm}
\end{table}

\begin{table}[htpb]
\centering
\caption{Maximum non-dimensionalized deflections and stresses for a four-layered anti-symmetric curvilinear fibre composite panels $( \langle -45^\circ,45^\circ \rangle, \langle 45^\circ,-45^\circ \rangle, \langle -45^\circ,45^\circ \rangle, \langle 45^\circ,-45^\circ \rangle)$ under sinusoidally distributed thermal load.}
\begin{tabular}{crrrrr}
\hline
 $a/h$ & Structural & $\overline{u}(-a/2,0,h/2)$ & $\overline{w}(0,0,h/2)$ & $\overline{\sigma}_{xx}(0,0,h/2)$ & $\overline{\tau}_{xz}(-a/2,0,-h/6)$ \\
\hline
\multirow{5}{*}{5} & HSDT13 & -18.956 & 28.981 & -166.942 & -136.980 \\
& HSDT11A & -16.059 & 18.141 & -279.092 & -124.888 \\
& HSDT11B & -19.015 & 28.861 & -169.577 & -128.414 \\
& TSDT7 & -13.978 & 15.957 & -509.318 & -104.787 \\
& FSDT5& -7.079 & 11.576 & -25.160 & -52.125 \\
\cline{2-6}
\multirow{5}{*}{10} & HSDT13 & -16.698 & 16.368 & -61.095 & -58.213 \\
& HSDT11A & -15.916 & 13.610 & -90.983 & -58.089 \\
& HSDT11B & -16.747 & 16.178 & -76.937 & -60.139 \\
& TSDT7 & -15.300 & 12.717 & -193.895 & -54.290 \\
& FSDT5& -12.498 & 10.791 & -0.795 & 28.503 \\
\cline{2-6}
\multirow{5}{*}{50} & HSDT13 & -14.283 & 10.432 & -75.392 & -6.851 \\
& HSDT11A & -14.255 & 10.325 & -76.578 & -6.977 \\
& HSDT11B & -14.266 & 10.398 & -75.355 & -6.955 \\
& TSDT7 & -14.165 & 10.232 & -73.767 & -7.103 \\
& FSDT5& -13.499 & 9.851 & -41.047 & -0.448 \\ 
\hline
\end{tabular}
\label{table:4LthermAntiSymm}
\end{table}

\section{Conclusion}
The performance of various structural models in predicting the static response of curvilinear fibre composite laminates is examined considering various parameters such as ply-angle, lay-up sequence, plate side to thickness ratio and loadings. From a detailed investigation, the following observations can be made:
\begin{itemize}
\item HSDT11A predicts accurately the in-plane and transverse deflection of the curvilinear laminates subjected to mechanical loading, whereas, for thermal loading, HSDT11B model yields accurate solutions in comparison with those of the full structural model considered here, viz., HSDT13.
\item The choice of HSDT11A or HSDT11B in evaluating the stress values and the variation of the stresses through the thickness highly depend on the plate side to thickness ratio and the curvilinear ply angle.
\item The performance of higher-order models for anti-symmetric lay-up case, in general, significantly noticeable over the lower-order models.
\item In-plane relative displacement has slope discontinuity at the layer interface whereas the transverse displacement varies quadratically through the thickness.
\item Higher-order model HSDT13 may be required for accurate prediction of displacements and stress variation though the thickness, irrespective of loading situation, ply-angle and lay-up sequence, in particular, for curvilinear case.
\end{itemize}

\section*{Acknowledgement}
S Natarajan would like to acknowledge the financial support of the School of Civil and Environmental Engineering, The University of New South Wales, for his research fellowship for the period September 2012 onwards. 

\section*{References}
\bibliographystyle{elsarticle-num}
\bibliography{curvifiber}

\begin{thebibliography}{10}
\expandafter\ifx\csname url\endcsname\relax
  \def\url#1{\texttt{#1}}\fi
\expandafter\ifx\csname urlprefix\endcsname\relax\def\urlprefix{URL }\fi
\expandafter\ifx\csname href\endcsname\relax
  \def\href#1#2{#2} \def\path#1{#1}\fi

\bibitem{botelhosilva2006}
E.~C. Botelho, R.~A. Silva, L.~C. Pardini, M.~C. Rezende, A review on the
  development and properties of continuous fibre/epoxy/aluminum hybrid
  composites for aircraft structures, Materials Research 9 (2006) 247--256.

\bibitem{hyerlee1991}
M.~Hyer, H.~Lee, The use of curvilinear fiber format to improve buckling
  resistance of composite plates with central circular holes, Composite
  Structures 18 (1991) 239--261.

\bibitem{setoodehabdalla2008}
S.~Setoodeh, M.~Abdalla, S.~Ijsselmuiden, Z.~G\"urdal, Design of variable
  stiffness composite panels for maximum buckling load, Composite Structures 87
  (2008) 109--117.

\bibitem{lopesgurdal2010}
C.~Lopes, Z.~G\"urdal, P.~Camanho, Tailoring for strength of composite steered
  fibre panels with cutouts, Composites Part A: Applied Science and
  Manufacturing 41 (2010) 1760--1767.

\bibitem{khaniijsselmuiden2011}
A.~Khani, S.~IJsselmuiden, M.~Abdalla, Z.~G\"urdal, Design of variable
  stiffness panels for maximum strength using lamination parameters, Composites
  Part B: Engineering 42 (2011) 546--552.

\bibitem{muruganflores2012}
S.~Murugan, E.~I.~S. Flores, S.~Adhikari, M.~Friswell, Optimal design of
  variable fiber spacing composites for morphing aircraft skins, Composite
  Structures 94 (2012) 1626--1633.

\bibitem{mukherjeevarughese2001}
A.~Mukherjee, B.~Varughese, Design guidelines for ply drop-off in laminated
  composite structures, Composites Part B: Engineering 32 (2001) 153--164.

\bibitem{mejdiatalla2012}
A.~Mejdi, N.~Atalla, Vibroacoustic analyses of laminated composite panels
  stiffened by complex laminated composite stiffeners, International Journal of
  Mechanical Sciences 58 (2012) 13--26.

\bibitem{gurdalolmedo1993}
Z.~G\"urdal, R.~Olmedo, In-plane response of laminates with spatially varying
  fiber orientations: variable stiffness concept, AIAA Journal 31 (1993)
  751--758.

\bibitem{abdallagurdal2009}
M.~Abdalla, Z.~G\"urdal, G.~Abdelal, Thermomechanical response of variable
  stiffness composite panels, Journal of Thermal Stress 32 (2009) 187--208.

\bibitem{akhavanribeiro2011}
H.~Akhavan, P.~Ribeiro, Natural modes of vibration of variable stiffness
  composite laminates with curvilinear fibers, Composite Structures 93 (2011)
  3040--3047.

\bibitem{hondanarita2011}
S.~Honda, Y.~Narita, Vibration design of laminated fibrous composite plates
  with local anisotropy induced by short fibers and curvilinear fibers,
  Composite Structures 93 (2011) 902--910.

\bibitem{houmat2013}
A.~Houmat, Nonlinear free vibration of laminated composite rectangular plates
  with curvilinear fibres, Composite Structures 106 (2013) 211--224.

\bibitem{grohweaver2014}
R.~Groh, P.~Weaver, Buckling analysis of variable angle tow, variable thickness
  panels with transverse shear effects, Composite Structures 107 (2014)
  482--493.

\bibitem{rajuwu2012}
G.~Raju, Z.~Wu, B.~C. Kim, P.~Weaver, Prebuckling and buckling analysis of
  variable angle tow plates with general boundary conditions, Composite
  Structures 94 (2012) 2961--2970.

\bibitem{akhavanribeiro2013}
H.~Akhavan, P.~Ribeiro, M.~de~Moura, Large deflection and stresses in variable
  stiffness composite laminates with curvilinear fibers, International Journal
  of Mechanical Sciences 73 (2013) 14--26.

\bibitem{kimpotter2012}
B.~C. Kim, K.~Potter, P.~Weaver, Continuous tow shearing for manufacturing
  variable angle tow composites, Composites Part A: Applied Science and
  Manufacturing 43 (2012) 1347--1356.

\bibitem{kimweaver2014}
B.~C. Kim, P.~Weaver, K.~Potter, Manufacturing characteristics of the
  continuous tow shearing method for manufacturing of variable angle tow
  composites, Composites Part A: Applied Science and Manufacturing.

\bibitem{falcomayugo2014}
O.~Falc\'o, J.~Mayugo, C.~Lopes, N.~Gascons, A.~Turon, J.~Costa, {Varaible
  stiffness composite panels: As-manufactured modeling and its influence on the
  failure behavior}, Composites Part B: Engineering 56 (2014) 660--669.

\bibitem{hondanarita2012}
S.~Honda, Y.~Narita, Natural frequencies and vibration modes of laminated
  composite plates reinforced with arbitrary curvilinear fiber shape paths,
  Journal of Sound and Vibration 331 (2012) 180--191.

\bibitem{rajuwu2013}
G.~Raju, Z.~Wu, P.~Weaver, Postbuckling analysis of varaible angle tow plates
  using differential quadrature method, Composite Structures 106 (2013) 74--84.

\bibitem{grohweaver2013}
R.~Groh, P.~Weaver, S.~White, G.~Raju, Z.~Wu, {A 2D equivalent single-layer
  formulation for the effect of transverse shear on laminated plates with
  curvilinear fibres}, Composite Structures 100 (2013) 464--478.

\bibitem{akhavanribeiro2013a}
H.~Akhavan, P.~Ribeiro, M.~de~Moura, Composites laminates with linear varying
  fiber angles under static and dynamic loads, in: Structures, Structural
  Dynamics and Materials and Co-located conferences, 2013.

\bibitem{murukami1986}
H.~Murukami, Laminated composite plate theory with improved in-plane responses,
  J Appl Mech 53 (1986) 661--666.

\bibitem{alibhaskar1999}
J.~Ali, K.~Bhaskar, T.~Varadan, A new theory for accurate thermal/mechanical
  flexural analysis of symmetrically laminated plates, Composite Structures 45
  (1999) 227--232.

\bibitem{makhechaganapathi2001}
D.~Makhecha, M.~Ganapathi, B.~Patel, Dynamic analysis of laminated composite
  plates subjected to thermal/mechanical loads using an accurate theory,
  Composite Structures 51 (2001) 221--236.

\bibitem{carrera2004}
E.~Carrera, {On the use of the Murakami's zig-zag function in the modeling of
  layered plates and shells}, Computers \& Structures 82 (2004) 541--554.

\bibitem{carrera2003}
E.~Carrera, Theories and finite elements for multilayered plates and shells: a
  unified compact formulation with numerical assessment and benchmarking, Arch
  Comput Meth Eng 10 (2003) 215--296.

\bibitem{waldhar1996}
C.~Waldhar, Analysis of tow-placed variable-stiffness laminates, Master's
  thesis, Virginia Tech, Blacksburg (1996).

\bibitem{roderiguesroque2011}
J.~Roderigues, C.~Roque, A.~Ferreira, E.~Carrera, M.~Cinfera, {Radial basis
  functions-finite differences collocation and a unified formulation for
  bending, vibration and buckling analysis of laminated plates according to
  Murakami's zig-zag theory}, Composite Structures 93 (2011) 1613--1620.

\bibitem{nosierkapania1993}
A.~Nosier, R.~Kapania, J.~Reddy, Free vibration analysis of laminated plates
  using a layerwise theory, AIAA Journal 31 (1993) 2335--2346.

\bibitem{ferreira2005}
A.~Ferreira, Analysis of composite plates using a layerwise deformation theory
  and multiquadratics discretization, Mech. Adv. Mater. Struct 12 (2005)
  99--112.

\bibitem{hughes2000}
T.~J.~R. Hughes, {The finite element method: Linear static and Dynamic finite
  element analysis}, Dover Publications, 2000.

\bibitem{bhaskarvaradan1996}
K.~Bhaskar, T.~Varadan, J.~Ali, Thermoelastic solutions for orthotropic and
  anisotropic composite laminates, Composites Part B: Engineering 27 (1996)
  415--420.

\end{thebibliography}

\end{document}